\pdfoutput=1
\documentclass[nofootinbib,aps]{revtex4}
\usepackage{amssymb,amsmath,graphicx,color,microtype}
\usepackage{epsf}
\usepackage{epstopdf}
\usepackage {amssymb}
\usepackage{color}
\newcommand{\nc}{\newcommand}
\nc{\ba}{\begin{eqnarray}}
\nc{\ea}{\end{eqnarray}}
\newcommand\be{\begin{equation}}
\newcommand\ee{\end{equation}}

\definecolor{darkgreen}{rgb}{0,0.5423,0.1}
\definecolor{darkblue}{rgb}{0,0.3,0.943}
\definecolor{darkred}{rgb}{0.6,0,0}

\definecolor{AliceBlue}{rgb}{0.94,0.97,1.00}
\definecolor{AntiqueWhite1}{rgb}{1.00,0.94,0.86}
\definecolor{AntiqueWhite2}{rgb}{0.93,0.87,0.80}
\definecolor{AntiqueWhite3}{rgb}{0.80,0.75,0.69}
\definecolor{AntiqueWhite4}{rgb}{0.55,0.51,0.47}
\definecolor{AntiqueWhite}{rgb}{0.98,0.92,0.84}
\definecolor{BlanchedAlmond}{rgb}{1.00,0.92,0.80}
\definecolor{BlueViolet}{rgb}{0.54,0.17,0.89}
\definecolor{CadetBlue1}{rgb}{0.60,0.96,1.00}
\definecolor{CadetBlue2}{rgb}{0.56,0.90,0.93}
\definecolor{CadetBlue3}{rgb}{0.48,0.77,0.80}
\definecolor{CadetBlue4}{rgb}{0.33,0.53,0.55}
\definecolor{CadetBlue}{rgb}{0.37,0.62,0.63}
\definecolor{CornflowerBlue}{rgb}{0.39,0.58,0.93}
\definecolor{DarkBlue}{rgb}{0.00,0.00,0.55}
\definecolor{DarkCyan}{rgb}{0.00,0.55,0.55}
\definecolor{DarkGoldenrod1}{rgb}{1.00,0.73,0.06}
\definecolor{DarkGoldenrod2}{rgb}{0.93,0.68,0.05}
\definecolor{DarkGoldenrod3}{rgb}{0.80,0.58,0.05}
\definecolor{DarkGoldenrod4}{rgb}{0.55,0.40,0.03}
\definecolor{DarkGoldenrod}{rgb}{0.72,0.53,0.04}
\definecolor{DarkGray}{rgb}{0.66,0.66,0.66}
\definecolor{DarkGreen}{rgb}{0.00,0.39,0.00}
\definecolor{DarkGrey}{rgb}{0.66,0.66,0.66}
\definecolor{DarkKhaki}{rgb}{0.74,0.72,0.42}
\definecolor{DarkMagenta}{rgb}{0.55,0.00,0.55}
\definecolor{DarkOliveGreen1}{rgb}{0.79,1.00,0.44}
\definecolor{DarkOliveGreen2}{rgb}{0.74,0.93,0.41}
\definecolor{DarkOliveGreen3}{rgb}{0.64,0.80,0.35}
\definecolor{DarkOliveGreen4}{rgb}{0.43,0.55,0.24}
\definecolor{DarkOliveGreen}{rgb}{0.33,0.42,0.18}
\definecolor{DarkOrange1}{rgb}{1.00,0.50,0.00}
\definecolor{DarkOrange2}{rgb}{0.93,0.46,0.00}
\definecolor{DarkOrange3}{rgb}{0.80,0.40,0.00}
\definecolor{DarkOrange4}{rgb}{0.55,0.27,0.00}
\definecolor{DarkOrange}{rgb}{1.00,0.55,0.00}
\definecolor{DarkOrchid1}{rgb}{0.75,0.24,1.00}
\definecolor{DarkOrchid2}{rgb}{0.70,0.23,0.93}
\definecolor{DarkOrchid3}{rgb}{0.60,0.20,0.80}
\definecolor{DarkOrchid4}{rgb}{0.41,0.13,0.55}
\definecolor{DarkOrchid}{rgb}{0.60,0.20,0.80}
\definecolor{DarkRed}{rgb}{0.55,0.00,0.00}
\definecolor{DarkSalmon}{rgb}{0.91,0.59,0.48}
\definecolor{DarkSeaGreen1}{rgb}{0.76,1.00,0.76}
\definecolor{DarkSeaGreen2}{rgb}{0.71,0.93,0.71}
\definecolor{DarkSeaGreen3}{rgb}{0.61,0.80,0.61}
\definecolor{DarkSeaGreen4}{rgb}{0.41,0.55,0.41}
\definecolor{DarkSeaGreen}{rgb}{0.56,0.74,0.56}
\definecolor{DarkSlateBlue}{rgb}{0.28,0.24,0.55}
\definecolor{DarkSlateGray1}{rgb}{0.59,1.00,1.00}
\definecolor{DarkSlateGray2}{rgb}{0.55,0.93,0.93}
\definecolor{DarkSlateGray3}{rgb}{0.47,0.80,0.80}
\definecolor{DarkSlateGray4}{rgb}{0.32,0.55,0.55}
\definecolor{DarkSlateGray}{rgb}{0.18,0.31,0.31}
\definecolor{DarkSlateGrey}{rgb}{0.18,0.31,0.31}
\definecolor{DarkTurquoise}{rgb}{0.00,0.81,0.82}
\definecolor{DarkViolet}{rgb}{0.58,0.00,0.83}
\definecolor{DeepPink1}{rgb}{1.00,0.08,0.58}
\definecolor{DeepPink2}{rgb}{0.93,0.07,0.54}
\definecolor{DeepPink3}{rgb}{0.80,0.06,0.46}
\definecolor{DeepPink4}{rgb}{0.55,0.04,0.31}
\definecolor{DeepPink}{rgb}{1.00,0.08,0.58}
\definecolor{DeepSkyBlue1}{rgb}{0.00,0.75,1.00}
\definecolor{DeepSkyBlue2}{rgb}{0.00,0.70,0.93}
\definecolor{DeepSkyBlue3}{rgb}{0.00,0.60,0.80}
\definecolor{DeepSkyBlue4}{rgb}{0.00,0.41,0.55}
\definecolor{DeepSkyBlue}{rgb}{0.00,0.75,1.00}
\definecolor{DimGray}{rgb}{0.41,0.41,0.41}
\definecolor{DimGrey}{rgb}{0.41,0.41,0.41}
\definecolor{DodgerBlue1}{rgb}{0.12,0.56,1.00}
\definecolor{DodgerBlue2}{rgb}{0.11,0.53,0.93}
\definecolor{DodgerBlue3}{rgb}{0.09,0.45,0.80}
\definecolor{DodgerBlue4}{rgb}{0.06,0.31,0.55}
\definecolor{DodgerBlue}{rgb}{0.12,0.56,1.00}
\definecolor{FloralWhite}{rgb}{1.00,0.98,0.94}
\definecolor{ForestGreen}{rgb}{0.13,0.55,0.13}
\definecolor{GhostWhite}{rgb}{0.97,0.97,1.00}
\definecolor{GreenYellow}{rgb}{0.68,1.00,0.18}
\definecolor{HotPink1}{rgb}{1.00,0.43,0.71}
\definecolor{HotPink2}{rgb}{0.93,0.42,0.65}
\definecolor{HotPink3}{rgb}{0.80,0.38,0.56}
\definecolor{HotPink4}{rgb}{0.55,0.23,0.38}
\definecolor{HotPink}{rgb}{1.00,0.41,0.71}
\definecolor{IndianRed1}{rgb}{1.00,0.42,0.42}
\definecolor{IndianRed2}{rgb}{0.93,0.39,0.39}
\definecolor{IndianRed3}{rgb}{0.80,0.33,0.33}
\definecolor{IndianRed4}{rgb}{0.55,0.23,0.23}
\definecolor{IndianRed}{rgb}{0.80,0.36,0.36}
\definecolor{LavenderBlush1}{rgb}{1.00,0.94,0.96}
\definecolor{LavenderBlush2}{rgb}{0.93,0.88,0.90}
\definecolor{LavenderBlush3}{rgb}{0.80,0.76,0.77}
\definecolor{LavenderBlush4}{rgb}{0.55,0.51,0.53}
\definecolor{LavenderBlush}{rgb}{1.00,0.94,0.96}
\definecolor{LawnGreen}{rgb}{0.49,0.99,0.00}
\definecolor{LemonChiffon1}{rgb}{1.00,0.98,0.80}
\definecolor{LemonChiffon2}{rgb}{0.93,0.91,0.75}
\definecolor{LemonChiffon3}{rgb}{0.80,0.79,0.65}
\definecolor{LemonChiffon4}{rgb}{0.55,0.54,0.44}
\definecolor{LemonChiffon}{rgb}{1.00,0.98,0.80}
\definecolor{LightBlue1}{rgb}{0.75,0.94,1.00}
\definecolor{LightBlue2}{rgb}{0.70,0.87,0.93}
\definecolor{LightBlue3}{rgb}{0.60,0.75,0.80}
\definecolor{LightBlue4}{rgb}{0.41,0.51,0.55}
\definecolor{LightBlue}{rgb}{0.68,0.85,0.90}
\definecolor{LightCoral}{rgb}{0.94,0.50,0.50}
\definecolor{LightCyan1}{rgb}{0.88,1.00,1.00}
\definecolor{LightCyan2}{rgb}{0.82,0.93,0.93}
\definecolor{LightCyan3}{rgb}{0.71,0.80,0.80}
\definecolor{LightCyan4}{rgb}{0.48,0.55,0.55}
\definecolor{LightCyan}{rgb}{0.88,1.00,1.00}
\definecolor{LightGoldenrod1}{rgb}{1.00,0.93,0.55}
\definecolor{LightGoldenrod2}{rgb}{0.93,0.86,0.51}
\definecolor{LightGoldenrod3}{rgb}{0.80,0.75,0.44}
\definecolor{LightGoldenrod4}{rgb}{0.55,0.51,0.30}
\definecolor{LightGoldenrodYellow}{rgb}{0.98,0.98,0.82}
\definecolor{LightGoldenrod}{rgb}{0.93,0.87,0.51}
\definecolor{LightGray}{rgb}{0.83,0.83,0.83}
\definecolor{LightGreen}{rgb}{0.56,0.93,0.56}
\definecolor{LightGrey}{rgb}{0.83,0.83,0.83}
\definecolor{LightPink1}{rgb}{1.00,0.68,0.73}
\definecolor{LightPink2}{rgb}{0.93,0.64,0.68}
\definecolor{LightPink3}{rgb}{0.80,0.55,0.58}
\definecolor{LightPink4}{rgb}{0.55,0.37,0.40}
\definecolor{LightPink}{rgb}{1.00,0.71,0.76}
\definecolor{LightSalmon1}{rgb}{1.00,0.63,0.48}
\definecolor{LightSalmon2}{rgb}{0.93,0.58,0.45}
\definecolor{LightSalmon3}{rgb}{0.80,0.51,0.38}
\definecolor{LightSalmon4}{rgb}{0.55,0.34,0.26}
\definecolor{LightSalmon}{rgb}{1.00,0.63,0.48}
\definecolor{LightSeaGreen}{rgb}{0.13,0.70,0.67}
\definecolor{LightSkyBlue1}{rgb}{0.69,0.89,1.00}
\definecolor{LightSkyBlue2}{rgb}{0.64,0.83,0.93}
\definecolor{LightSkyBlue3}{rgb}{0.55,0.71,0.80}
\definecolor{LightSkyBlue4}{rgb}{0.38,0.48,0.55}
\definecolor{LightSkyBlue}{rgb}{0.53,0.81,0.98}
\definecolor{LightSlateBlue}{rgb}{0.52,0.44,1.00}
\definecolor{LightSlateGray}{rgb}{0.47,0.53,0.60}
\definecolor{LightSlateGrey}{rgb}{0.47,0.53,0.60}
\definecolor{LightSteelBlue1}{rgb}{0.79,0.88,1.00}
\definecolor{LightSteelBlue2}{rgb}{0.74,0.82,0.93}
\definecolor{LightSteelBlue3}{rgb}{0.64,0.71,0.80}
\definecolor{LightSteelBlue4}{rgb}{0.43,0.48,0.55}
\definecolor{LightSteelBlue}{rgb}{0.69,0.77,0.87}
\definecolor{LightYellow1}{rgb}{1.00,1.00,0.88}
\definecolor{LightYellow2}{rgb}{0.93,0.93,0.82}
\definecolor{LightYellow3}{rgb}{0.80,0.80,0.71}
\definecolor{LightYellow4}{rgb}{0.55,0.55,0.48}
\definecolor{LightYellow}{rgb}{1.00,1.00,0.88}
\definecolor{LimeGreen}{rgb}{0.20,0.80,0.20}
\definecolor{MediumAquamarine}{rgb}{0.40,0.80,0.67}
\definecolor{MediumBlue}{rgb}{0.00,0.00,0.80}
\definecolor{MediumOrchid1}{rgb}{0.88,0.40,1.00}
\definecolor{MediumOrchid2}{rgb}{0.82,0.37,0.93}
\definecolor{MediumOrchid3}{rgb}{0.71,0.32,0.80}
\definecolor{MediumOrchid4}{rgb}{0.48,0.22,0.55}
\definecolor{MediumOrchid}{rgb}{0.73,0.33,0.83}
\definecolor{MediumPurple1}{rgb}{0.67,0.51,1.00}
\definecolor{MediumPurple2}{rgb}{0.62,0.47,0.93}
\definecolor{MediumPurple3}{rgb}{0.54,0.41,0.80}
\definecolor{MediumPurple4}{rgb}{0.36,0.28,0.55}
\definecolor{MediumPurple}{rgb}{0.58,0.44,0.86}
\definecolor{MediumSeaGreen}{rgb}{0.24,0.70,0.44}
\definecolor{MediumSlateBlue}{rgb}{0.48,0.41,0.93}
\definecolor{MediumSpringGreen}{rgb}{0.00,0.98,0.60}
\definecolor{MediumTurquoise}{rgb}{0.28,0.82,0.80}
\definecolor{MediumVioletRed}{rgb}{0.78,0.08,0.52}
\definecolor{MidnightBlue}{rgb}{0.10,0.10,0.44}
\definecolor{MintCream}{rgb}{0.96,1.00,0.98}
\definecolor{MistyRose1}{rgb}{1.00,0.89,0.88}
\definecolor{MistyRose2}{rgb}{0.93,0.84,0.82}
\definecolor{MistyRose3}{rgb}{0.80,0.72,0.71}
\definecolor{MistyRose4}{rgb}{0.55,0.49,0.48}
\definecolor{MistyRose}{rgb}{1.00,0.89,0.88}
\definecolor{NavajoWhite1}{rgb}{1.00,0.87,0.68}
\definecolor{NavajoWhite2}{rgb}{0.93,0.81,0.63}
\definecolor{NavajoWhite3}{rgb}{0.80,0.70,0.55}
\definecolor{NavajoWhite4}{rgb}{0.55,0.47,0.37}
\definecolor{NavajoWhite}{rgb}{1.00,0.87,0.68}
\definecolor{NavyBlue}{rgb}{0.00,0.00,0.50}
\definecolor{OldLace}{rgb}{0.99,0.96,0.90}
\definecolor{OliveDrab1}{rgb}{0.75,1.00,0.24}
\definecolor{OliveDrab2}{rgb}{0.70,0.93,0.23}
\definecolor{OliveDrab3}{rgb}{0.60,0.80,0.20}
\definecolor{OliveDrab4}{rgb}{0.41,0.55,0.13}
\definecolor{OliveDrab}{rgb}{0.42,0.56,0.14}
\definecolor{OrangeRed1}{rgb}{1.00,0.27,0.00}
\definecolor{OrangeRed2}{rgb}{0.93,0.25,0.00}
\definecolor{OrangeRed3}{rgb}{0.80,0.22,0.00}
\definecolor{OrangeRed4}{rgb}{0.55,0.15,0.00}
\definecolor{OrangeRed}{rgb}{1.00,0.27,0.00}
\definecolor{PaleGoldenrod}{rgb}{0.93,0.91,0.67}
\definecolor{PaleGreen1}{rgb}{0.60,1.00,0.60}
\definecolor{PaleGreen2}{rgb}{0.56,0.93,0.56}
\definecolor{PaleGreen3}{rgb}{0.49,0.80,0.49}
\definecolor{PaleGreen4}{rgb}{0.33,0.55,0.33}
\definecolor{PaleGreen}{rgb}{0.60,0.98,0.60}
\definecolor{PaleTurquoise1}{rgb}{0.73,1.00,1.00}
\definecolor{PaleTurquoise2}{rgb}{0.68,0.93,0.93}
\definecolor{PaleTurquoise3}{rgb}{0.59,0.80,0.80}
\definecolor{PaleTurquoise4}{rgb}{0.40,0.55,0.55}
\definecolor{PaleTurquoise}{rgb}{0.69,0.93,0.93}
\definecolor{PaleVioletRed1}{rgb}{1.00,0.51,0.67}
\definecolor{PaleVioletRed2}{rgb}{0.93,0.47,0.62}
\definecolor{PaleVioletRed3}{rgb}{0.80,0.41,0.54}
\definecolor{PaleVioletRed4}{rgb}{0.55,0.28,0.36}
\definecolor{PaleVioletRed}{rgb}{0.86,0.44,0.58}
\definecolor{PapayaWhip}{rgb}{1.00,0.94,0.84}
\definecolor{PeachPuff1}{rgb}{1.00,0.85,0.73}
\definecolor{PeachPuff2}{rgb}{0.93,0.80,0.68}
\definecolor{PeachPuff3}{rgb}{0.80,0.69,0.58}
\definecolor{PeachPuff4}{rgb}{0.55,0.47,0.40}
\definecolor{PeachPuff}{rgb}{1.00,0.85,0.73}
\definecolor{PowderBlue}{rgb}{0.69,0.88,0.90}
\definecolor{RosyBrown1}{rgb}{1.00,0.76,0.76}
\definecolor{RosyBrown2}{rgb}{0.93,0.71,0.71}
\definecolor{RosyBrown3}{rgb}{0.80,0.61,0.61}
\definecolor{RosyBrown4}{rgb}{0.55,0.41,0.41}
\definecolor{RosyBrown}{rgb}{0.74,0.56,0.56}
\definecolor{RoyalBlue1}{rgb}{0.28,0.46,1.00}
\definecolor{RoyalBlue2}{rgb}{0.26,0.43,0.93}
\definecolor{RoyalBlue3}{rgb}{0.23,0.37,0.80}
\definecolor{RoyalBlue4}{rgb}{0.15,0.25,0.55}
\definecolor{RoyalBlue}{rgb}{0.25,0.41,0.88}
\definecolor{SaddleBrown}{rgb}{0.55,0.27,0.07}
\definecolor{SandyBrown}{rgb}{0.96,0.64,0.38}
\definecolor{SeaGreen1}{rgb}{0.33,1.00,0.62}
\definecolor{SeaGreen2}{rgb}{0.31,0.93,0.58}
\definecolor{SeaGreen3}{rgb}{0.26,0.80,0.50}
\definecolor{SeaGreen4}{rgb}{0.18,0.55,0.34}
\definecolor{SeaGreen}{rgb}{0.18,0.55,0.34}
\definecolor{SkyBlue1}{rgb}{0.53,0.81,1.00}
\definecolor{SkyBlue2}{rgb}{0.49,0.75,0.93}
\definecolor{SkyBlue3}{rgb}{0.42,0.65,0.80}
\definecolor{SkyBlue4}{rgb}{0.29,0.44,0.55}
\definecolor{SkyBlue}{rgb}{0.53,0.81,0.92}
\definecolor{SlateBlue1}{rgb}{0.51,0.44,1.00}
\definecolor{SlateBlue2}{rgb}{0.48,0.40,0.93}
\definecolor{SlateBlue3}{rgb}{0.41,0.35,0.80}
\definecolor{SlateBlue4}{rgb}{0.28,0.24,0.55}
\definecolor{SlateBlue}{rgb}{0.42,0.35,0.80}
\definecolor{SlateGray1}{rgb}{0.78,0.89,1.00}
\definecolor{SlateGray2}{rgb}{0.73,0.83,0.93}
\definecolor{SlateGray3}{rgb}{0.62,0.71,0.80}
\definecolor{SlateGray4}{rgb}{0.42,0.48,0.55}
\definecolor{SlateGray}{rgb}{0.44,0.50,0.56}
\definecolor{SlateGrey}{rgb}{0.44,0.50,0.56}
\definecolor{SpringGreen1}{rgb}{0.00,1.00,0.50}
\definecolor{SpringGreen2}{rgb}{0.00,0.93,0.46}
\definecolor{SpringGreen3}{rgb}{0.00,0.80,0.40}
\definecolor{SpringGreen4}{rgb}{0.00,0.55,0.27}
\definecolor{SpringGreen}{rgb}{0.00,1.00,0.50}
\definecolor{SteelBlue1}{rgb}{0.39,0.72,1.00}
\definecolor{SteelBlue2}{rgb}{0.36,0.67,0.93}
\definecolor{SteelBlue3}{rgb}{0.31,0.58,0.80}
\definecolor{SteelBlue4}{rgb}{0.21,0.39,0.55}
\definecolor{SteelBlue}{rgb}{0.27,0.51,0.71}
\definecolor{VioletRed1}{rgb}{1.00,0.24,0.59}
\definecolor{VioletRed2}{rgb}{0.93,0.23,0.55}
\definecolor{VioletRed3}{rgb}{0.80,0.20,0.47}
\definecolor{VioletRed4}{rgb}{0.55,0.13,0.32}
\definecolor{VioletRed}{rgb}{0.82,0.13,0.56}
\definecolor{WhiteSmoke}{rgb}{0.96,0.96,0.96}
\definecolor{YellowGreen}{rgb}{0.60,0.80,0.20}
\definecolor{aliceblue}{rgb}{0.94,0.97,1.00}
\definecolor{antiquewhite}{rgb}{0.98,0.92,0.84}
\definecolor{aquamarine1}{rgb}{0.50,1.00,0.83}
\definecolor{aquamarine2}{rgb}{0.46,0.93,0.78}
\definecolor{aquamarine3}{rgb}{0.40,0.80,0.67}
\definecolor{aquamarine4}{rgb}{0.27,0.55,0.45}
\definecolor{aquamarine}{rgb}{0.50,1.00,0.83}
\definecolor{azure1}{rgb}{0.94,1.00,1.00}
\definecolor{azure2}{rgb}{0.88,0.93,0.93}
\definecolor{azure3}{rgb}{0.76,0.80,0.80}
\definecolor{azure4}{rgb}{0.51,0.55,0.55}
\definecolor{azure}{rgb}{0.94,1.00,1.00}
\definecolor{beige}{rgb}{0.96,0.96,0.86}
\definecolor{bisque1}{rgb}{1.00,0.89,0.77}
\definecolor{bisque2}{rgb}{0.93,0.84,0.72}
\definecolor{bisque3}{rgb}{0.80,0.72,0.62}
\definecolor{bisque4}{rgb}{0.55,0.49,0.42}
\definecolor{bisque}{rgb}{1.00,0.89,0.77}
\definecolor{black}{rgb}{0.00,0.00,0.00}
\definecolor{blanchedalmond}{rgb}{1.00,0.92,0.80}
\definecolor{blue1}{rgb}{0.00,0.00,1.00}
\definecolor{blue2}{rgb}{0.00,0.00,0.93}
\definecolor{blue3}{rgb}{0.00,0.00,0.80}
\definecolor{blue4}{rgb}{0.00,0.00,0.55}
\definecolor{blueviolet}{rgb}{0.54,0.17,0.89}
\definecolor{blue}{rgb}{0.00,0.00,1.00}
\definecolor{brown1}{rgb}{1.00,0.25,0.25}
\definecolor{brown2}{rgb}{0.93,0.23,0.23}
\definecolor{brown3}{rgb}{0.80,0.20,0.20}
\definecolor{brown4}{rgb}{0.55,0.14,0.14}
\definecolor{brown}{rgb}{0.65,0.16,0.16}
\definecolor{burlywood1}{rgb}{1.00,0.83,0.61}
\definecolor{burlywood2}{rgb}{0.93,0.77,0.57}
\definecolor{burlywood3}{rgb}{0.80,0.67,0.49}
\definecolor{burlywood4}{rgb}{0.55,0.45,0.33}
\definecolor{burlywood}{rgb}{0.87,0.72,0.53}
\definecolor{cadetblue}{rgb}{0.37,0.62,0.63}
\definecolor{chartreuse1}{rgb}{0.50,1.00,0.00}
\definecolor{chartreuse2}{rgb}{0.46,0.93,0.00}
\definecolor{chartreuse3}{rgb}{0.40,0.80,0.00}
\definecolor{chartreuse4}{rgb}{0.27,0.55,0.00}
\definecolor{chartreuse}{rgb}{0.50,1.00,0.00}
\definecolor{chocolate1}{rgb}{1.00,0.50,0.14}
\definecolor{chocolate2}{rgb}{0.93,0.46,0.13}
\definecolor{chocolate3}{rgb}{0.80,0.40,0.11}
\definecolor{chocolate4}{rgb}{0.55,0.27,0.07}
\definecolor{chocolate}{rgb}{0.82,0.41,0.12}
\definecolor{coral1}{rgb}{1.00,0.45,0.34}
\definecolor{coral2}{rgb}{0.93,0.42,0.31}
\definecolor{coral3}{rgb}{0.80,0.36,0.27}
\definecolor{coral4}{rgb}{0.55,0.24,0.18}
\definecolor{coral}{rgb}{1.00,0.50,0.31}
\definecolor{cornflowerblue}{rgb}{0.39,0.58,0.93}
\definecolor{cornsilk1}{rgb}{1.00,0.97,0.86}
\definecolor{cornsilk2}{rgb}{0.93,0.91,0.80}
\definecolor{cornsilk3}{rgb}{0.80,0.78,0.69}
\definecolor{cornsilk4}{rgb}{0.55,0.53,0.47}
\definecolor{cornsilk}{rgb}{1.00,0.97,0.86}
\definecolor{cyan1}{rgb}{0.00,1.00,1.00}
\definecolor{cyan2}{rgb}{0.00,0.93,0.93}
\definecolor{cyan3}{rgb}{0.00,0.80,0.80}
\definecolor{cyan4}{rgb}{0.00,0.55,0.55}
\definecolor{cyan}{rgb}{0.00,1.00,1.00}
\definecolor{darkblue}{rgb}{0.00,0.00,0.55}
\definecolor{darkcyan}{rgb}{0.00,0.55,0.55}
\definecolor{darkgoldenrod}{rgb}{0.72,0.53,0.04}
\definecolor{darkgray}{rgb}{0.66,0.66,0.66}
\definecolor{darkgreen}{rgb}{0.00,0.39,0.00}
\definecolor{darkgrey}{rgb}{0.66,0.66,0.66}
\definecolor{darkkhaki}{rgb}{0.74,0.72,0.42}
\definecolor{darkmagenta}{rgb}{0.55,0.00,0.55}
\definecolor{darkolive}{rgb}{0.33,0.42,0.18}
\definecolor{darkorange}{rgb}{1.00,0.55,0.00}
\definecolor{darkorchid}{rgb}{0.60,0.20,0.80}
\definecolor{darkred}{rgb}{0.55,0.00,0.00}
\definecolor{darksalmon}{rgb}{0.91,0.59,0.48}
\definecolor{darksea}{rgb}{0.56,0.74,0.56}
\definecolor{darkslate}{rgb}{0.18,0.31,0.31}
\definecolor{darkslate}{rgb}{0.18,0.31,0.31}
\definecolor{darkslate}{rgb}{0.28,0.24,0.55}
\definecolor{darkturquoise}{rgb}{0.00,0.81,0.82}
\definecolor{darkviolet}{rgb}{0.58,0.00,0.83}
\definecolor{deeppink}{rgb}{1.00,0.08,0.58}
\definecolor{deepsky}{rgb}{0.00,0.75,1.00}
\definecolor{dimgray}{rgb}{0.41,0.41,0.41}
\definecolor{dimgrey}{rgb}{0.41,0.41,0.41}
\definecolor{dodgerblue}{rgb}{0.12,0.56,1.00}
\definecolor{firebrick1}{rgb}{1.00,0.19,0.19}
\definecolor{firebrick2}{rgb}{0.93,0.17,0.17}
\definecolor{firebrick3}{rgb}{0.80,0.15,0.15}
\definecolor{firebrick4}{rgb}{0.55,0.10,0.10}
\definecolor{firebrick}{rgb}{0.70,0.13,0.13}
\definecolor{floralwhite}{rgb}{1.00,0.98,0.94}
\definecolor{forestgreen}{rgb}{0.13,0.55,0.13}
\definecolor{gainsboro}{rgb}{0.86,0.86,0.86}
\definecolor{ghostwhite}{rgb}{0.97,0.97,1.00}
\definecolor{gold1}{rgb}{1.00,0.84,0.00}
\definecolor{gold2}{rgb}{0.93,0.79,0.00}
\definecolor{gold3}{rgb}{0.80,0.68,0.00}
\definecolor{gold4}{rgb}{0.55,0.46,0.00}
\definecolor{goldenrod1}{rgb}{1.00,0.76,0.15}
\definecolor{goldenrod2}{rgb}{0.93,0.71,0.13}
\definecolor{goldenrod3}{rgb}{0.80,0.61,0.11}
\definecolor{goldenrod4}{rgb}{0.55,0.41,0.08}
\definecolor{goldenrod}{rgb}{0.85,0.65,0.13}
\definecolor{gold}{rgb}{1.00,0.84,0.00}
\definecolor{gray0}{rgb}{0.00,0.00,0.00}
\definecolor{gray100}{rgb}{1.00,1.00,1.00}
\definecolor{gray10}{rgb}{0.10,0.10,0.10}
\definecolor{gray11}{rgb}{0.11,0.11,0.11}
\definecolor{gray12}{rgb}{0.12,0.12,0.12}
\definecolor{gray13}{rgb}{0.13,0.13,0.13}
\definecolor{gray14}{rgb}{0.14,0.14,0.14}
\definecolor{gray15}{rgb}{0.15,0.15,0.15}
\definecolor{gray16}{rgb}{0.16,0.16,0.16}
\definecolor{gray17}{rgb}{0.17,0.17,0.17}
\definecolor{gray18}{rgb}{0.18,0.18,0.18}
\definecolor{gray19}{rgb}{0.19,0.19,0.19}
\definecolor{gray1}{rgb}{0.01,0.01,0.01}
\definecolor{gray20}{rgb}{0.20,0.20,0.20}
\definecolor{gray21}{rgb}{0.21,0.21,0.21}
\definecolor{gray22}{rgb}{0.22,0.22,0.22}
\definecolor{gray23}{rgb}{0.23,0.23,0.23}
\definecolor{gray24}{rgb}{0.24,0.24,0.24}
\definecolor{gray25}{rgb}{0.25,0.25,0.25}
\definecolor{gray26}{rgb}{0.26,0.26,0.26}
\definecolor{gray27}{rgb}{0.27,0.27,0.27}
\definecolor{gray28}{rgb}{0.28,0.28,0.28}
\definecolor{gray29}{rgb}{0.29,0.29,0.29}
\definecolor{gray2}{rgb}{0.02,0.02,0.02}
\definecolor{gray30}{rgb}{0.30,0.30,0.30}
\definecolor{gray31}{rgb}{0.31,0.31,0.31}
\definecolor{gray32}{rgb}{0.32,0.32,0.32}
\definecolor{gray33}{rgb}{0.33,0.33,0.33}
\definecolor{gray34}{rgb}{0.34,0.34,0.34}
\definecolor{gray35}{rgb}{0.35,0.35,0.35}
\definecolor{gray36}{rgb}{0.36,0.36,0.36}
\definecolor{gray37}{rgb}{0.37,0.37,0.37}
\definecolor{gray38}{rgb}{0.38,0.38,0.38}
\definecolor{gray39}{rgb}{0.39,0.39,0.39}
\definecolor{gray3}{rgb}{0.03,0.03,0.03}
\definecolor{gray40}{rgb}{0.40,0.40,0.40}
\definecolor{gray41}{rgb}{0.41,0.41,0.41}
\definecolor{gray42}{rgb}{0.42,0.42,0.42}
\definecolor{gray43}{rgb}{0.43,0.43,0.43}
\definecolor{gray44}{rgb}{0.44,0.44,0.44}
\definecolor{gray45}{rgb}{0.45,0.45,0.45}
\definecolor{gray46}{rgb}{0.46,0.46,0.46}
\definecolor{gray47}{rgb}{0.47,0.47,0.47}
\definecolor{gray48}{rgb}{0.48,0.48,0.48}
\definecolor{gray49}{rgb}{0.49,0.49,0.49}
\definecolor{gray4}{rgb}{0.04,0.04,0.04}
\definecolor{gray50}{rgb}{0.50,0.50,0.50}
\definecolor{gray51}{rgb}{0.51,0.51,0.51}
\definecolor{gray52}{rgb}{0.52,0.52,0.52}
\definecolor{gray53}{rgb}{0.53,0.53,0.53}
\definecolor{gray54}{rgb}{0.54,0.54,0.54}
\definecolor{gray55}{rgb}{0.55,0.55,0.55}
\definecolor{gray56}{rgb}{0.56,0.56,0.56}
\definecolor{gray57}{rgb}{0.57,0.57,0.57}
\definecolor{gray58}{rgb}{0.58,0.58,0.58}
\definecolor{gray59}{rgb}{0.59,0.59,0.59}
\definecolor{gray5}{rgb}{0.05,0.05,0.05}
\definecolor{gray60}{rgb}{0.60,0.60,0.60}
\definecolor{gray61}{rgb}{0.61,0.61,0.61}
\definecolor{gray62}{rgb}{0.62,0.62,0.62}
\definecolor{gray63}{rgb}{0.63,0.63,0.63}
\definecolor{gray64}{rgb}{0.64,0.64,0.64}
\definecolor{gray65}{rgb}{0.65,0.65,0.65}
\definecolor{gray66}{rgb}{0.66,0.66,0.66}
\definecolor{gray67}{rgb}{0.67,0.67,0.67}
\definecolor{gray68}{rgb}{0.68,0.68,0.68}
\definecolor{gray69}{rgb}{0.69,0.69,0.69}
\definecolor{gray6}{rgb}{0.06,0.06,0.06}
\definecolor{gray70}{rgb}{0.70,0.70,0.70}
\definecolor{gray71}{rgb}{0.71,0.71,0.71}
\definecolor{gray72}{rgb}{0.72,0.72,0.72}
\definecolor{gray73}{rgb}{0.73,0.73,0.73}
\definecolor{gray74}{rgb}{0.74,0.74,0.74}
\definecolor{gray75}{rgb}{0.75,0.75,0.75}
\definecolor{gray76}{rgb}{0.76,0.76,0.76}
\definecolor{gray77}{rgb}{0.77,0.77,0.77}
\definecolor{gray78}{rgb}{0.78,0.78,0.78}
\definecolor{gray79}{rgb}{0.79,0.79,0.79}
\definecolor{gray7}{rgb}{0.07,0.07,0.07}
\definecolor{gray80}{rgb}{0.80,0.80,0.80}
\definecolor{gray81}{rgb}{0.81,0.81,0.81}
\definecolor{gray82}{rgb}{0.82,0.82,0.82}
\definecolor{gray83}{rgb}{0.83,0.83,0.83}
\definecolor{gray84}{rgb}{0.84,0.84,0.84}
\definecolor{gray85}{rgb}{0.85,0.85,0.85}
\definecolor{gray86}{rgb}{0.86,0.86,0.86}
\definecolor{gray87}{rgb}{0.87,0.87,0.87}
\definecolor{gray88}{rgb}{0.88,0.88,0.88}
\definecolor{gray89}{rgb}{0.89,0.89,0.89}
\definecolor{gray8}{rgb}{0.08,0.08,0.08}
\definecolor{gray90}{rgb}{0.90,0.90,0.90}
\definecolor{gray91}{rgb}{0.91,0.91,0.91}
\definecolor{gray92}{rgb}{0.92,0.92,0.92}
\definecolor{gray93}{rgb}{0.93,0.93,0.93}
\definecolor{gray94}{rgb}{0.94,0.94,0.94}
\definecolor{gray95}{rgb}{0.95,0.95,0.95}
\definecolor{gray96}{rgb}{0.96,0.96,0.96}
\definecolor{gray97}{rgb}{0.97,0.97,0.97}
\definecolor{gray98}{rgb}{0.98,0.98,0.98}
\definecolor{gray99}{rgb}{0.99,0.99,0.99}
\definecolor{gray9}{rgb}{0.09,0.09,0.09}
\definecolor{gray}{rgb}{0.75,0.75,0.75}
\definecolor{green1}{rgb}{0.00,1.00,0.00}
\definecolor{green2}{rgb}{0.00,0.93,0.00}
\definecolor{green3}{rgb}{0.00,0.80,0.00}
\definecolor{green4}{rgb}{0.00,0.55,0.00}
\definecolor{greenyellow}{rgb}{0.68,1.00,0.18}
\definecolor{green}{rgb}{0.00,1.00,0.00}
\definecolor{grey0}{rgb}{0.00,0.00,0.00}
\definecolor{grey100}{rgb}{1.00,1.00,1.00}
\definecolor{grey10}{rgb}{0.10,0.10,0.10}
\definecolor{grey11}{rgb}{0.11,0.11,0.11}
\definecolor{grey12}{rgb}{0.12,0.12,0.12}
\definecolor{grey13}{rgb}{0.13,0.13,0.13}
\definecolor{grey14}{rgb}{0.14,0.14,0.14}
\definecolor{grey15}{rgb}{0.15,0.15,0.15}
\definecolor{grey16}{rgb}{0.16,0.16,0.16}
\definecolor{grey17}{rgb}{0.17,0.17,0.17}
\definecolor{grey18}{rgb}{0.18,0.18,0.18}
\definecolor{grey19}{rgb}{0.19,0.19,0.19}
\definecolor{grey1}{rgb}{0.01,0.01,0.01}
\definecolor{grey20}{rgb}{0.20,0.20,0.20}
\definecolor{grey21}{rgb}{0.21,0.21,0.21}
\definecolor{grey22}{rgb}{0.22,0.22,0.22}
\definecolor{grey23}{rgb}{0.23,0.23,0.23}
\definecolor{grey24}{rgb}{0.24,0.24,0.24}
\definecolor{grey25}{rgb}{0.25,0.25,0.25}
\definecolor{grey26}{rgb}{0.26,0.26,0.26}
\definecolor{grey27}{rgb}{0.27,0.27,0.27}
\definecolor{grey28}{rgb}{0.28,0.28,0.28}
\definecolor{grey29}{rgb}{0.29,0.29,0.29}
\definecolor{grey2}{rgb}{0.02,0.02,0.02}
\definecolor{grey30}{rgb}{0.30,0.30,0.30}
\definecolor{grey31}{rgb}{0.31,0.31,0.31}
\definecolor{grey32}{rgb}{0.32,0.32,0.32}
\definecolor{grey33}{rgb}{0.33,0.33,0.33}
\definecolor{grey34}{rgb}{0.34,0.34,0.34}
\definecolor{grey35}{rgb}{0.35,0.35,0.35}
\definecolor{grey36}{rgb}{0.36,0.36,0.36}
\definecolor{grey37}{rgb}{0.37,0.37,0.37}
\definecolor{grey38}{rgb}{0.38,0.38,0.38}
\definecolor{grey39}{rgb}{0.39,0.39,0.39}
\definecolor{grey3}{rgb}{0.03,0.03,0.03}
\definecolor{grey40}{rgb}{0.40,0.40,0.40}
\definecolor{grey41}{rgb}{0.41,0.41,0.41}
\definecolor{grey42}{rgb}{0.42,0.42,0.42}
\definecolor{grey43}{rgb}{0.43,0.43,0.43}
\definecolor{grey44}{rgb}{0.44,0.44,0.44}
\definecolor{grey45}{rgb}{0.45,0.45,0.45}
\definecolor{grey46}{rgb}{0.46,0.46,0.46}
\definecolor{grey47}{rgb}{0.47,0.47,0.47}
\definecolor{grey48}{rgb}{0.48,0.48,0.48}
\definecolor{grey49}{rgb}{0.49,0.49,0.49}
\definecolor{grey4}{rgb}{0.04,0.04,0.04}
\definecolor{grey50}{rgb}{0.50,0.50,0.50}
\definecolor{grey51}{rgb}{0.51,0.51,0.51}
\definecolor{grey52}{rgb}{0.52,0.52,0.52}
\definecolor{grey53}{rgb}{0.53,0.53,0.53}
\definecolor{grey54}{rgb}{0.54,0.54,0.54}
\definecolor{grey55}{rgb}{0.55,0.55,0.55}
\definecolor{grey56}{rgb}{0.56,0.56,0.56}
\definecolor{grey57}{rgb}{0.57,0.57,0.57}
\definecolor{grey58}{rgb}{0.58,0.58,0.58}
\definecolor{grey59}{rgb}{0.59,0.59,0.59}
\definecolor{grey5}{rgb}{0.05,0.05,0.05}
\definecolor{grey60}{rgb}{0.60,0.60,0.60}
\definecolor{grey61}{rgb}{0.61,0.61,0.61}
\definecolor{grey62}{rgb}{0.62,0.62,0.62}
\definecolor{grey63}{rgb}{0.63,0.63,0.63}
\definecolor{grey64}{rgb}{0.64,0.64,0.64}
\definecolor{grey65}{rgb}{0.65,0.65,0.65}
\definecolor{grey66}{rgb}{0.66,0.66,0.66}
\definecolor{grey67}{rgb}{0.67,0.67,0.67}
\definecolor{grey68}{rgb}{0.68,0.68,0.68}
\definecolor{grey69}{rgb}{0.69,0.69,0.69}
\definecolor{grey6}{rgb}{0.06,0.06,0.06}
\definecolor{grey70}{rgb}{0.70,0.70,0.70}
\definecolor{grey71}{rgb}{0.71,0.71,0.71}
\definecolor{grey72}{rgb}{0.72,0.72,0.72}
\definecolor{grey73}{rgb}{0.73,0.73,0.73}
\definecolor{grey74}{rgb}{0.74,0.74,0.74}
\definecolor{grey75}{rgb}{0.75,0.75,0.75}
\definecolor{grey76}{rgb}{0.76,0.76,0.76}
\definecolor{grey77}{rgb}{0.77,0.77,0.77}
\definecolor{grey78}{rgb}{0.78,0.78,0.78}
\definecolor{grey79}{rgb}{0.79,0.79,0.79}
\definecolor{grey7}{rgb}{0.07,0.07,0.07}
\definecolor{grey80}{rgb}{0.80,0.80,0.80}
\definecolor{grey81}{rgb}{0.81,0.81,0.81}
\definecolor{grey82}{rgb}{0.82,0.82,0.82}
\definecolor{grey83}{rgb}{0.83,0.83,0.83}
\definecolor{grey84}{rgb}{0.84,0.84,0.84}
\definecolor{grey85}{rgb}{0.85,0.85,0.85}
\definecolor{grey86}{rgb}{0.86,0.86,0.86}
\definecolor{grey87}{rgb}{0.87,0.87,0.87}
\definecolor{grey88}{rgb}{0.88,0.88,0.88}
\definecolor{grey89}{rgb}{0.89,0.89,0.89}
\definecolor{grey8}{rgb}{0.08,0.08,0.08}
\definecolor{grey90}{rgb}{0.90,0.90,0.90}
\definecolor{grey91}{rgb}{0.91,0.91,0.91}
\definecolor{grey92}{rgb}{0.92,0.92,0.92}
\definecolor{grey93}{rgb}{0.93,0.93,0.93}
\definecolor{grey94}{rgb}{0.94,0.94,0.94}
\definecolor{grey95}{rgb}{0.95,0.95,0.95}
\definecolor{grey96}{rgb}{0.96,0.96,0.96}
\definecolor{grey97}{rgb}{0.97,0.97,0.97}
\definecolor{grey98}{rgb}{0.98,0.98,0.98}
\definecolor{grey99}{rgb}{0.99,0.99,0.99}
\definecolor{grey9}{rgb}{0.09,0.09,0.09}
\definecolor{grey}{rgb}{0.75,0.75,0.75}
\definecolor{honeydew1}{rgb}{0.94,1.00,0.94}
\definecolor{honeydew2}{rgb}{0.88,0.93,0.88}
\definecolor{honeydew3}{rgb}{0.76,0.80,0.76}
\definecolor{honeydew4}{rgb}{0.51,0.55,0.51}
\definecolor{honeydew}{rgb}{0.94,1.00,0.94}
\definecolor{hotpink}{rgb}{1.00,0.41,0.71}
\definecolor{indianred}{rgb}{0.80,0.36,0.36}
\definecolor{ivory1}{rgb}{1.00,1.00,0.94}
\definecolor{ivory2}{rgb}{0.93,0.93,0.88}
\definecolor{ivory3}{rgb}{0.80,0.80,0.76}
\definecolor{ivory4}{rgb}{0.55,0.55,0.51}
\definecolor{ivory}{rgb}{1.00,1.00,0.94}
\definecolor{khaki1}{rgb}{1.00,0.96,0.56}
\definecolor{khaki2}{rgb}{0.93,0.90,0.52}
\definecolor{khaki3}{rgb}{0.80,0.78,0.45}
\definecolor{khaki4}{rgb}{0.55,0.53,0.31}
\definecolor{khaki}{rgb}{0.94,0.90,0.55}
\definecolor{lavenderblush}{rgb}{1.00,0.94,0.96}
\definecolor{lavender}{rgb}{0.90,0.90,0.98}
\definecolor{lawngreen}{rgb}{0.49,0.99,0.00}
\definecolor{lemonchiffon}{rgb}{1.00,0.98,0.80}
\definecolor{lightblue}{rgb}{0.68,0.85,0.90}
\definecolor{lightcoral}{rgb}{0.94,0.50,0.50}
\definecolor{lightcyan}{rgb}{0.88,1.00,1.00}
\definecolor{lightgoldenrod}{rgb}{0.93,0.87,0.51}
\definecolor{lightgoldenrod}{rgb}{0.98,0.98,0.82}
\definecolor{lightgray}{rgb}{0.83,0.83,0.83}
\definecolor{lightgreen}{rgb}{0.56,0.93,0.56}
\definecolor{lightgrey}{rgb}{0.83,0.83,0.83}
\definecolor{lightpink}{rgb}{1.00,0.71,0.76}
\definecolor{lightsalmon}{rgb}{1.00,0.63,0.48}
\definecolor{lightsea}{rgb}{0.13,0.70,0.67}
\definecolor{lightsky}{rgb}{0.53,0.81,0.98}
\definecolor{lightslate}{rgb}{0.47,0.53,0.60}
\definecolor{lightslate}{rgb}{0.47,0.53,0.60}
\definecolor{lightslate}{rgb}{0.52,0.44,1.00}
\definecolor{lightsteel}{rgb}{0.69,0.77,0.87}
\definecolor{lightyellow}{rgb}{1.00,1.00,0.88}
\definecolor{limegreen}{rgb}{0.20,0.80,0.20}
\definecolor{linen}{rgb}{0.98,0.94,0.90}
\definecolor{magenta1}{rgb}{1.00,0.00,1.00}
\definecolor{magenta2}{rgb}{0.93,0.00,0.93}
\definecolor{magenta3}{rgb}{0.80,0.00,0.80}
\definecolor{magenta4}{rgb}{0.55,0.00,0.55}
\definecolor{magenta}{rgb}{1.00,0.00,1.00}
\definecolor{maroon1}{rgb}{1.00,0.20,0.70}
\definecolor{maroon2}{rgb}{0.93,0.19,0.65}
\definecolor{maroon3}{rgb}{0.80,0.16,0.56}
\definecolor{maroon4}{rgb}{0.55,0.11,0.38}
\definecolor{maroon}{rgb}{0.69,0.19,0.38}
\definecolor{mediumaquamarine}{rgb}{0.40,0.80,0.67}
\definecolor{mediumblue}{rgb}{0.00,0.00,0.80}
\definecolor{mediumorchid}{rgb}{0.73,0.33,0.83}
\definecolor{mediumpurple}{rgb}{0.58,0.44,0.86}
\definecolor{mediumsea}{rgb}{0.24,0.70,0.44}
\definecolor{mediumslate}{rgb}{0.48,0.41,0.93}
\definecolor{mediumspring}{rgb}{0.00,0.98,0.60}
\definecolor{mediumturquoise}{rgb}{0.28,0.82,0.80}
\definecolor{mediumviolet}{rgb}{0.78,0.08,0.52}
\definecolor{midnightblue}{rgb}{0.10,0.10,0.44}
\definecolor{mintcream}{rgb}{0.96,1.00,0.98}
\definecolor{mistyrose}{rgb}{1.00,0.89,0.88}
\definecolor{moccasin}{rgb}{1.00,0.89,0.71}
\definecolor{navajowhite}{rgb}{1.00,0.87,0.68}
\definecolor{navyblue}{rgb}{0.00,0.00,0.50}
\definecolor{navy}{rgb}{0.00,0.00,0.50}
\definecolor{oldlace}{rgb}{0.99,0.96,0.90}
\definecolor{olivedrab}{rgb}{0.42,0.56,0.14}
\definecolor{orange1}{rgb}{1.00,0.65,0.00}
\definecolor{orange2}{rgb}{0.93,0.60,0.00}
\definecolor{orange3}{rgb}{0.80,0.52,0.00}
\definecolor{orange4}{rgb}{0.55,0.35,0.00}
\definecolor{orangered}{rgb}{1.00,0.27,0.00}
\definecolor{orange}{rgb}{1.00,0.65,0.00}
\definecolor{orchid1}{rgb}{1.00,0.51,0.98}
\definecolor{orchid2}{rgb}{0.93,0.48,0.91}
\definecolor{orchid3}{rgb}{0.80,0.41,0.79}
\definecolor{orchid4}{rgb}{0.55,0.28,0.54}
\definecolor{orchid}{rgb}{0.85,0.44,0.84}
\definecolor{palegoldenrod}{rgb}{0.93,0.91,0.67}
\definecolor{palegreen}{rgb}{0.60,0.98,0.60}
\definecolor{paleturquoise}{rgb}{0.69,0.93,0.93}
\definecolor{paleviolet}{rgb}{0.86,0.44,0.58}
\definecolor{papayawhip}{rgb}{1.00,0.94,0.84}
\definecolor{peachpuff}{rgb}{1.00,0.85,0.73}
\definecolor{peru}{rgb}{0.80,0.52,0.25}
\definecolor{pink1}{rgb}{1.00,0.71,0.77}
\definecolor{pink2}{rgb}{0.93,0.66,0.72}
\definecolor{pink3}{rgb}{0.80,0.57,0.62}
\definecolor{pink4}{rgb}{0.55,0.39,0.42}
\definecolor{pink}{rgb}{1.00,0.75,0.80}
\definecolor{plum1}{rgb}{1.00,0.73,1.00}
\definecolor{plum2}{rgb}{0.93,0.68,0.93}
\definecolor{plum3}{rgb}{0.80,0.59,0.80}
\definecolor{plum4}{rgb}{0.55,0.40,0.55}
\definecolor{plum}{rgb}{0.87,0.63,0.87}
\definecolor{powderblue}{rgb}{0.69,0.88,0.90}
\definecolor{purple1}{rgb}{0.61,0.19,1.00}
\definecolor{purple2}{rgb}{0.57,0.17,0.93}
\definecolor{purple3}{rgb}{0.49,0.15,0.80}
\definecolor{purple4}{rgb}{0.33,0.10,0.55}
\definecolor{purple}{rgb}{0.63,0.13,0.94}
\definecolor{red1}{rgb}{1.00,0.00,0.00}
\definecolor{red2}{rgb}{0.93,0.00,0.00}
\definecolor{red3}{rgb}{0.80,0.00,0.00}
\definecolor{red4}{rgb}{0.55,0.00,0.00}
\definecolor{red}{rgb}{1.00,0.00,0.00}
\definecolor{rosybrown}{rgb}{0.74,0.56,0.56}
\definecolor{royalblue}{rgb}{0.25,0.41,0.88}
\definecolor{saddlebrown}{rgb}{0.55,0.27,0.07}
\definecolor{salmon1}{rgb}{1.00,0.55,0.41}
\definecolor{salmon2}{rgb}{0.93,0.51,0.38}
\definecolor{salmon3}{rgb}{0.80,0.44,0.33}
\definecolor{salmon4}{rgb}{0.55,0.30,0.22}
\definecolor{salmon}{rgb}{0.98,0.50,0.45}
\definecolor{sandybrown}{rgb}{0.96,0.64,0.38}
\definecolor{seagreen}{rgb}{0.18,0.55,0.34}
\definecolor{seashell1}{rgb}{1.00,0.96,0.93}
\definecolor{seashell2}{rgb}{0.93,0.90,0.87}
\definecolor{seashell3}{rgb}{0.80,0.77,0.75}
\definecolor{seashell4}{rgb}{0.55,0.53,0.51}
\definecolor{seashell}{rgb}{1.00,0.96,0.93}
\definecolor{sienna1}{rgb}{1.00,0.51,0.28}
\definecolor{sienna2}{rgb}{0.93,0.47,0.26}
\definecolor{sienna3}{rgb}{0.80,0.41,0.22}
\definecolor{sienna4}{rgb}{0.55,0.28,0.15}
\definecolor{sienna}{rgb}{0.63,0.32,0.18}
\definecolor{skyblue}{rgb}{0.53,0.81,0.92}
\definecolor{slateblue}{rgb}{0.42,0.35,0.80}
\definecolor{slategray}{rgb}{0.44,0.50,0.56}
\definecolor{slategrey}{rgb}{0.44,0.50,0.56}
\definecolor{snow1}{rgb}{1.00,0.98,0.98}
\definecolor{snow2}{rgb}{0.93,0.91,0.91}
\definecolor{snow3}{rgb}{0.80,0.79,0.79}
\definecolor{snow4}{rgb}{0.55,0.54,0.54}
\definecolor{snow}{rgb}{1.00,0.98,0.98}
\definecolor{springgreen}{rgb}{0.00,1.00,0.50}
\definecolor{steelblue}{rgb}{0.27,0.51,0.71}
\definecolor{tan1}{rgb}{1.00,0.65,0.31}
\definecolor{tan2}{rgb}{0.93,0.60,0.29}
\definecolor{tan3}{rgb}{0.80,0.52,0.25}
\definecolor{tan4}{rgb}{0.55,0.35,0.17}
\definecolor{tan}{rgb}{0.82,0.71,0.55}
\definecolor{thistle1}{rgb}{1.00,0.88,1.00}
\definecolor{thistle2}{rgb}{0.93,0.82,0.93}
\definecolor{thistle3}{rgb}{0.80,0.71,0.80}
\definecolor{thistle4}{rgb}{0.55,0.48,0.55}
\definecolor{thistle}{rgb}{0.85,0.75,0.85}
\definecolor{tomato1}{rgb}{1.00,0.39,0.28}
\definecolor{tomato2}{rgb}{0.93,0.36,0.26}
\definecolor{tomato3}{rgb}{0.80,0.31,0.22}
\definecolor{tomato4}{rgb}{0.55,0.21,0.15}
\definecolor{tomato}{rgb}{1.00,0.39,0.28}
\definecolor{turquoise1}{rgb}{0.00,0.96,1.00}
\definecolor{turquoise2}{rgb}{0.00,0.90,0.93}
\definecolor{turquoise3}{rgb}{0.00,0.77,0.80}
\definecolor{turquoise4}{rgb}{0.00,0.53,0.55}
\definecolor{turquoise}{rgb}{0.25,0.88,0.82}
\definecolor{violetred}{rgb}{0.82,0.13,0.56}
\definecolor{violet}{rgb}{0.93,0.51,0.93}
\definecolor{wheat1}{rgb}{1.00,0.91,0.73}
\definecolor{wheat2}{rgb}{0.93,0.85,0.68}
\definecolor{wheat3}{rgb}{0.80,0.73,0.59}
\definecolor{wheat4}{rgb}{0.55,0.49,0.40}
\definecolor{wheat}{rgb}{0.96,0.87,0.70}
\definecolor{whitesmoke}{rgb}{0.96,0.96,0.96}
\definecolor{white}{rgb}{1.00,1.00,1.00}
\definecolor{yellow1}{rgb}{1.00,1.00,0.00}
\definecolor{yellow2}{rgb}{0.93,0.93,0.00}
\definecolor{yellow3}{rgb}{0.80,0.80,0.00}
\definecolor{yellow4}{rgb}{0.55,0.55,0.00}
\definecolor{yellowgreen}{rgb}{0.60,0.80,0.20}
\definecolor{yellow}{rgb}{1.00,1.00,0.00}

\nc{\e}{{\bf{e}}}
\nc{\kk}{{\bf{k}}}
\nc{\pp}{{\bf{p}}}

\nc{\bfk}{{\bf{k}}}
\nc{\bfx}{{\bf{x}}}
\nc{\bfp}{{\bf{p}}}

\nc{\eH}{{\epsilon_H}}
\nc{\calP}{{\cal P}}
\nc{\im}{{ \mathrm{Im} } }

\begin{document}

%%%%%%%%%%%%%%%%%%%%%%%%%%%%%%%%%%%%%%%%%%%%%%%%%%

\title{  The TT, TB, EB and BB  correlations in  anisotropic inflation}

\author{Xingang Chen $^1$}
\email{Xingang.Chen-AT-utdallas.edu}

\author{Razieh Emami $^2$}
\email{emami-AT-ipm.ir}

\author{Hassan Firouzjahi$^3$}
\email{firouz-AT-ipm.ir}

\author{Yi Wang $^4$}
\email{yw366-AT-cam.ac.uk}

\affiliation{$^1$Department of Physics, The University of Texas at Dallas, Richardson, TX 75083, USA}

\affiliation{$^2$School of Physics, Institute for Research in
Fundamental Sciences (IPM)
P.~O.~Box 19395-5531,
Tehran, Iran}

\affiliation{$^3$School of Astronomy, Institute for Research in
Fundamental Sciences (IPM)
P.~O.~Box 19395-5531,
Tehran, Iran}

\affiliation{$^4$Centre for Theoretical Cosmology, DAMTP, University of Cambridge, Cambridge CB3 0WA, UK}

\begin{abstract}
%\vspace{0.3cm}

The ongoing and future experiments will measure the B-mode from different sky coverage and frequency bands, with the potential to reveal non-trivial features in  polarization map.
In this work we study the TT, TB, EB and BB correlations associated with the B-mode polarization of CMB map in models of charged anisotropic inflation. The model contains a chaotic-type large field  complex inflaton  which is charged under the $U(1)$ gauge field. We calculate the statistical
anisotropies generated in the power spectra of the  curvature perturbation, the tensor perturbation and their cross-correlation. It is shown that the
asymmetry in tensor power spectrum is a very sensitive probe of the gauge coupling. While the level of statistical  anisotropy in temperature power spectrum can be small and satisfy the observational bounds, the interactions from the gauge coupling can induce large directional dependence
in tensor modes. This will leave interesting anisotropic fingerprints in various correlations involving the B-mode polarization such as the TB cross-correlation which may be detected in upcoming Planck polarization data. In addition, the TT correlation receives an anisotropic contribution from the tensor sector which naturally decays after $l \gtrsim 100$. We expect that the mechanism of
using tensor sector to induce  asymmetry  at low $l$ to be generic which can also be applied to address other low $l$ CMB anomalies.

\vspace{0.3cm}

\end{abstract}

\maketitle

%%%%%%%%%%%%%%%%%%%%%%%%%%%%%%%%%%%%%%%%%%%%%%%%%%%%%

\section{Introduction}

The apparent detection of B-mode polarization of the Cosmic Microwave Background (CMB) on $l \sim 100$ by the BICEP2 observation has stirred significant interests in primordial gravitational waves  \cite{Ade:2014xna}. If confirmed to be primordial \cite{Flauger:2014qra, Mortonson:2014bja},  this detection implies the existence of primordial gravitational waves with the scalar-to-tensor ratio  $r= 0.2^{\, +0.07}_{\, -0.05}$.
In the near future, the BICEP2 detection of $r$ will be cross-checked by many on going and forthcoming experiments. For example, Planck \cite{Planck16}, SPTPol \cite{Austermann12}, ACTPol \cite{Niemack10}, PolarBear \cite{PolarBear} and CLASS \cite{Eimer12}.

These forthcoming experiments can  probe more detailed properties for the tensor mode.  For example, when the statistical features of the primordial perturbations are not isotropic, a large number of new observables arises, including the correlation functions with different multipole moment $l$, and TB and EB cross-correlations, which have been forbidden by the isotropic statistics of the primordial fluctuations \cite{Watanabe:2009ct, Chen:2013eaa, new paper}.

Motivated by the future data, in this paper we study inflationary dynamics producing statistical
 anisotropies in details. We investigate the anisotropic inflation scenario with a charged scalar inflaton coupled to the gauge field. This model is studied originally in \cite{Emami:2010rm} at the background level and its predictions for statistical anisotropies in curvature perturbation power spectrum were studied in \cite{Emami:2013bk}. However,  the perturbations in tensor sectors and their correlation with the scalar perturbations  have  not been explored so far.  We show that the gauge coupling induces large statistical anisotropies in tensor perturbations as compared to model of anisotropic inflation with no charge coupling \cite{Ohashi:2013qba}. We show that while statistical anisotropies in temperature power spectrum can be small as required by the Planck data \cite{Kim:2013gka}, the tensor mode can develop significant statistical anisotropies.  Therefore it is important for the forthcoming experiments to look for the statistical anisotropies in the B-mode polarization even though the statistical anisotropies in temperature map is well-constrained. For this purpose, in this work we calculate the primordial correlations of the curvature perturbation $\zeta$, and the two tensor modes $h_+$ and $h_\times$. The CMB temperature and polarization correlations TT, TE, EE, TB, EB and BB are then calculated from these primordial perturbations, for the same and different multipole moments.

A novel result in our analysis is that  the anisotropies in the tensor sector also contribute to the TT correlation on the CMB. However, the transfer function from primordial tensor to CMB temperature  decays towards large $l$. Thus the TT anisotropy coming from the primordial tensor has a decaying amplitude and is highly suppressed after $l \gtrsim 100$. As a result, we naturally obtain anisotropies at low multipoles of TT without modifying the high multipoles. This scale-dependent
anisotropies will have a better fit to the CMB anomalies.

Another motivation of the present work is that, the Planck experiment puts a limit of tensor-to-scalar ratio $r<0.12$. This result is in tension with the BICEP2 detection. The tension is reported to be as unlikely as $0.1\%$. Recently, it is proposed \cite{Contaldi:2014zua} that with anti-correlated curvature and tensor perturbations, the tension between Planck and BICEP2 may be reconciled. However, our detailed study show that such a mechanism does not work. As we shall show, there is indeed an anti-correlation between the curvature and tensor perturbations. However the contribution to the TT power spectrum is cancelled when summing over the angular mode $m$.

The rest of the paper is organized as follows. In Section \ref{model} we present our model of anisotropic inflation and review the background dynamics and the perturbations. In Section
\ref{correlations} we calculate the anisotropic scalar and the tensor power spectra and the scalar-tensor
cross-correlations. In Section \ref{numerical} we present the predictions of our model for CMB.
We relegate the technical details into appendices.

%%%%%%%%%%%%%%%%%%%%%%%%%%%%%%%%%%%%%%%%%%%%%%%%%%%%%

%%%%%%%%%%%%%%%%%%%%%%%%%%%%%%%%%%%%%%%%%%%%%%%%%%%%%

\section{Anisotropic Inflation}
\label{model}

Here we review anisotropic inflation.   For earlier works on various aspects of anisotropic inflation see
\cite{Watanabe:2009ct, Ohashi:2013pca, Ohashi:2013qba, Kim:2013gka,
Ohashi:2013mka, Emami:2010rm, Thorsrud:2012mu,
Dulaney:2010sq, Gumrukcuoglu:2010yc, Watanabe:2010fh, Bartolo:2012sd,
Funakoshi:2012ym, Yamamoto:2012sq, Emami:2013bk,
Shiraishi:2013vja, Abolhasani:2013zya,
Abolhasani:2013bpa, Ramazanov:2013wea, Nurmi:2013gpa, Urban:2013spa, Fujita:2013qxa, Shiraishi:2013oqa, Jain:2012vm, Kanno:2010nr, Murata:2011wv, Bhowmick:2011em, Hervik:2011xm,  Boehmer:2007ut, Koivisto:2008xf, Yokoyama:2008xw, Baghram:2013lxa, Watanabe:2010bu, Yamamoto:2012tq, Emami:2011yi, Emami:2014tpa, BeltranAlmeida:2011db, Rodriguez:2013cj, Lyth:2013sha, TuanQ}, for a review of anisotropic inflation see \cite{Soda:2012zm, Maleknejad:2012fw}. Also see \cite{Pereira:2007yy, Pitrou:2008gk, Pitrou:2012ge} for related works on primordial anisotropies.

As mentioned before, the model contains  a $U(1)$ gauge field  $A_\mu$, as in Maxwell theory, which is turned on at the background level.  However, it is well-known that the Maxwell theory suffers from the conformal invariance on expanding backgrounds.
Therefore, the background gauge field $A_\mu(t)$ is diluted exponentially during inflation. Furthermore,  the quantum excitations of the gauge field $\delta A_\mu (t, \bfx)$ are not scale-invariant. One interesting mechanism to break the conformal invariance is to couple the gauge field to the inflaton field non-trivially.

With these general discussions in mind we present our model of anisotropic inflation.  The model contains  a complex inflaton field $\phi$ which is charged under the $U(1)$ gauge field $A_\mu$ with the electric charge (coupling) $\e$. The action is given by
\ba
 \label{action} S= \int
d^4 x  \sqrt{-g} \left [ \frac{M_P^2}{2} R - \frac{1}{2} D_\mu \varphi
\,  \overline {D^\mu \varphi} -   \frac{f^{2}(\varphi)}{4} F_{\mu \nu} F^{\mu
\nu}  - V(\varphi, \overline \varphi) \right] \, ,
\ea
where $M_P$ is the reduced Planck mass.

As usual, $F_{\mu \nu}$ is  the gauge field strength given by
\ba F_{\mu \nu} = \nabla_\mu A_\nu
- \nabla_\nu A_\mu  = \partial_\mu A_\nu - \partial_\nu A_\mu \, .
\ea
In addition,  the covariant derivative associated with the $U(1)$ gauge field is given by
\ba
D_\mu \varphi = \partial_\mu  \varphi + i \e \,  \varphi  \, A_\mu \, .
\ea
where $\e$ represents  the electric gauge coupling. The model of anisotropic inflation based on the above action was studied at the background level
 in \cite{Emami:2010rm} and its  analysis for curvature perturbations were performed in
 \cite{Emami:2013bk}.

Note that in Maxwell theory $f(\varphi)=1$. However, as explained above,  we need a time-dependent gauge kinetic coupling $f(\varphi)$ in order to break the conformal invariance such that the background gauge field survives the exponential expansion and the gauge field excitations acquire a nearly scale-invariant power spectrum.
As we shall verify below one requires $f \propto a^{-2}$  in order to obtain a scale-invariant power spectrum for the  gauge field quantum excitations.  This choice of the gauge kinetic coupling corresponds to a constant background electric field energy density during inflation.

The complex scalar field can be decomposed into a radial part and an axial part via $\varphi(x) = \phi(x) \,  e^{i \theta(x)}$. As usual, we assume the model is axially symmetric in field space so the potential
$V$ and $f(\varphi)$ are only functions of $\varphi \bar \varphi=  \phi ^2$. It is convenient to go to the unitary gauge in which $\theta=0$. In this gauge, $\varphi$ becomes real-valued and in the analysis below we  take $\varphi=\varphi^* = \phi$.

We consider the coordinate system in which the background gauge field is turned on along the $x$-direction so  $A_\mu = (0, A_x(t), 0,0)$. With the background gauge field in the $x$-direction the
background space-time becomes anisotropic, taking the form of  type I Bianchi Universe, with the metric
\ba
\label{bian0}
\label{metric} ds^2 = - dt^2 + e^{2\alpha(t)}\left( e^{-4\sigma(t)}d x^2
+e^{2\sigma(t)}(d y^2 +d z^2) \right) \, .
\ea
Note that the metric (\ref{bian0}) still has a residue two-dimensional rotational symmetry on the $y-z$ plane. In this convention $\dot \alpha$ measures the averaged Hubble expansion while $\dot \sigma(t)$ measures the level of anisotropic expansion.  However, on the observational grounds, as we shall see in next Section, the level of anisotropy in curvature perturbations is not more than few percent which subsequently is translated  into the conclusion that
 $\dot \sigma /\dot \alpha \ll 1$. As a result, we can treat the analysis prturbatively in term of background  anisotropy.

The background fields equations are  \cite{Emami:2010rm}
\ba
\label{back-A-eq}
\partial_t{\left(  f^2(\phi) e^{\alpha + 4 \sigma} \dot A_x        \right)}& =& - \e^2 \phi^2 e^{\alpha + 4 \sigma}  A_x \\
%\ddotA_{x}+\left[\dot\alpha+4\dot\sigma+2\frac{f_\phi(\phi)}{f(\phi)}\dot \phi
%\right]\dot A_{x}+e^2\phi^2f^{-2}(\phi)A_{x}&=&0 \, , \\
\label{back-rho-eq}
\ddot\phi+3\dot \alpha\dot \phi+ V_{,\phi}+ \left(
-f(\phi)f_{,\phi}(\phi)\dot A_x^2 +\e^2 \phi A_x^2   \right) e^{-2\alpha+4\sigma}&=&0  \\
\label{Ein1-eq}
\frac{1}{2}\dot
\phi^2+V(\phi)+ \left(   \frac{1}{2}f^2(\phi)\dot
A_x^2 +\frac{\e^2\phi^2}{2}A_x^2 \right) e^{-2\alpha+4\sigma}
&=&
3 M_P^2 \left(   \dot \alpha^2-\dot \sigma^2 \right)  \\
\label{Ein2-eq}
V(\phi)+  \left(  \frac{1}{6}f^2(\phi)\dot
A_x^2+\frac{\e^2\phi^2}{3}A_x^2  \right)e^{-2\alpha+4\sigma}
&=& M_P^2 \left( \ddot \alpha    + 3 \dot \alpha^2 \right)  \\
\label{anisotropy-eq}
\left(\frac{1}{3}f^2(\phi)\dot A_x^2  -\frac{\e^2\phi^2}{3}A_x^2    \right) e^{-2\alpha+4\sigma}
&=& M_P^2\left( 3\dot \alpha \dot \sigma+ \ddot \sigma      \right)\, ,
\ea
in which a dot indicates derivative with respect to $t$ and $f_{, \phi}= \partial_\phi f$ and so on.

Here we summarize the main features of the above equations. Eq. (\ref{back-A-eq}) is the Maxwell equation in the inflationary background with the gauge coupling $\e$ appearing in the source term.
Eq. (\ref{back-rho-eq}) is the modified Klein-Gordon equation in the presence of the gauge field.
As we see the scalar field dynamics are affected by the gauge field via the last two terms in the bracket. The first term in the bracket in Eq. (\ref{back-rho-eq}) comes from the gauge kinetic term while the last term  comes from the charge coupling $\e$.
As we shall see, these two terms play crucial roles both at the background and the perturbation levels. The remaining equations are the Einstein equations with Eq. (\ref{anisotropy-eq}) controlling the dynamics of the anisotropy $\dot \sigma$.

In general it is not easy to solve the above set of equations analytically, even in the slow-roll limit. However, as we mentioned before,  the level of anisotropy in curvature perturbations is small
which also yields $\dot \sigma / \dot{\alpha} \ll 1$.  This means that the background expansion is nearly
isotropic.  Therefore,  as in in conventional models of inflation, the background expansion is controlled  by
the potential term $V$. However, we expect the gauge field also to contribute in the background expansion in the form of electric field energy density $E_x^2$ where $E_x = F_{0 x}$.
In order for the background to be nearly isotropic  we require that the electric field energy density to be very small compared to $V$. It is convenient to parameterize the gauge field (electric) energy density
by the parameter $\Omega_A$ via
\ba
\label{Omega-def}
\Omega_A \equiv \frac{\dot A_x^2 f(\phi)^2 e^{-2 \alpha}}{2 V} \, .
\ea
In order for the anisotropy to be small we assume $\Omega_A \ll 1$.

Alternatively, for the perturbation analysis, it is more convenient to express the background metric (\ref{bian0}) in the following form
\ba
\label{Bianchi-metric}
ds^2 = a(\eta)^2 (d \eta^2 + d x^2) + b(\eta)^2 ( d y^2 + d z^2)
\ea
where $a= e^{\alpha - 2 \sigma}$ and $b=e^{\alpha + \sigma}$ and the conformal time  $\eta$ is defined via via $dt = a(\eta) d \eta$.

We will work in the slow roll limit where the change in expansion rates in all directions are small. Defining the average Hubble expansion rate via $\dot \alpha \equiv H$, the
slow-roll parameters are given by
\ba
\epsilon_H \equiv - \frac{\dot H}{H^2}  \quad , \quad
\eta_H \equiv \epsilon_H - \frac{\ddot H}{2 H \dot H}
 %\quad , \quad \dot \epsilon_H = 2 H \epsilon_H (2 \epsilon_H - \eta_H)
\ea
We work in  the slow-roll limit where $\epsilon_H , \eta_H \ll 1$ and  to leading order in slow-roll parameters and anisotropy $a \simeq b \simeq -1/H \eta$.

Before we study the attractor solution in next subsection, let us discuss the effects of the gauge coupling $\e$. The main effect of the gauge coupling is captured by the third term in bracket in Eq. (\ref{back-rho-eq}). From this term we see that the interaction $\e^2 \phi^2 A_\mu A^\mu$ induces a
time-dependent mass for the inflaton. Since this induced mass is exponentially time-dependent, we expect it becomes important only towards the end of inflation where the exponential growth
of the gauge field becomes significant.  As studied in details in
\cite{Emami:2010rm} inflation ends when the induced mass from the  back-reaction $\e^2 \phi^2 A_\mu A^\mu$ becomes comparable to the bare inflaton mass $m$. Therefore, in order for inflation to sustain  long enough, the back-reaction $\e^2 \phi^2 A_\mu A^\mu$ is negligible during much of the period of inflation and it only controls the mechanism of end of inflation.  As studied  in \cite{ Emami:2010rm} the end of inflation depends logarithmically on $\e$ where $
\alpha_e \sim -\frac{\ln \e}{2} + ...
$
where dots indicate the dependence on other parameters such as the inflaton value and its mass.
Therefore, the larger is the gauge coupling $\e$, the shorter is the period of inflation.
As explained above this is easily understood from the fact that the induced mass  for the inflaton field from  the Higgs mechanism, $\e^2 A_\mu A^\mu $, becomes comparable to the bare inflaton mass.

In the remaining analysis of the background dynamics we neglect the effects of $\e$ during
inflation. In particular, one can neglect the source term in Maxwell equation  (\ref{back-A-eq})
during much of period of inflation.   In this approximation one can easily solve the Maxwell equation (\ref{back-A-eq}) to get
\ba
\label{gaugefield}
\dot{A_{x}}= f(\phi)^{-2}e^{-\alpha(t)-4\sigma(t)}p_{A} \, ,
\ea
where  $p_{A}$ is a constant of integration.  Of course, this approximation breaks down near the end of inflation where the induced mass term from the interaction $\e^2 A_\mu A^\mu \phi^2$ becomes important.

%%%%%%%%%%%%%%%%%%%%%%%%%%%%%%%%%%%%%%%%%%%%%%%%%%
\subsection{The Attractor Solution}
\label{background formulas}

As we discussed above  the anisotropy is small and  the average Hubble expansion rate in Eq. (\ref{Ein1-eq}) mainly comes from the isotropic potential term. However, the back-reactions of the gauge field on the inflaton field induce an effective mass term for the inflaton  as captured by the last term in Eq. (\ref{back-rho-eq}). Therefore, the dynamics of the  inflaton field is affected by the  back-reactions of the gauge field.  A key  observation was made in  \cite{Watanabe:2009ct} where it is shown that
for a general form of slow-roll potential $V$ and  with an appropriate choice of $f(\phi)$ the system reaches an attractor regime where $ \Omega_A \sim \eH$.  As a result,  the modified Klein-Gordon equation still admits a slow-roll solution but now with  a modified effective mass for the inflaton which is  induced from the back-reactions of the gauge field.

Let us see under what condition $\Omega_A$ reaches a near constant value during the attractor regime. Combining Eqs. (\ref{gaugefield})  and (\ref{Omega-def}) we obtain
\ba
\label{R-def2}
\Omega_A = \frac{p_A^2}{2 V} f(\phi)^{-2} e^{-4 \alpha - 8 \sigma} \, .
\ea
In order for $\Omega_A$ to reach a nearly constant value, and neglecting the contribution of $\sigma$
in the small anisotropy limit,   one has to choose $f(\phi)$ such that $f(\phi) \propto e^{-2 \alpha(t)} = a(t) ^{-2}$. Now to find $f$ as a function of $\phi$, our job is to solve the background expansion equation and  express $a(t)$ in terms of $\phi$. In the small anisotropy limit, and for a given potential $V(\phi)$,  the background isotropic expansion is given by
\ba
\label{a-scale}
a(t) \propto \exp \left[ - \int d \phi \frac{V}{ V_{,\phi}} \right] \, .
\ea
This indicates that  if one chooses
\ba
\label{f-scale}
f \propto \exp \left[ -n  \int d \phi \frac{V}{V_\phi} \right] \, ,
\ea
then one obtains $f \propto a^{n}$.

The exact form of $f(\phi)$  depends on $V(\phi)$.  To be specific, in this work we consider the  simple  chaotic potential
\ba
\label{Chaoticpotential}
V=  \frac{1}{2} m^2 | \varphi|^2 =
\frac{1}{2} m^2 \phi^2 \, .
\ea
Plugging this form of potential into Eq. (\ref{f-scale}) yields   \cite{Watanabe:2009ct}
\ba
\label{f-form0}
f(\phi) = \exp {\left( \frac{c\phi^2}{2 M_P^2}  \right)} \, ,
\ea
where $c$ is a constant.

Now we examine under what conditions the system allows for an attractor solution where the anisotropy reaches a sub-dominant but constant value.  In the small anisotropy limit, the background equation is given by the potential term  and we have
\ba
\label{Friedmann}
3 M_P^2 \dot \alpha^2 \simeq V = \frac{m^2 \phi^2}{2} \, .
\ea
On the other hand, in the slow-roll regime, the scalar field equation, Eq. (\ref{back-rho-eq}), is given by
\ba
\label{KG-eq1}
3 \dot \alpha \dot \phi = -m^2 \phi + \frac{c\,  p_A^2 \,  \phi }{M_P^2} f(\phi)^{-2} e^{-4\alpha} \, .
\ea
Note that in order to get the second term in the right hand side  the solution for $\dot A_x$ and
 the form of $f(\phi)$, given respectively in  Eq. (\ref{gaugefield}) and  Eq. (\ref{f-form0}), have been used.  Eliminating $\dot \alpha$ from  Eqs. (\ref{Friedmann})  and (\ref{KG-eq1}) we obtain
an equation for $\phi$ in terms of the number of e-folds $\alpha$ as
\ba
\phi \frac{d \phi}{d \alpha} = - 2 M_P^2 + \frac{2 c p_A^2}{m^2} e^{- c \phi^2/M_P^2} e^{- 4 \alpha} \, .
\ea
One can easily solve  this differential equation to get
\ba
\label{sol-general}
e^{-4 \alpha} e^{-c \phi^2/M_P^2} = \frac{m^2 (c-1) M_P^2}{c^2 p_A^2} \left[ 1+ C\,  e^{-4 (c-1) \alpha} \right]^{-1} \, ,
\ea
where $C $ is a constant of integration. Now the important point to note  is that for
$c\ge 1$, the second term in the square bracket above decays exponentially  during inflation
and we quickly reaches the attractor regime
\ba
\label{att1}
e^{-4 \alpha} e^{-c \phi^2/M_P^2} \simeq \frac{m^2 (c-1) M_P^2}{c^2 p_A^2} \, .
\ea
This indicates that $\Omega_A$,  defined in Eq. (\ref{R-def2}), reaches a constant value during the attractor regime as claimed.  More specifically, from Eq. (\ref{R-def2}) we obtain
\ba
\Omega_A = \frac{m^2 M_P^2}{2 V} \frac{c-1}{c} \, .
\ea
On the other hand, using  the remaining Einstein equation,  one can also verify that the slow-roll parameter is
\ba
\epsilon_H \equiv -\frac{\dot H}{H^2} = -\frac{\ddot \alpha}{\dot \alpha^2} = \frac{2 M_P^2}{c \phi^2} \, .
\ea
As a result, we conclude
\ba
\label{R-app}
\Omega_A = \frac{c-1}{2c}\epsilon_{H} = \frac{1}{2}I\epsilon_{H} \, ,
\ea
where we have introduced the anisotropy parameter $I$ via
\ba
 I \equiv\frac{c-1}{c} \, .
 \ea
 Note that in order for the gauge field to survive the exponential expansion and the system reaches the attractor regime we require $I \ge 0$.

It is also instructive to look at the scalar field equation in the attractor regime.  Plugging  Eq. (\ref{att1}) back into the scalar field equation  (\ref{KG-eq1})  we get
\ba
\label{klinGordon}
M_P^{-2}\frac{d \phi}{d \alpha} \simeq -\frac{V_\phi}{ V} +
\frac{c-1}{c}\frac{V_\phi}{ V}  = \frac{-2 c}{\phi}
\ea
This equation implies  that the back-reaction of the gauge field has reduced the effective mass of the inflaton such that    $m \rightarrow m - \frac{c-1}{c}m = \frac{m}{c}$.

The above equation can be easily solved to find $\phi$ as a function of the number of e-foldings $\alpha$ as
\ba
\label{klinN}
\phi_{e}^2 - \phi^2 = 4 M_P^2 \alpha (1-I) \, ,
\ea
where $\phi_e$ represents  the value of $\phi$ at the end of inflation. Note that the above solution is valid until end of inflation when the effects of the gauge coupling $\e$ starts to dominate. As we discussed before, there is additional induced mass from the term $\e^2 A_\mu A^\mu \phi^2$ which
terminate inflation quickly when it becomes comparable to the bare mass $m$.

%%%%%%%%%%%%%%%%%%%%%%%%%%%%%%%%%%%%%%%%%%
\subsection{Perturbations}

Here we present cosmological perturbations in anisotropic inflation. The perturbation analysis
for various models of inflation were studied in
\cite{Dulaney:2010sq, Gumrukcuoglu:2010yc, Watanabe:2010fh, Bartolo:2012sd,
Funakoshi:2012ym, Yamamoto:2012sq, Emami:2013bk,
Shiraishi:2013vja, Abolhasani:2013zya,
Abolhasani:2013bpa, Ohashi:2013qba}.
The general form of the metric and matter perturbations have been studied in \cite{Emami:2013bk}, see also \cite{Watanabe:2010fh}.  For the perturbations in  the metric we note that there are some non-dynamical degrees of freedom, namely $\delta g_{0 \mu}$, which should be integrated out in order to calculate the dynamical action. Integrating out these non-dynamical metric degrees of freedom  lead us to extra terms in the action. Therefore, one important question is what the leading contributions in the final action are. Are they from the metric sector perturbations or from the matter sector perturbations?
Fortunately, as it has been verified  in \cite{Emami:2013bk}, it turns out that the metric sector perturbations are  either slow-roll suppressed or would cancel with each other. As a result,  we conclude that  the leading terms in the interactions come from the matter sector.  In other word,  we do not need to consider the perturbations from the non-dynamical degrees of freedom  in the metric. To simplify the situation further, we go to flat gauge where the curvature perturbations is given by the
inflaton perturbations $\zeta = - \frac{H}{\dot \phi} \delta \phi$. As a result, the metric has no scalar perturbations and we are left with the simple form of metric perturbations
\ba
\label{dynamical metric}
ds^2 = a(\eta)^2 \left(-d\eta^2 + \left[\delta_{ij}+ h_{ij} \right]dx^i dx^j \right ) \, .
\ea
Note that since we work in small anisotropy limit, we can set $a=b$ to leading order.
The corrections in our results below from using this assumption
will be suppressed by additional factors of $I$ or $\epsilon_H$
which are small.  The perturbations $h_{ij} $ represents the tensor modes subject to the transverse and traceless conditions  $\partial_i h_{ij} = 0$ and $ h_{ii}=0$ where the repeated indices are summed.  We denote the two independent polarizations of the metric by $h_\times$ and  $h_+$.

As for the perturbations in gauge field  sector  there is one non-dynamical degree of freedom,  $\delta A_{0}$, which must be integrated out from the action. However, similar to the case of   non-dynamical degrees of freedom from the metric perturbations, it turns out that the new terms from integrating out $\delta A_{0}$ are also sub-leading. As a result, the leading interaction terms in the total Lagrangian come from  the dynamical degrees of freedom $\delta A_i$.

In order to simplify the analysis further, we can use the remaining  two-dimensional rotational symmetry on the $y-z$ plane to set $k_{z} \equiv 0$ so $\overrightarrow{k} = (k_x , k_{y} , 0) =
k \, (\cos \theta, \sin \theta, 0 )$ in Fourier space. In addition, since the gauge field  has three polarizations in the unitary gauge,
two transverse and one longitudinal polarizations, we  can choose the following ansatz for the gauge field perturbations, i.e. $\delta A_i$,
\ba
\delta A_\mu^{(S)} = (\delta A_0, \delta A_1, \partial_y M, 0) \quad \quad , \quad \quad
\delta A_\mu^{(V)} =(0, 0, 0, D) \, .
\ea
Where we have defined, $\delta A_1, \delta A_2$ and $\delta A_3$ to be $\delta A_x, \partial_y M$ and $D$ respectively. Here $A_\mu^{(V)}$ refers to one transverse polarizations in the vector sector  while $A_\mu^{(S)}$
represents the two polarizations in the scalar sector.
Furthermore, we can decompose the two polarizations in $A_\mu^{(S)}$ into  one transverse and one longitudinal polarizations
as follows \cite{Emami:2013bk}
\begin{align}
\label{D12}
D_{1}&\equiv \delta A_{1} -ik \cos{\theta}M \\
D_{2}&\equiv \cos{\theta} \delta A_{1} + ik \sin^2{\theta}M \, .
\end{align}
In this decomposition $D_{1}$ represents  the transverse polarization while $D_{2}$ refers to the longitudinal polarization of the gauge field.  However, as it has been demonstrated in \cite{Emami:2013bk}, the interactions containing the longitudinal mode are  exponentially suppressed during inflation and can be neglected from the analysis. Physically, this is understandable since the interactions containing the
longitudinal mode $D_2$ originate from the ``Higgs mechanism'' via the interaction $\e^2 A_\mu A^\mu \phi^2$ which are exponentially suppressed during much of the period of inflation as discussed before.

We can quantize the curvature perturbation and the gauge field perturbations as usual.
For the curvature perturbation, note that we work in the flat gauge so
\ba
\label{zeta-phi}
\zeta = - \frac{H}{\dot \phi} \delta \phi  = \frac{\delta \phi}{M_P \sqrt{2 \epsilon_H}}.
\ea
Expanding the quantum operator $\widehat \zeta$ in terms of the annihilation and the creation operator
$a(\bfk)$ and $a^{\dagger}(\bfk)$ we have
\ba
\widehat \zeta (\bfx , \eta) =  \int \frac{d^3k}{(2\pi)^{3/2}} e^{i \mathbf{k}.\mathbf{x}}
\widehat\zeta(\mathbf{k},\eta) \quad , \quad
\widehat\zeta(\mathbf{k},\eta) = \zeta(k, \eta) a(\bfk) +   \zeta^*(k, \eta)  a^{\dagger}(- \bfk)
\ea
where the creation and the annihilation operators satisfy the usual commutation relation
$ [ a(\bfk),    a^\dagger  (\bfk')] = \delta^{(3)} (\bfk-\bfk' )$.

The wave function  of the curvature perturbation has the standard form of  the excitations of
a  massless scalar field on a dS background
\ba
\label{zeta-wave}
\zeta_k(\eta) = \frac{i H \eta }{M_{P}\sqrt{2\epsilon_{H}k} }  \left( 1 - \frac{i}{k\eta} \right) e^{-ik\eta} \, .
\ea
The power spectrum of the curvature perturbations is given by
\ba
\langle   \widehat{\zeta}(\mathbf{k_1})  \widehat{\zeta} (\mathbf{k_2}) \rangle
= (2 \pi)^3 \delta^{(3)} (\bfk_1 + \bfk_2) P_\zeta (k_1) \quad , \quad
\calP_\zeta \equiv   \frac{ k_1^3  }{ 2 \pi^2}  P_\zeta(k_1)
\ea
In particular, the power spectrum for the free isotropic theory is
\ba
\label{calP0}
\calP_\zeta^{(0)} = \frac{H^2}{8 \pi^2 \epsilon_H M_P^2}\, .
\ea

Similarly, the quantum excitations of the gauge field perturbations $\widehat D_{1\bf k}(\eta)$ and $\widehat D_\bfk(\eta)$ can be expanded in terms of their annihilation and creation operators with the wave functions
\ba
\label{D-wave}
\sin{\theta} D_{1 k}(\eta)  = D_k(\eta) =
 \frac{i}{f \sqrt{2k}} \left( 1 - \frac{i}{k\eta} \right) e^{-ik\eta} \, .
\ea

Now we present our decomposition of the tensor perturbations  $h_{ij}$ into $h_\times$ and $h_+$ polarizations following the method of  \cite{Ohashi:2013qba}.  Decomposing
$h_{ij}$ into $e_{ij}^{(s)}(\bfk)$ in Fourier space, the traceless  and transverse conditions, $h_{ii} = h_{ij,j} =0$,
yields
\ba
e_{i i}^{(s)}(\bfk) = 0 \quad , \quad  k_j e_{i j}^{(s)}(\bfk) = 0  \, ,
\ea
with $s= \times, +$ representing the two polarizations. In addition, we choose the following normalization
\ba
e^{(s)}_{ij}(\mathbf{k}) e^{*(s')}_{ij}(\mathbf{k}) = \delta_{ss'} \, ,
\ea
where $*$ represents the complex-conjugation. Note that we also have $e^{(s)}_{ij}(\mathbf{k}) = e^{*(s)}_{ij}(\mathbf{-k})$.

The  quantum operators $\widehat{h}_{ij}(\mathbf{k},\eta) $ in Fourier space
are represented in terms  of the annihilation and creation operators by
\ba
\label{tensor}
\widehat{h}_{ij}(\mathbf{k},\eta) = \sum _{s=+,\times}  \widehat{h}_{s}(\mathbf{k},\eta)
e_{ij}^{(s)}(\bfk)
 \quad , \quad \widehat{h}_{s}(\mathbf{k},\eta)=
h_{s}(k, \eta)a_{s}(\mathbf{k})+ h^{*}_{s}(k, \eta)a^{\dag}_{s}(-\mathbf{k}) \, ,
\ea
with  the commutation relations  $ [ a_{s}(\bfk),    a_{s}^\dagger  (\bfk')] = \delta_{s s'} \delta^{(3)} (\bfk-\bfk' )$.

As we mentioned before, we chose the convention that
$\mathbf{k} = k(\cos{\theta}, \sin{\theta}, 0)$. With this choice, the polarizations
$e^{+}_{ij}(\mathbf{k})$ and $e^{\times}_{ij}(\mathbf{k})$ become
\begin{align}
\label{polarization2}
e^{+}_{ij}(\mathbf{k}) =\frac{1}{\sqrt{2}} \left( \begin{array}{ccc}
 \sin^2{\theta} & -\sin{\theta}\cos{\theta} & 0 \\
  -\sin{\theta}\cos{\theta} & \cos^2{\theta} & 0 \\
  0 & 0 & -1 \\
\end{array} \right) ~~~,~~~
e^{\times}_{ij}(\mathbf{k}) = \frac{i}{\sqrt{2}}\left( \begin{array}{ccc}
  0 & 0 & -\sin{\theta} \\
  0 & 0 & \cos{\theta} \\
  -\sin{\theta} & \cos{\theta} & 0 \\
\end{array} \right) ~.
\end{align}
Using Eq. (\ref{tensor}) and Eq. (\ref{polarization2}), we find the following expression for the Fourier mode of the tensor field
\begin{align}
\label{polarization3}
\widehat{h}_{ij}(\mathbf{k}) =\frac{1}{\sqrt{2}} \left( \begin{array}{ccc}
 \widehat{h}_{+}\sin^2{\theta} & -\widehat{h}_{+}\sin{\theta}\cos{\theta} & -i\widehat{h}_{\times}\sin{\theta} \\
  -\widehat{h}_{+}\sin{\theta}\cos{\theta} & \widehat{h}_{+}\cos^2{\theta} & i\widehat{h}_{\times}\cos{\theta} \\
  -i\widehat{h}_{\times}\sin{\theta} &  i\widehat{h}_{\times}\cos{\theta}& -\widehat{h}_{+} \\
\end{array} \right) \, .
\end{align}
We will use this expression later on when  calculating the cross-correlation between the tensor mode and the curvature as well as the gauge field.

The profile of the tensor excitations has the standard form
\ba
\label{hs-wave}
{h}_{s}(k,\eta) = \frac{2 i H\eta }{M_{P}\sqrt{2 k}}\left(1-\frac{i}{ k\eta} \right)e^{-ik\eta} ~~~,~~~ (s= +, \times) \, .
\ea
The power spectrum of the tensor perturbations is given by
\ba
\langle   \widehat{h}_{ij}(\mathbf{k_1})  \widehat{h}_{ij}(\mathbf{k_2}) \rangle
= (2 \pi)^3 \delta^{(3)} (\bfk_1 + \bfk_2) P_h(k_1) \quad , \quad
\calP_h \equiv   \frac{ k_1^3  }{ 2 \pi^2}  P_h(k_1)
\ea
In the absence of anisotropy the power spectrum has the standard form
\ba
\label{calPh0}
\calP_h^{(0)} = \frac{2 H^2}{\pi^2 M_P^2}  = 16 \epsilon_H  \calP_\zeta^{(0)} \, .
\ea
Therefore, defining the tensor-to scalar ratio $r \equiv \calP_h/\calP_\zeta  $ we have $r = 16 \epsilon_H$
for the isotropic theory.

%%%%%%%%%%%%%%%%%%%%%%%%%%%%%%%%%%%%%%%%%%%%%%%%%%%%%

\subsection{The Interaction Lagrangian}
\label{interactions}

Having presented the background in some details, here we separate the Lagrangian into the free field part and interaction part. Here and below we call the latter the interaction Lagrangian.
The starting Lagrangian from the action (\ref{action}) is
\ba
\label{interaction Lagrangian1}
L = -\frac{a^4}{4}f(\phi)^2 F_{\mu \nu} F^{\mu \nu} - \frac{a^4}{2}  \e^2 \phi^2 A_{\mu} A^{\mu} \, .
\ea
Expanding the above action around the background values,  neglecting the contributions of the non-dynamical field $\delta A_0$ which are sub-leading as discussed before, and using the relation
$\left(\frac{\partial f^2}{\partial \phi}\right)\delta \phi = 4 f^2 \zeta$,
the interaction Lagrangians in the Fourier space is calculated as (see Apendix for further details)
\ba
\label{Lzetah}
L_{\zeta h_+} &=&
 - \frac{3\sqrt{2}}{2} I \epsilon_{H} M_{P}^2\sin^2{\theta}a^2\left(-\eta\right)^{-2}\left( \zeta^{*} {h}_{+} + c.c.  \right)  + \frac{\e^2 \sqrt{2}}{6}  I \epsilon_{H}M_{P}^4 \sin^2{\theta} \left(\frac{a ^4}{f^2}\right)\left( \zeta^{*} {h}_{+} + c.c. \right)  \\
\label{LzetaA}
 L_{\zeta D_1} &=& -2 M_{P}\sqrt{3I \epsilon_{H}}\sin^2{\theta} \left(\frac{a f}{\eta}\right) \left( \zeta ^{*} D'_{1} + c.c.\right) - 2 \e^2 M_{P}^3 \sqrt{\frac{I \epsilon_{H}}{3}}\sin^2{\theta} \left(\frac{a^3}{f}\right) \left( \zeta ^{*} D_{1} + c.c.\right) \\
 \label{Lh+D1}
 L_{h_+  D_1} &=& \frac{M_{P}}{2} \sqrt{\frac{3I\epsilon_{H}}{2}}\sin^2{\theta}\left( \frac{fa}{\eta}\right) \left( D^{'*}_{1} {h}_{+} +  c.c.\right) + \sqrt{\frac{I}{6\epsilon_{H}}} \e^2 M_{P}^3 \sin^2{\theta}\left( \frac{a^3}{f}\right)\left( D^{*}_{1} {h}_{+} + c.c.\right) \\
\label{LhxzD}
 L_{h_\times  D} &=&
 \frac{M_{P}}{2} \sqrt{\frac{3I \epsilon_{H}}{2}}\sin{\theta} \left(\frac{fa}{\eta}\right)\left( i D^{'} {h}^{*}_{\times} + c.c.\right) + \sqrt{\frac{I}{6\epsilon_{H}}}\e^2 M_{P}^3 \sin{\theta} \left(\frac{a^3}{f}\right)\left( i D{h}^{*}_{\times} + c.c. \right)
\ea
where c.c stands for complex conjugation.

The above interaction Lagrangians are needed in order to calculate the anisotropy corrections in
$\langle \zeta \zeta \rangle,  \langle h_s  h_{s'} \rangle$ and the cross-correlations $\langle \zeta h_s \rangle$.  Note that in the free (isotropic) theory with $I=\e =0$ there is no anisotropy corrections in power spectra and $\langle \zeta h_s \rangle$ as expected.

%%%%%%%%%%%%%%%%%%%%%%%%%%%%%%%%%%%%%%%%%%%%%%%%%%%%%%%%%%%%
\section{Anisotropic Correlations  }
\label{correlations}

Having calculated the interaction Lagrangians as given in Eqs. (\ref{Lzetah})-(\ref{LhxzD})
now we are ready to calculate the anisotropic correlation functions by using the in-in formalism.
For this purpose, we need to obtain the interaction Hamiltonian from the interaction Lagrangian.
One should notice that $H_{int} = -L_{int}$ is not necessary true with kinetically coupled interactions.
So it is worth to check it in this model before proceeding with the in-in calculation of the correlation functions. We have calculated it in the Appendix \ref{int-hamilton}. It turns out that the above formula is true for the whole of the interactions except $H_{\zeta h_{+}}$. So in the following we use $H_{int} = -L_{int}$ everywhere except that in $H_{\zeta h_{+}}$, special care is taken of.

We are interested in anisotropic contributions in $\langle \zeta_\bfk \zeta_\bfk^* \rangle  $, $\langle h_{s \, \bfk} h_{s' \, \bfk}^* \rangle  $  and $ \langle \zeta_{ \bfk} h_{s' \, \bfk}^* \rangle$. We calculate each term in turn.
Note that the wave function of the free theory for $\zeta_k, D_k, D_{1 k}$ and $h_{s k}$ are given in Eqs. (\ref{zeta-wave}),  (\ref{D-wave}) and (\ref{hs-wave}).

%%%%%%%%%%%%%%%%%%%%%%%%%%%%%%%%%%%%%%%%%%%%%%%%%%%%%
\subsection{Anisotropies in curvature power spectrum}

\label{zeta-power}

Here we calculate the anisotropic contributions in curvature perturbation power spectrum $\langle \zeta_\bfk \zeta_\bfk^* \rangle  $.  We denote the change in curvature perturbation power spectrum from the anisotropic sources by $\delta \langle \zeta_\bfk \zeta_\bfk^* \rangle  $.  This analysis were performed in  \cite{Emami:2013bk} and here we outline the analysis briefly.

We use the in-in formalism to take care of the corrections from the bilinear coupling terms \cite{Chen:2009zp, Weinberg:2005vy, Chen:2010xka, Wang:2013zva}. The leading order corrections in curvature perturbation power spectrum are given by
\ba
\label{delta-P-zeta}
\delta \langle \zeta_\bfk \zeta_\bfk^* \rangle  = - \int_{\eta_{0}}^{\eta_{e}} d\eta_{1} \int_{\eta_{0}}^{\eta_{1}}d\eta_{2} \left \langle
\bigg{[} L_{I}(\eta_{2}) , \bigg{[} L_{I}(\eta_{1}) ,
 \zeta_\bfk (\eta_e) \zeta_\bfk^*(\eta_e)  \bigg{]}\bigg{]} \right \rangle  \, ,
\ea
where $L_I$ represents the interaction Lagrangian.  The lower limit of the integral should be set
$k \eta_0 \rightarrow -\infty$  corresponding to initial  modes being deep inside the horizon. However, as studied in \cite{Bartolo:2012sd, Emami:2013bk}, the interactions responsible for anisotropies operate on super-horizon scales so to a good approximation one can safely take $k \eta_0 =-1$ corresponding to the time when the mode leaves the horizon.
The upper limit of the above integral as usual corresponds to $k \eta_e \simeq 0$.

The interaction Lagrangians relevant to
$\delta \langle \zeta_\bfk \zeta_\bfk^* \rangle$ are $L_{\zeta h_+}$ and  $L_{\zeta D_1}$  given
in Eqs. (\ref{Lzetah}) and (\ref{LzetaA}).  A look at these two equations show that $L_{\zeta h_+}$
is suppressed compared to $L_{\zeta D_1}$ by the factor $\sqrt{I \eH} \ll 1$. Therefore, the leading order anisotropic corrections in  curvature perturbation power spectrum comes from  $L_{\zeta D_1}$.
In addition, $L_{\zeta D_1}$ has two independent terms denoted by $L^{(1)}_{\zeta D_1}$ and
$L^{(2)}_{\zeta D_1}$:
\ba
L^{(1)}_{\zeta D_1} \equiv
 -2 M_{P}\sqrt{3I \epsilon_{H}}\sin^2{\theta} \left(\frac{a f}{\eta}\right) \left( \zeta ^{*} D'_{1} + c.c.\right)
  \quad , \quad
L^{(2)}_{\zeta D_1} \equiv
- 2 \e^2 M_{P}^3 \sqrt{\frac{I \epsilon_{H}}{3}}\sin^2{\theta} \left(\frac{a^3}{f}\right) \left( \zeta ^{*} D_{1} + c.c.\right)
\ea
Depending on whether one chooses either $L^{(1)}_{\zeta D_1}$ or $L^{(2)}_{\zeta D_1}$ in place of  $L_I(\eta_1)$ and $L_I(\eta_2)$ in the   integral in Eq. (\ref{delta-P-zeta}), there are four possible contributions in  $\delta \langle \zeta_\bfk \zeta_\bfk^* \rangle$ denoted by $\delta \langle \zeta_\bfk \zeta_\bfk^* \rangle_{ij}$ where $i, j=1, 2$ with the  assumption  that
$L_I(\eta_1) = L^{(i)}_{\zeta D_1}$  and  $L_I(\eta_2) = L^{(j)}_{\zeta D_1}$.
For example,    $\delta \langle \zeta_\bfk \zeta_\bfk^* \rangle_{12}$ means
$L_I(\eta_1) = L^{(1)}_{\zeta D_1}$  and  $L_I(\eta_2) = L^{(2)}_{\zeta D_1}$. With this identification we have
\ba
\label{correction zeta}
\delta  \bigg{\langle} \zeta_{\bfk}(\eta_{e}) \zeta_{\bfk}^*(\eta_{e})\bigg{\rangle} =
\delta \bigg{\langle} \zeta_{\bfk}(\eta_{e}) \zeta_{\bfk}^*(\eta_{e})\bigg{\rangle} _{11}
+\delta \bigg{\langle} \zeta_{\bfk}(\eta_{e}) \zeta_{\bfk}^*(\eta_{e})\bigg{\rangle} _{12}
+ \delta \bigg{\langle} \zeta_{\bfk}(\eta_{e}) \zeta_{\bfk}^*(\eta_{e})\bigg{\rangle} _{21}
+ \delta \bigg{\langle} \zeta_{\bfk}(\eta_{e}) \zeta_{\bfk}^*(\eta_{e})\bigg{\rangle} _{22}
\ea
The details of the in-in analysis are presented in Appendix \ref{in-in}. As a sample analysis, here we present the integral form of
$\delta  \bigg{\langle} \zeta_{\mathbf{k}_{1}}(\eta_{e}) \zeta_{\mathbf{k}_{1}}(\eta_{e})^*\bigg{\rangle}_{11} $
which is  (here and hence after, the momentum conservation $\delta$-function is omitted to save writing)
\ba
\delta  \bigg{\langle} \zeta_{\mathbf{k}_{1}}(\eta_{e}) \zeta_{\mathbf{k}_{1}}^*(\eta_{e})\bigg{\rangle}_{11}  &&=  384 I \eH M_P^2  \sin^4{\theta} \
\int_{\eta_{0}}^{\eta_{e}} d\eta_{1} \left(\frac{a f}{\eta} \right)_{\eta_1}
\im \left[ \zeta_k (\eta_{1})\zeta_k^{*}(\eta_{e})\right]  \\
&& ~~~~~~~~~~~~~~~~~~~~~~~\times \int_{\eta_{0}}^{\eta_{1}}d\eta_{2}
 \left(\frac{a f}{\eta} \right)_{\eta_2}
\im \left[ \zeta_k(\eta_{2})\zeta_k^{*}(\eta_{e})  D_{1k}^{' *} (\eta_{1}) D_{1k}^{'}(\eta_{2})
\right] \, .
\ea
Expanding the integrand for small $k \eta$ arguments and assuming $k \eta_0 = -1$ and
$k \eta_e =0$ as explained above, the above integral yields
\ba
\delta  \bigg{\langle} \zeta_{\mathbf{k}_{1}}(\eta_{e}) \zeta_{\mathbf{k}_{1}}(\eta_{e})\bigg{\rangle}_{11}&=& \frac{6 I N^2}{k_{1}^3 \epsilon_{H}} \left(\frac{H}{M_{P}} \right)^2 \sin^2{\theta}
\ea
in which $N= - \ln (-k \eta_e)$ represents the number of e-folds when the mode
$k$ has left the horizon. Taking $k$ to be the CMB scales we need $N \sim 60$ in order to solve
the flatness and the horizon problem.

Performing the same procedure for other integrals, we obtain  \cite{Emami:2013bk}
\ba
\delta  \bigg{\langle} \zeta_{\mathbf{k}}(\eta_{e}) \zeta_{\mathbf{k}}^*(\eta_{e})\bigg{\rangle}_{12}&=& -\frac{31 }{490}  \frac{\e^2 I}{k^3 {\epsilon_{H}}} \sin^2{\theta} \\
\delta  \bigg{\langle} \zeta_{\mathbf{k}}(\eta_{e}) \zeta_{\mathbf{k}}^*(\eta_{e})\bigg{\rangle}_{21}&=& -\frac{ I \e^2 N}{7 k^3 \epsilon_H}   \sin^2{\theta} \\
\delta  \bigg{\langle} \zeta_{\mathbf{k}}(\eta_{e}) \zeta_{\mathbf{k}}^*(\eta_{e})\bigg{\rangle}_{22}&=& \frac{9}{2156}  \frac{\e^4 I}{k_{1}^3}\left(\frac{M_{P}}{m} \right)^2 \sin^2{\theta}
\ea
So combining these four contributions, and assuming $N \gg 1$,  we have
\ba
\label{correction zeta final}
\delta  \bigg{\langle} \zeta_{\mathbf{k}}(\eta_{e}) \zeta_{\mathbf{k}}^*(\eta_{e})\bigg{\rangle}
&\simeq&
\left( \frac{6 I N^2}{ \epsilon_{H}} \frac{H^2}{M_{P}^2}
-\frac{ I \e^2 N}{7  \epsilon_H}
 +\frac{9 \, \e^4 I}{2156} \frac{M_{P}^2}{m^2}  \right) \left(\frac{\sin^2{\theta}}{k^3}\right)\\
 &=&   \frac{6 I N^2}{ \epsilon_{H}} \frac{H^2}{M_{P}^2}    \frac{\sin^2{\theta}}{k^3} F (\beta)
\ea
where we have defined
\ba
\label{beta-def}
\beta \equiv \frac{\e^2 }{42 N}  \left(\frac{M_P}{H} \right)^2  \quad , \quad
F(\beta) \equiv 1- \beta + \frac{9}{22} \beta^2 \, .
\ea
We have also written $m$ in terms of $H$ and $\epsilon_H$ by using $m^2 = \left(3 \epsilon_H  H^2 \right)$.\\
Note that $\beta$ is a measure of the gauge field coupling $\e^2$. In particular, in the model of \cite{Watanabe:2010fh, Bartolo:2012sd, Ohashi:2013qba} with $\e=0$ we have  $F(\beta)=1$.
With $M_P/H \sim 10^{5}$, and with $\e \gtrsim 10^{-4}$ we obtain $\beta \gtrsim 1$. For larger value of $\e$ we see that  $F(\beta)$ grows  like $\beta^2$.

The anisotropic power spectrum $\delta \calP_\zeta= \frac{k^3}{2 \pi^2} \delta {\langle} \zeta_{\mathbf{k}}(\eta_{e}) \zeta_{\mathbf{k}}^*(\eta_{e}) {\rangle}$ therefore is
\ba
\label{delta-calP-zeta}
\delta \calP_\zeta= \frac{3 I N^2 H^2 }{\eH \pi^2 M_P^2} F(\beta) \sin^2 \theta \, .
\ea
Correspondingly, the total anisotropic power spectrum  $\calP_\zeta$ is
\ba
\calP_\zeta =  \calP_\zeta^{(0)} \left[ 1+ 24 I N^2  F(\beta)  \sin^2 \theta
\right] \, ,
\ea
where $\calP_\zeta^{(0)}$ represents the isotropic power spectrum for the free theory. Note that in the limit  when $\e=\beta=0$ so $F(\beta)=1$, our result for  $\delta \calP_\zeta$ agrees with the result
in  \cite{Watanabe:2010fh, Bartolo:2012sd, Ohashi:2013qba}.

Now defining the anisotropy estimator $g_*$ via
\ba
\label{g*}
\calP_\zeta =  \calP_\zeta^{(0)} \left[ 1+  g_* \left( \widehat\bfk \cdot \widehat\bfp \right)^2 \right]
\ea
where $\bfp$ is the preferred anisotropic direction in the sky (the $x$-direction in our example), we
obtain
\ba
\label{g*2}
g_* = -24 I N^2 F(\beta) \,
\ea
It is important to note that the form of $g_*$ we have defined here is with respect to the primordial curvature power spectrum. In the TT and other correlation functions, the anisotropy not only comes from the $g_*$ here, but also comes from an ``effective $g_*$'' contribution from the tensor sector. Such an ``effective $g_*$'' has a scale dependence on the TT and other correlations because the tensor mode is decaying after it returns to the horizon. We will return to this issue later.
Also we see that the gauge coupling $\e$ appears  in $g_*$ via the parameter $\beta$.
Taking $| g_*| \lesssim 10^{-2}$ from the Planck data  constraint  \cite{Kim:2013gka} we require
$I F(\beta) \lesssim 10^{-6}$.

\begin{figure}[!t]
  \centering
  \includegraphics[width=0.6\textwidth]{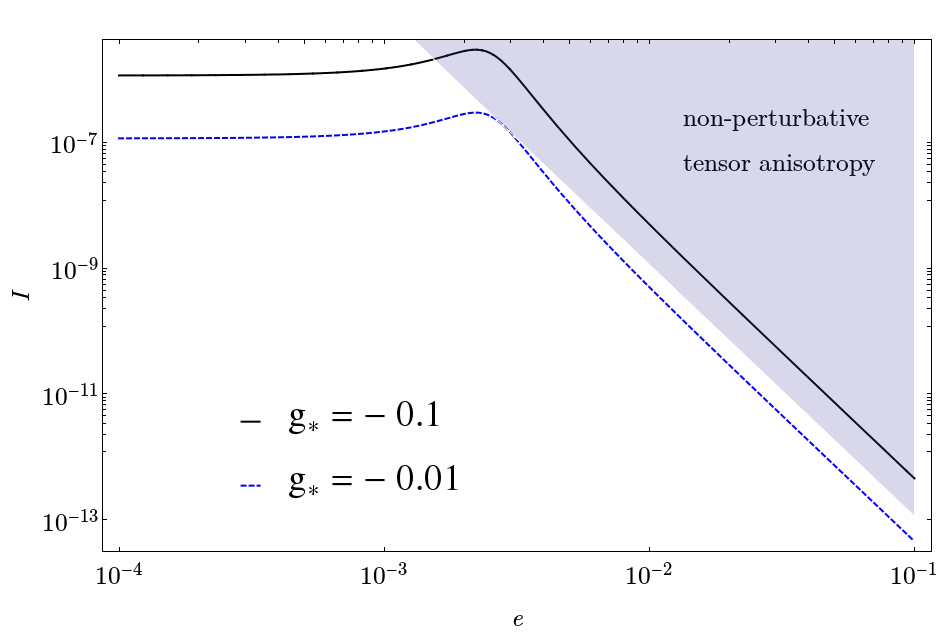}
  \caption{The  allowed range of $I$ and $\e$ for different values of $g_*$.  The shaded region
  is beyond our current scope because a large value of $\e$ induces too much anisotropy in tensor power spectrum
  which undermines our perturbative approach, i.e. $\delta \calP_h $ in Eq. (\ref{delta-Ph})
  becomes comparable to the   isotropic power spectrum $\calP_h^{(0)}$. In addition, note that we require $\e < 0.1$ in order to get large enough number of e-folds. As a result the range  $\e < 0.1$ is consistent both at the background and perturbation levels.}
  \label{Ie-plot}
\end{figure}

One may ask what the theoretical limits on the value of $\e$ or the parameter $\beta$ are. First, we have to make sure that we get enough number of e-folds of inflation at the background level. As we mentioned before, the number of e-folds depends logarithmically on $\e$
so as studied in \cite{Emami:2010rm} one can take say $\e <0. 1$ to get  a long enough
period of inflation. In addition, our assumption in parametrizing the anisotropy was that
the anisotropic  power spectrum  is smaller than the isotropic power spectrum, i.e.
$|g_*| < 1$ so our perturbative approach using the leading order in-in formalism is valid.
Therefore, demanding $|g_*| < 1$ we need  $I F(\beta) < 10^{-4}$. We have presented the contour plot of the allowed range of $I$ and $\e$ in Fig. \ref{Ie-plot}. As can be seen, we need $\e \lesssim 10^{-3}$ in order not to produce too much anisotropy in tensor perturbations (to be discussed in next subsection). Therefore, with $\e \lesssim 10^{-3}$, and $M_P/H \sim 10^5$ we have $\beta \lesssim 3$.

Checking the behavior of the function $F(\beta)$ for the approximate allowed range
$\beta \lesssim 3$ indicates that  $g_*$ has  a weak dependence on $\e$.
One can check that  for $0\le  \beta \le 3$, $F(\beta)$
takes the value in the range $0.4 \lesssim    F(\beta) \lesssim 1.7 $.  As we shall see this conclusion plays important roles for the predictions of our model for various cross-correlations.  While the anisotropy in curvature perturbation is under control for the above range  of $\beta$, the tensor perturbations become highly anisotropic when $\beta \gtrsim 1$.

Here we pause to mention one conceptual problem associated with anisotropic inflation. As seen from
Eq. (\ref{g*2}) the amplitude of anisotropy in scalar power spectrum scales like $N^2$. If inflation is prolonged in the past, this yields a large value of $g_*$ and the system becomes highly anisotropic. Therefore, our treatment of taking the anisotropies as small corrections to the isotropic
FRW background will be invalid. As pointed out in  \cite{Bartolo:2012sd} this corresponds to IR gauge field fluctuations which have left the horizon in the past inflationary history and contributed to the classical background trajectory. Therefore, a prolonged period of inflation will bring more contributions from these anisotropic IR modes which can destroy the near isotropy of the background.
In order to prevent this to happen, we demand that the total period of anisotropic inflation is under control, say less than few hundreds of e-folds.

%%%%%%%%%%%%%%%%%%%%%%%%%%%%%%%%%%%%%%%%%%%%%%%%%%%%%
%%%%%%%%%%%%%%%%%%%%%%%%%%%%%%%%%%%%%%%%%%%%%%%%%%%%%

%%%%%%%%%%%%%%%%%%%%%%%%%%%%%%%%%%%%%%%%%%%%%%%%%%%%%
\subsection{Correction to the tensor power spectrum}

Now we calculate the anisotropy in tensor power spectra
 $\langle {h}_{\bfk \times } {h}_{\bfk \times }^*\rangle$ and $\langle {h}_{\bfk + } {h}_{\bfk + }^*\rangle$.

Let us  start with  $\langle {h}_{\times } {h}_{\times }^*\rangle$.  The relevant  interaction Lagrangian is
$L_{D {h}_{\times } } =  L_{D {h}_{\times } }^{(1)} +  L_{D {h}_{\times } }^{(2)} $ where
$ L_{D {h}_{\times } }^{(1)}$ and  $ L_{D {h}_{\times } }^{(2)}$ respectively are the first and the second terms in Eq. (\ref{LhxzD}).   Following the same convention as  in our analysis for
anisotropies in curvature perturbation power spectrum we have
\ba
\delta \bigg{\langle} {h}_{\times } {h}_{\times }^*\bigg{\rangle} &=& - \int_{\eta_{0}}^{\eta_{e}} d\eta_{1}\int_{\eta_{0}}^{\eta_{1}} d\eta_{2}\bigg{[}L_{D {h}_{\times }  } ,\bigg{[}L_{D {h}_{\times}}, {h}_{\times\mathbf{k}} {h}_{\times\mathbf{k}}\bigg{]}\bigg{]} \nonumber\\
&=& \delta \bigg{\langle} \widehat{h}_{\times }\widehat{h}_{\times }\bigg{\rangle}_{11} +  \delta \bigg{\langle} \widehat{h}_{\times }\widehat{h}_{\times }\bigg{\rangle}_{12} + \delta \bigg{\langle} \widehat{h}_{\times }\widehat{h}_{\times }\bigg{\rangle}_{21} + \delta \bigg{\langle} \widehat{h}_{\times }\widehat{h}_{\times }\bigg{\rangle}_{22}
\ea
The details of the in-in analysis are presented in Appendix \ref{in-in}. The results for each contribution are
\ba
\label{tensor times 1}
\delta \bigg{\langle} {h}_{\times } {h}_{\times }^*\bigg{\rangle}_{11}
&=& \left(\frac{12}{k^3}\right)\left(\frac{H}{M_{P}}\right)^2 I \epsilon_{H} N^2\sin^2{\theta}
\\
\label{tensor times 2}
\delta \bigg{\langle} {h}_{\times}  {h}_{\times}^*\bigg{\rangle}_{12}&=&  -\left(\frac{4}{7 k^3}\right)N I\e^2\sin^2{\theta}\\
\label{tensor times 3}
\delta \bigg{\langle} {h}_{\times}  {h}_{\times}^*\bigg{\rangle}_{21}&=&  -\left(\frac{62}{245 k^3}\right) I \e^2 \sin^2{\theta}
\\
\label{tensor times 4}
\delta \bigg{\langle} {h}_{\times}  {h}_{\times}^*\bigg{\rangle}_{22}
&=& \left(\frac{6 }{539 k^3 }\right) \left(\frac{M_{P}}{H}\right)^2 \left(\frac{I \e^4}{\epsilon_{H}}\right) \sin^2{\theta}
\ea

Summing up those four terms we get
\ba
\label{correction hh times final}
\delta \bigg{\langle} \widehat{h}_{\times\mathbf{k}_{1}}  \widehat{h}_{\times\mathbf{k}_{2}}\bigg{\rangle}
&\simeq& \left( 12I \epsilon_{H}N^2 \frac{H^2}{M_{P}^2} -\frac{4}{7}N I\e^2 + \frac{6 I \e^4}{539 \eH} \frac{M_{P}^2 }{H^2 } \right) \frac{\sin^2{\theta}}{k^3} \nonumber\\
&=&  12I \epsilon_{H}N^2 \frac{H^2}{M_{P}^2} F(\hat \beta)  \frac{\sin^2{\theta}}{k^3}
\ea
where the function $F(x)$ is defined in Eq. (\ref{beta-def}) and $\hat \beta$ is related to
$\beta$ via
\ba
\hat \beta \equiv \frac{2 \beta}{\epsilon_H} \, .
\ea
Note the crucial point that $\hat \beta$ is enhanced compared to $\beta$ by the factor $1/\epsilon_H$.
As we have discussed before, from the observational and the theoretical constraints on $\beta$ we have
$\beta \lesssim 1$. Now, from the above relation between $\hat \beta$ and $\beta$, we see that
$\hat \beta$ can be as large as $100$ for $\epsilon_H \sim 0.01$. As we shall see shortly, the anisotropy in tensor power spectra becomes very strong for large $\e$ so there will be  upper bound on $\e$ and
$\hat \beta$.

Now we calculate   $\langle {h}_{+ }{h}_{+}^*\rangle$. In this case the relevant interaction Lagrangians are  $L_{D_{1} {h}_{+ } }$ and $ L_{\zeta {h}_{+ }  }$ so we have
\ba
\label{tensor plus}
\delta \bigg{\langle} {h}_{+} {h}_{+}\bigg{\rangle} &=& - \int_{\eta_{0}}^{\eta_{e}} d\eta_{1}\int_{\eta_{0}}^{\eta_{1}} d\eta_{2}\bigg{[}L_{D_{1} {h}_{+ }  } ,\bigg{[}L_{D_{1} {h}_{+}}, {h}_{+} {h}_{+}\bigg{]}\bigg{]} - \int_{\eta_{0}}^{\eta_{e}} d\eta_{1}\int_{\eta_{0}}^{\eta_{1}} d\eta_{2}\bigg{[}L_{\zeta {h}_{+ }  } ,\bigg{[}L_{\zeta {h}_{+}}, {h}_{+} {h}_{+}\bigg{]}\bigg{]}
\ea
Comparing $L_{D_{1} {h}_{+ } }$ and $ L_{\zeta {h}_{+ }  }$ we see that $ L_{\zeta {h}_{+ }  }$ is suppressed compared to $L_{D_{1} {h}_{+ } }$ by a factor $\sqrt{I} \ll 1$
so to leading order in anisotropy we can neglect the contribution from  $ L_{\zeta {h}_{+ }  }$ in
$\langle {h}_{+ }{h}_{+}^*\rangle$. Now the analysis is exactly the same as what we performed
in  $\langle {h}_{\times }{h}_{\times}^*\rangle$ and therefore
\ba
\label{tensor plus final}
\delta \bigg{\langle} {h}_{+} {h}_{+}\bigg{\rangle} =
\delta \bigg{\langle} {h}_{\times} {h}_{\times}\bigg{\rangle} \, .
\ea

To summarize,  the anisotropy in total tensor power spectrum is
\ba
\label{delta-Ph}
\delta \calP_h &=& 2 \left( \frac{k^3}{2 \pi^2}   \right)
  \delta \bigg{\langle} {h}_{\times} {h}_{\times}^*\bigg{\rangle} \nonumber\\
&=&   24 I \epsilon_{H}N^2 \frac{H^2}{M_{P}^2} F(\hat \beta)  \sin^2{\theta} ~.
\ea
so the total tensor power spectrum is
\ba
\calP_h = \calP_h^{(0)} \left( 1 +  6   I \epsilon_{H}N^2 F(\hat \beta) \sin^2 \theta ~.
\right)
\ea
This is an interesting formula indicating that the effects of the gauge coupling is very strong
in tensor power spectrum anisotropy. This is because $\hat \beta = 2 \beta/\epsilon_H$ so with
$\beta \sim 1$ we gate  $\hat \beta \sim 100$  and therefore
$\delta \calP_h /\calP_h^{(0)} \simeq 24 I N^2 (\beta^2/\epsilon_H) = |g_*| \beta^2/\epsilon_H$.
With $| g_*| \sim O(\epsilon_H)$, which is consistent with the observational constraints and with
$\beta \sim 1$, one easily gets to the regime in which  $\delta \calP_h /\calP_h^{(0)} \simeq1 $.
Note that the epsilon enhancement for the charged interaction is a very special feature of this model, from the specific form of the potential. Explicitly, since our interaction includes $\e^2 \phi^2 A^2$, the charged contribution in $\langle \zeta \zeta\rangle$ comes from $(\e^2 \phi A \delta\phi \delta A )^2$, while the charged contribution in $\langle hh\rangle$ comes from $(\e^2 \phi^2 A h_{ij} \delta A)^2$. So we see that the ratio between these two effects controls with $\phi^2$.
In the chaotic inflation, $\phi^2$ is proportional to $1/\epsilon$. \footnote{On the other hand, for the symmetry breaking potential, the ratio would be $\epsilon$, so no hope to see any enhancement for this case.}
This signals the strong dependence of the tensor anisotropies to the gauge coupling. Of course, we can not trust our analysis when we approach the limit $\delta \calP_h /\calP_h^{(0)} \simeq1 $. This is because, we have followed a perturbative approach and only kept the leading interaction terms in
our in-in analysis. Our situation  is in contrast to models of anisotropic inflation with a real inflaton field, as studied in  \cite{Ohashi:2013qba}, where $\e= \beta=0$,  so $F(\beta) =1$ and
$\delta \calP_h /\calP_h^{(0)} = - g_* \eH /4$ which is highly suppressed.

Demanding that $\delta \calP_h < \calP_h^{(0)}$ so our theoretical analysis is under perturbative control, we obtain the following upper bound
on the parameter $\beta$
\ba
\label{beta-cons}
\beta \lesssim  \sqrt{\frac{\eH}{ |g_* | }} \, .
\ea
With $\epsilon_H \sim 10^{-2}$  and $| g_*| \lesssim 10^{-2}$ we conclude that $\beta \lesssim 1$
in order for the anisotropic contribution in tensor power spectrum to be under control, corresponding to  $\e \lesssim 10^{-3}$.  The contour plot of  $I $ versus $\e$ is shown in Fig. \ref{Ie-plot}. The strong constraints on the allowed range of $\e$ comes from the tensor power spectrum.

%%%%%%%%%%%%%%%%%%%%%%%%%%%%%%%%%%%%%%%%%%%%%%%%%%%%%
\subsection{Cross-Correlation between $\zeta$ and $h_{ij}$}

Here we calculate the cross-correlation $\zeta$ and $h_{ij}$. We should calculate the following terms,
\ba
\bigg{\langle} \zeta_{\mathbf{k}_{1}}(\eta_{e}) {h}_{+\mathbf{k}_{2}}(\eta_{e}) \bigg{\rangle} &=& i \int_{\eta_{0}}^{\eta_{e}} d\eta_{1} \bigg{\langle}   \bigg{[} H_{\zeta {h}_{+}} , \zeta_{\mathbf{k}_{1}}{h}_{+\mathbf{k}_{2}}\bigg{]}   \bigg{\rangle}
- \int_{\eta_{0}}^{\eta_{e}} d\eta_{1}\int_{\eta_{0}}^{\eta_{1}} d\eta_{2}
\bigg{\langle} \bigg{[}L_{\zeta D_{1}} ,\bigg{[}L_{D_{1} {h}_{+}}, \zeta_{\mathbf{k}_{1}}{h}_{+\mathbf{k}_{2}}\bigg{]}\bigg{]}  \bigg{\rangle} \nonumber\\
&& -\int_{\eta_{0}}^{\eta_{e}} d\eta_{1}\int_{\eta_{0}}^{\eta_{1}} d\eta_{2}
\bigg{\langle} \bigg{[} L_{D_{1} {h}_{+}},\bigg{[} L_{\zeta D_{1}}, \zeta_{\mathbf{k}_{1}}{h}_{+\mathbf{k}_{2}}\bigg{]}\bigg{]}  \bigg{\rangle}
\nonumber\\
&\equiv& \bigg{\langle} \zeta_{\mathbf{k}_{1}}(\eta_{e}) {h}_{+\mathbf{k}_{2}}(\eta_{e})\bigg{\rangle}_{1} + \bigg{\langle} \zeta_{\mathbf{k}_{1}}(\eta_{e}) {h}_{+\mathbf{k}_{2}}(\eta_{e})\bigg{\rangle}_{2} + \bigg{\langle} \zeta_{\mathbf{k}_{1}}(\eta_{e}) {h}_{+\mathbf{k}_{2}}(\eta_{e})\bigg{\rangle}_{3}
\ea
where the indices 1, 2 and 3 indicate the above three integrals respectively. The nested integrals 2 and 3 each have four different contributions as in previous analysis so in total we have nine contributions in the above cross correlation.  The details of the analysis are given in Appendix \ref{in-in} and here we present the final result:
\ba
\label{final cross correlation}
\bigg{\langle} \zeta_{\mathbf{k}_{1}}(\eta_{e}) {h}_{+\mathbf{k}_{2}}(\eta_{e})\bigg{\rangle} &\simeq&
I \left(- 6\sqrt{2} \frac{ N^2 H^2 }{M_{P}^2}  + \frac{  \sqrt{2} \e^2 N}{7 \epsilon_{H} }  - \frac{3\sqrt{2} \e^4 }{1078 \epsilon_{H} } \frac{M_{P}^2}{H^2}  \right) \frac{\sin^2{\theta}}{k^3}
\ea
Eq. (\ref{final cross correlation}) is the final result for the cross-correlation.

Finally one can easily check that
\ba
\bigg{\langle} \zeta_{\mathbf{k}_{1}}(\eta_{e}) {h}_{\times\mathbf{k}_{2}}(\eta_{e})\bigg{\rangle}=0 \, .
\ea
This is because at the second order level $\zeta $ does not see ${h}_{\times}$.

The power spectrum of $\langle \zeta h \rangle$ cross-correlation is therefore
\ba
\label{calP-zh}
\calP_{\zeta h} &=& \frac{k^3}{2 \pi^2 } \bigg{\langle} \zeta_{\mathbf{k}}(\eta_{e}) {h}_{+\mathbf{k}^*}(\eta_{e})\bigg{\rangle}  \nonumber\\
&=& -24 \sqrt{2}  I N^2 \eH \calP_\zeta^{(0)} G(\beta) \sin^2 \theta
\ea
where the function $G(\beta)$ is defined via
\ba
\label{G-beta}
G(\beta) \equiv 1- \frac{\beta }{\epsilon_H} + \frac{9}{11} \frac{\beta^2 }{\epsilon_H} \, .
\ea
For typical value of $\epsilon_H \ll 1$, the function  $G(\beta)$ has two positive roots $\beta_1$ and $\beta_2$ where  $\beta_1  \ll 1$ and $\beta _2 \gtrsim  1$. Thus the function $G(\beta)$ is negative in the range $\beta_1 < \beta < \beta_2$ while it is positive beyond this region. Therefore, with appropriate
choice of $\e$ or $\beta$ the cross-correlation $\calP_{\zeta h} $ can have both signs, i.e. $\zeta$ and $h$
can be either correlated or anti-correlated.

Alternatively, one can also write $\calP_{\zeta h} $ in terms of $g_*$ as (note that $g_* <0$)
\ba
\calP_{\zeta h} =  \sqrt{2} g_*  \eH \calP_\zeta^{(0)} \frac{G(\beta)}{F(\beta)} \, .
\ea
Note that in the limit where $\e= \beta=0$ so $F(\beta) = G(\beta) =1$, the above expression coincides with the result obtained in  \cite{Ohashi:2013qba}. With $\beta \sim1$ the ratio $\calP_{\zeta h}/\calP_{\zeta}^{(0)} $ in our model is about one or two orders of magnitude bigger than the result in
 \cite{Ohashi:2013qba} in which $\beta =0$.

Before closing this Section and presenting our numerical results for various correlations, let us summarize the main results of our model.
The anisotropy in curvature perturbation power spectrum is given by Eq. (\ref{delta-calP-zeta})
with $g_*$ given in Eq. (\ref{g*}). As discussed below  Eq. (\ref{g*}), $\delta \calP_\zeta$ and $g_*$ depend weakly on $\beta$ so we do not get strong constraints on the value of $\e$ from the constraints on curvature perturbations anisotropies. On the other hand, the anisotropic tensor power spectrum is given in Eq. (\ref{delta-Ph}). The crucial point is that $\delta \calP_h$ scales
with $F (\hat \beta)$ where $\hat \beta = 2 \beta/\epsilon_H$. With $\beta \sim 1$  we get
$\hat \beta \sim 100$ which yields  an enhancement $\sim 10^4$ from the function  $F (\hat \beta)$.
This is a novel effect indicating that while the scalar perturbations are well-constrained to be statistically symmetric, the tensor perturbations show strong directional dependence. This is the motivation for careful scrutiny of B-mode polarizations for the TB, EB and BB correlations in the upcoming Planck polarization maps.
Finally the cross-correlation of scalar-tensor,  $\calP_{\zeta h}$,  is given in Eq. (\ref{calP-zh}). The situation here is a hybrid of the above two limits of $\delta \calP_h$ and $\delta \calP_\zeta$.
For large enough value of $\e$, i.e. with $  \beta \sim 1$, we get an enhancement of order 10-100 from
the function $G(\beta)$.

%%%%%%%%%%%%%%%%%%%%%%%%%%%%%%%%%%%%%%%%%%%%%%%%%%%%%

%%%%%%%%%%%%%%%%%%%%%%%%%%%%%%%%%%%%%%%%%%%%%%%%%%%%%

\section{From the primordial fluctuations to the CMB}
\label{numerical}
In this section, we shall relate the calculation of the above sections to CMB anisotropies, using the same method as in \cite{Watanabe:2010bu}. We use the spin weighted spherical harmonics in our analysis, \cite{Kamionkowski:1996zd, Hu:1997hp}.

To calculate the CMB anisotropies, one has to project the three-momentum onto two dimensional spherical harmonics. The $\int d^3k$ integral breaks up into two pieces with the presence of statistical anisotropy -- the radius part and the angular part. The correlators of CMB observables takes the form
\begin{align}\label{eq:aXaX}
\langle a^{X_1}_{l_1, m_1} a^{X_2}_{l_2, m_2}\rangle = 4 \pi
\int \frac{dk}{k} \Delta_{l_1}^{i_1 X_1}(k) \Delta_{l_2}^{i_2 X_2}(k)
\int  d\Omega ~  [{}_{i_1}Y^*_{l_1m_1}(\theta, \phi)] [{}_{i_2}Y_{l_2m_2}(\theta, \phi)] P^{i_1, i_2}(k, \theta, \phi)~,
\end{align}
where $X^i$ takes value (T, E, B), which are the temperature anisotropy, the E-mode and B-mode respectively. Here although the dimensionless power spectrum in principle depends on $k$, we approximate it as scale invariant. The effects on the plots are tinny\footnote{On the other hand, the scale dependence is needed once data analysis is to be performed for this model and the modification is straightforward if the scale dependence can be written in a factorizable form $P^{i_1, i_2}(k, \theta, \phi) = \sum_n f_n(k)g_n(\theta,\phi)$.}. Also, in our case there is no $\phi$-dependence for the power spectrum and the $\theta$-dependence takes the shape $P^{i_1, i_2} = P^{i_1, i_2}(\sin^2 \theta)$. With the absence of $\phi$-dependence, the momentum along the $\phi$-rotation is conserved and thus $\langle a^{X_1}_{l_1, m_1} a^{X_2}_{l_1, m_1}\rangle$ is non-vanishing only when $m_1 = m_2$.

However, the rotational symmetry on the $\theta$ direction is broken. As a result, the correlation $\langle a^{X_1}_{l_1, m_1} a^{X_2}_{l_1, m_1}\rangle$ is not restricted to $l_1=l_2$. Instead, other than those diagonal correlations, we also have $l_1=l_2\pm1$ for TB and EB, and $l_1=l_2\pm 2$ for TT, TE, EE and BB respectively.

In \eqref{eq:aXaX}, the $i_1$ and $i_2$ are indices indicating the spin of the component. And the ${}_{i}Y^*_{lm}(\theta, \phi)$ is the spin-$i$-weighted spherical harmonics. On the $P^{i_1, i_2}(\sin^2 \theta)$ side, the correlation functions on the spin bases takes the form
\begin{align}
  P^{0,0} = P^{\zeta\zeta}~, \quad P^{0,\pm2} = P^{\pm2,0} = \frac{1}{\sqrt2} P^{0,+}~, \quad
  P^{\pm2, \pm2} = \frac{1}{2} \left( P^{++} + P^{\times\times}\right)~, \quad P^{\pm2, \mp2} = \frac{1}{2} \left( P^{++} - P^{\times\times}\right)~.
\end{align}
Note that  $P^{\zeta\zeta} \equiv P_{\zeta}$, $P^{0,+} \equiv P_{\zeta h_+}$, $P^{++}\equiv P_{h_+}$ and $P^{\times\times} \equiv P_{h_\times}$ components are given by  equations \eqref{correction zeta final}, \eqref{final cross correlation}, \eqref{correction hh times final} and \eqref{tensor plus final} respectively.

Among those angular integrals $\int  d\Omega ~  [{}_{i_1}Y^*_{l_1m_1}(\theta, \phi)] [{}_{i_2}Y^*_{l_2m_2}(\theta, \phi)] P^{i_1, i_2}(\sin^2\theta)$, we would like to emphasize a particularly interesting integral:
\begin{align}
 \int  d\Omega ~  [{}_{0}Y^*_{lm}(\theta, \phi)] [{}_{2}Y_{lm}(\theta, \phi)] \sin^2\theta ~.
\end{align}
Note that this term picks up cross correlation of $\langle\zeta h_+\rangle$, and maps it onto TT or other CMB anisotropies. It is claimed in \cite{Contaldi:2014zua} that such a term, with anti-correlated $\langle\zeta h_+\rangle$ (which is indeed possible in our case), suppresses the TT power spectrum and thus reconciles the current tension between BICEP2 and Planck.

However, we shall show that this is unfortunately not true. By expanding ${}_{2}Y^*_{lm}(\theta, \phi)$ into the spin 0 spherical harmonics, the $\sin^2\theta$ factor is cancelled and the integral ends up to be
\begin{align}
  \langle T_l^\zeta T_l^h\rangle \propto \sum_{m=-l}^{l} \int  d\Omega ~  [{}_{0}Y^*_{lm}(\theta, \phi)] [{}_{2}Y_{lm}(\theta, \phi)] \sin^2\theta \propto \sum_{m=-l}^{l} [3m^2-l(l+1)] = 0~,
\end{align}
where $T_l^\zeta $ and $T_l^h$ denotes the contribution from $\zeta$ and $h$ to the temperature fluctuations, respectively. Thus no contribution from $\zeta h_+$ to $\langle TT\rangle$ could be observed once summing over $m$. On the other hand, before summing over $m$, the correlation does exist. But this does not help reconciling the tension between Planck and BICEP2 because the data analysis by the Planck team (for the purpose of tensor to scalar ratio) indeed sums over $m$. The above conclusion can actually be generalized to any primordial power spectrum, not restricted to isotropic inflation. The details of the general no-go result is presented in \cite{Emami:2014xga}.

The transfer function part encodes very complicated late time physics. Fortunately this part does not have angular dependence thus one can use the standard Boltzmann code for the calculation. Here ``the Cosmic Linear Anisotropy Solving System'' (CLASS) \cite{Blas:2011rf} is used for solving those transfer functions, where the cosmological parameters are chosen to be the same as those used by the BICEP2 group.

Now for the purpose of studying statistical anisotropies, we are not to sum over $m_1$ and $m_2$ (unlike the case above) because the summation would average away some signals of the anisotropy. Rather, we leave $m_1=m_2=m$ free and look into $m=0$ and $m=l$ cases respectively for illustration.

Three sets of parameters are examined numerically:
\begin{itemize}
\item Case I: Real Inflaton field with no electric gauge coupling:  In this class of models we plot $I=10^{-7}$ and $\e=0$. This case is the same as previous studies on the chargeless scalar field \cite{Watanabe:2010bu}. On the plots, Case I is shown in blue color.
\item Case II: Balanced: In this class of models we plot $I=10^{-7}$ and $\e=10^{-3}$. Such choice of parameters does not significantly modify $g_*$ of the scalar sector. However, the gravitational sector is largely modified because the two point correlation functions with tensors are more sensitive to the charge of the complex inflaton, as we have seen in the previous sections. On the plots, Case II is shown in black color.
\item Case III: Charge coupling dominated: In this class of models we plot $I=10^{-11}$ and $\e=0.025$. In this case  $g_*$, which is a measure of the scalar power spectrum anisotropy,
is considerably smaller than Cases I and II since $g_*$
is mostly sensitive to $I$ than $\e$. However, the charge contribution is large in the tensor sector and is marginally under control to calculate the statistical anisotropy perturbatively (where the anisotropic term is about 34\% of the isotropic term). On the plots, Case III is shown in red color.
\end{itemize}

By choosing the above parameters, we have taken into consideration that the anisotropies in the temperature correlations cannot be large, making use of Planck data in range $2 \leq l \leq 2000$ \cite{Kim:2013gka}. On the other hand, if scale dependence of the anisotropy is allowed, the constraint may not be so strong. In that case, we would have larger effects from anisotropies and our plots are shifted upwards.

The figures are plotted in logarithm scales because sometimes different cases differ by order-of-magnitude. However, we note that in various cases the correlations could go negative. To represent as detailed information as possible, we shall use solid lines to denote the logarithm of a quantity, where the quantity is positive (for example, $\log C_{l, l+2}^{TT}$) and dashed lines for multiplying $-1$ before taking logarithm (for example, $\log (- C_{l, l+2}^{TT})$), where the quantity is negative.

\subsection{The TT and BB modes}

\begin{figure}[!h]
  \centering
  \includegraphics[width=0.45\textwidth]{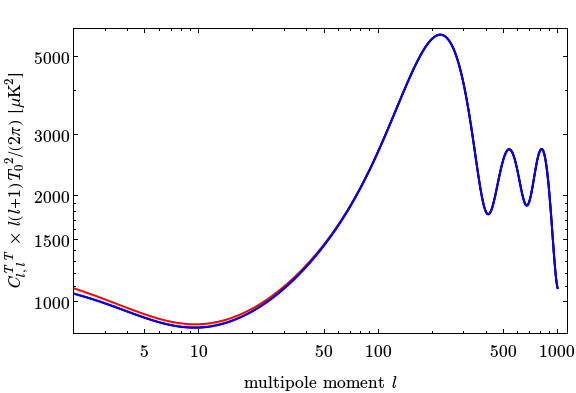}
  \hspace{0.05\textwidth}
  \includegraphics[width=0.45\textwidth]{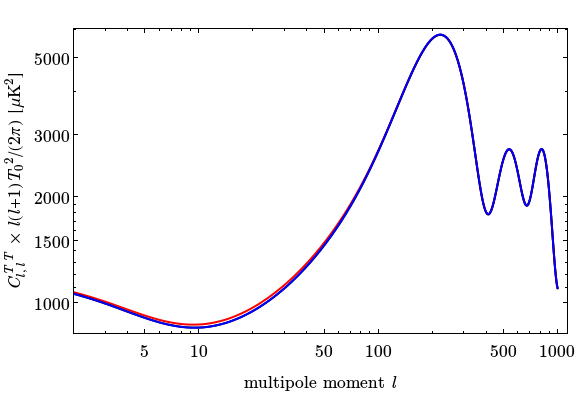}
  \caption{\label{fig:TT0} The TT correlation at $l_2=l_1$. Here and hence after, the blue curve denotes Case I, with $I=10^{-7}$ and $\e=0$; the black curve denotes Case II, with $I=10^{-7}$ and $\e=10^{-3}$. The red curve denotes Case III, with $I=10^{-11}$ and $\e=0.025$. The left panel is for $m=0$ and the right panel is for $m=l_1$. It is important to note that the anisotropic contribution in Case III is considerably larger than the anisotropic contribution in case I and II at low $l$. But the contribution decays at high $l$. Here the black and blue curves are not visibly distinguishable from each other, which also coincides with the isotropic power spectrum, because of the tight constraint on the $g_*$ of the scalar sector.}
\end{figure}

\begin{figure}[!h]
  \centering
  \includegraphics[width=0.45\textwidth]{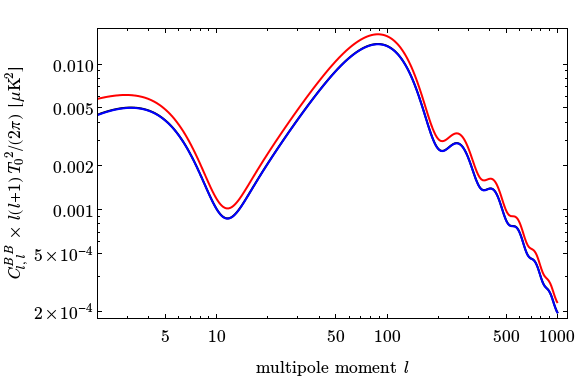}
  \hspace{0.05\textwidth}
  \includegraphics[width=0.45\textwidth]{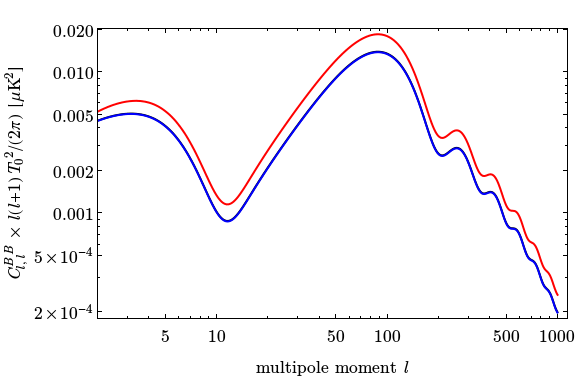}
  \caption{\label{fig:BB0} The $m=0$ (left) and $m=l_1$ (right) plots for BB correlation with $l_2=l_1$. }
\end{figure}

\begin{figure}[!h]
  \centering
  \includegraphics[width=0.45\textwidth]{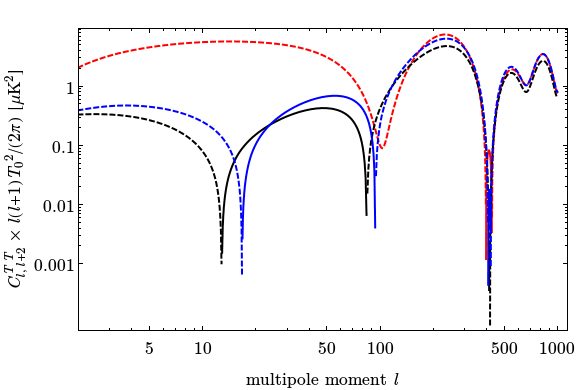}
  \hspace{0.05\textwidth}
  \includegraphics[width=0.45\textwidth]{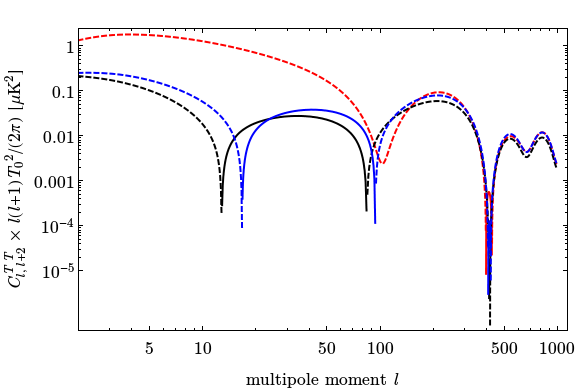}
  \caption{\label{fig:TT2} The $m=0$ (left) and $m=l_1$ (right) plots for TT correlation with $l_2=l_1+2$. Here and hence after, the dashed lines denote the plotted quantity (here $C_{l, l+2}^{TT}$) is negative along this line segment, and thus we plot $-C_{l, l+2}^{TT}$ on the logarithm scales.}
\end{figure}

\begin{figure}[!h]
  \centering
  \includegraphics[width=0.45\textwidth]{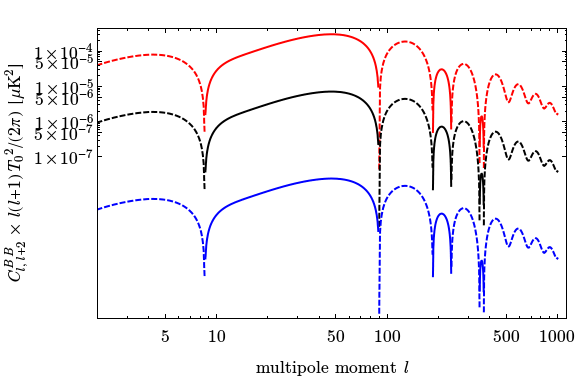}
  \hspace{0.05\textwidth}
  \includegraphics[width=0.45\textwidth]{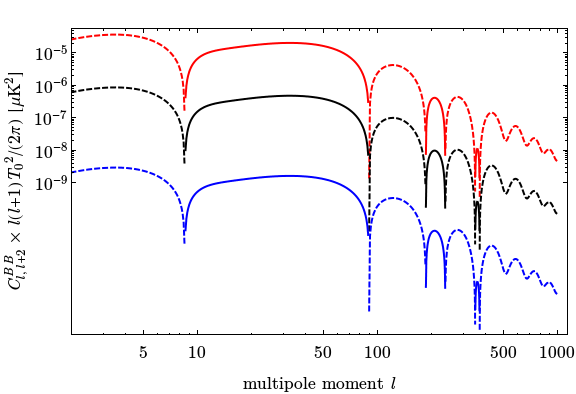}
  \caption{\label{fig:BB2} The $m=0$ (left) and $m=l_1$ (right) plots for BB correlation with $l_2=l_1+2$.}
\end{figure}

With the above choice of parameters and conventions, here we plot the TT and BB power spectra in Figs.~\ref{fig:TT0}, \ref{fig:BB0}, \ref{fig:TT2} and \ref{fig:BB2}. In Figs.~\ref{fig:TT0} and \ref{fig:BB0} the $l_1=l_2$ part of the correlation function is plotted and in Figs.~\ref{fig:TT2} and \ref{fig:BB2} with $l_2 = l_1 +2$. The left and right panels corresponds to $m=0$ and $m=l_1$ respectively. From Figs.~\ref{fig:TT0}, \ref{fig:BB0}, one observes that the anisotropic modification to the standard TT and BB power spectra are slight but still visible.

\subsection{TB and EB correlations}

\begin{figure}[!h]
  \centering
  \includegraphics[width=0.45\textwidth]{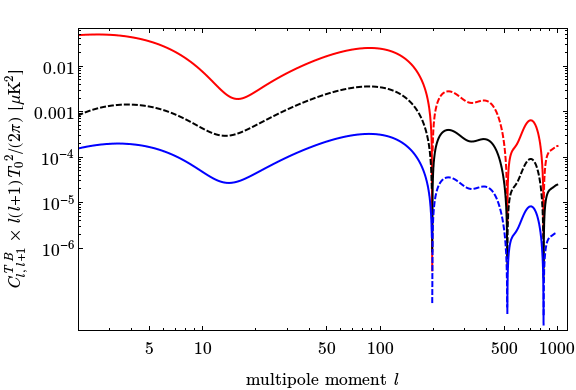}
  \hspace{0.05\textwidth}
  \includegraphics[width=0.45\textwidth]{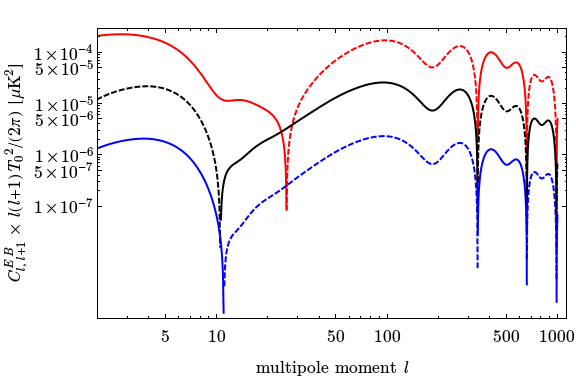}
  \caption{\label{fig:TBEB1} The $m=l_1$ plots for TB (left) and EB (right) correlation with $l_2=l_1+1$.}
\end{figure}

In the anisotropic case, the TB and EB correlations are opened up, with $l_2 = l_1 \pm 1$. In Fig.~\ref{fig:TBEB1}, those cross correlations are plotted.

\subsection{TE and EE correlations}

\begin{figure}[!h]
  \centering
  \includegraphics[width=0.45\textwidth]{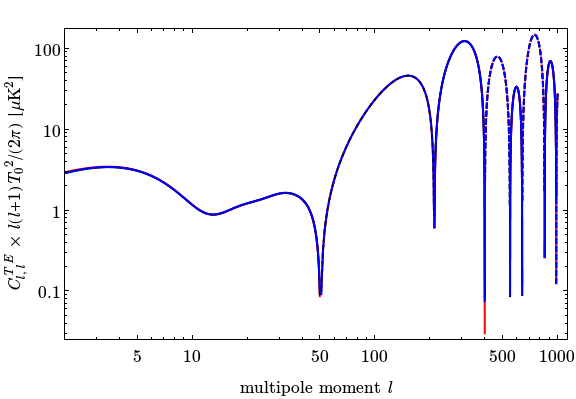}
  \hspace{0.05\textwidth}
  \includegraphics[width=0.45\textwidth]{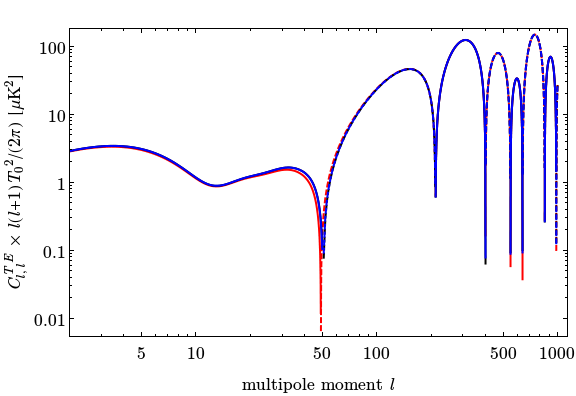}
  \caption{\label{fig:TE0} The $m=0$ (left) and $m=l_1$ (right) plots for TE correlation with $l_2=l_1$.}
\end{figure}

\begin{figure}[!h]
  \centering
  \includegraphics[width=0.45\textwidth]{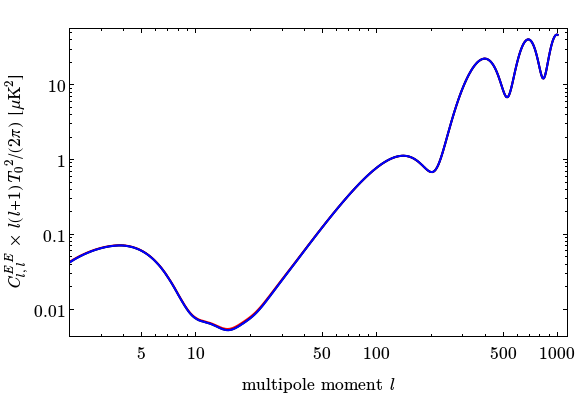}
  \hspace{0.05\textwidth}
  \includegraphics[width=0.45\textwidth]{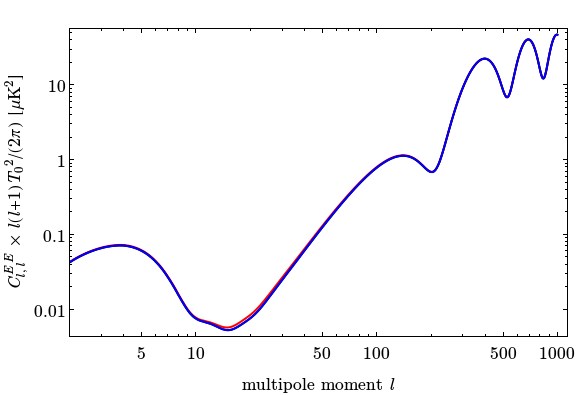}
  \caption{\label{fig:EE0} The $m=0$ (left) and $m=l_1$ (right) plots for EE correlation with $l_2=l_1$.}
\end{figure}

\begin{figure}[!h]
  \centering
  \includegraphics[width=0.45\textwidth]{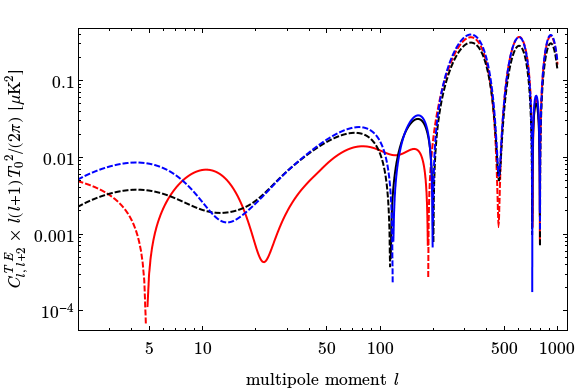}
  \hspace{0.05\textwidth}
  \includegraphics[width=0.45\textwidth]{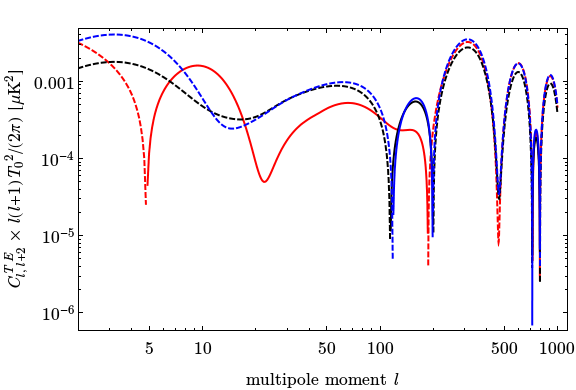}
  \caption{\label{fig:TE2} The $m=0$ (left) and $m=l_1$ (right) plots for TE correlation with $l_2=l_1+2$.}
\end{figure}

\begin{figure}[!h]
  \centering
  \includegraphics[width=0.45\textwidth]{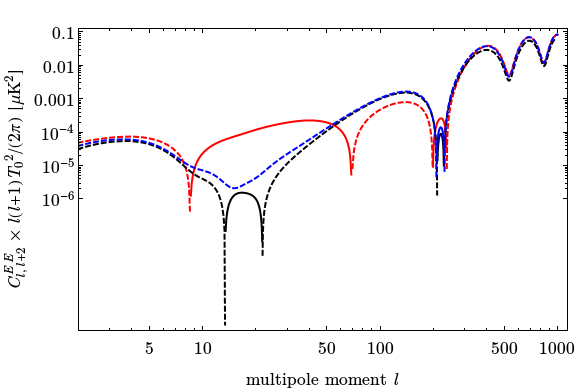}
  \hspace{0.05\textwidth}
  \includegraphics[width=0.45\textwidth]{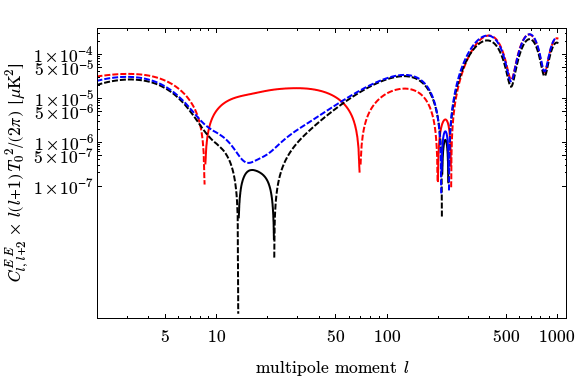}
  \caption{\label{fig:EE2} The $m=0$ (left) and $m=l_1$ (right) plots for EE correlation with $l_2=l_1+2$.}
\end{figure}

In Figs. \ref{fig:TE0}, \ref{fig:EE0}, \ref{fig:TE2} and \ref{fig:EE2}, the TE and EE correlations for $l_2=l_1$ and $l_2=l_1+2$ are plotted respectively.

%%%%%%%%%%%%%%%%%%%%%%%%%%%%%%%%%%%%%%%%%%%%%%%%%%%%%
\section{Conclusion and discussion}

In this work we have studied  statistical anisotropies in model of anisotropic inflation from a charged inflaton field. More specifically, we work in large field inflation model with the chaotic potential  $V(\varphi) = \frac{1}{2}m^2 |\varphi|^2$ with the conformal coupling $f $ given
in Eq. (\ref{f-form0}).  It is worth to emphasis that our results, especially the enhancement of the tensor mode, is because we work in large field model. We have calculated the anisotropies generated in $\langle \zeta \zeta \rangle $ and $\langle h_{s}  h_{s} \rangle$ with the tensor polarizations $s=+, \times$. In addition, we also calculated the cross-correlation $\langle \zeta  h_{s} \rangle$ which is sourced in the  anisotropic inflationary background. Our general conclusion is that
the anisotropy in curvature perturbation power spectrum $\delta \calP_\zeta$ can be small so one can easily satisfy the observational bound  \cite{Kim:2013gka} $|g_*| \lesssim 10^{-2} $.  However, the effects of gauge coupling $\e$ appears strongly in tensor perturbations. In particular, we have shown that while $g_*$ is small enough to be within the observational bound, $\delta \calP_h$ induced from anisotropies  can be comparable to the background tensor power spectrum $\calP_h$.

The induced anisotropy of TT spectrum from the primordial tensor fluctuations can obtain an ``effective $g_*$''  as large as a few percents at low $l$. This effective $g_*$ contribution is decaying towards high $l$, characterized by the tensor to temperature transfer function. It remains interesting to explore CMB anisotropies with this new profile of TT anisotropy.

We would also like to mention that the mechanism we have used to generate small $l$ CMB anomalies from the tensor sector is quite general. It may also be applied to other kinds of CMB anomalies such as non-Gaussianities, hemispherical asymmetry, etc. In particular, the possibility that
hemispherical asymmetry in tensor modes which are generated from the long mode modulations  \cite{Dai:2013kfa, Abolhasani:2013vaa} may be behind the tension between the BICEP2 and
Planck has been emphasized in \cite{Chluba:2014uba}.
It is also another interesting way to get out of this discrepancy as it has been shown by \cite{Kamionkowski:2014faa}.
We hope we can address these issues in a future work.

Note that we have assumed that the anisotropies are small and have expanded those anisotropies at linear order. This is a very good assumption for models with no electric charge since both the scalar sector and the tensor sector remains nearly isotropic as studied in \cite{Ohashi:2013qba}.
However, in the presence of the  gauge coupling, things become different. As described above
with large enough value of $\e$ the tensor sector becomes completely anisotropic. Thus our assumption of small anisotropy is now a purely technical simplification. It is interesting to generalize our perturbative technique in analyzing the tensor perturbations   into a non-perturbative calculation, in which case larger cross-correlations with more significant observational signatures can be calculated.

 Based on the correlations $\langle \zeta \zeta \rangle $, $\langle h_{s}  h_{s} \rangle$
 and $\langle \zeta h_s \rangle $ obtained in this model, we have calculated the observational signatures on the CMB temperature and polarization maps. Before the BICEP2 detection of B-modes polarization, the probe of statistical anisotropies in the cosmological perturbations were confined to the TT, TE and EE maps, which are dominated by the anisotropy of the scalar sector. However, this statement is now changed after  the recent detection of the B-modes. The TB, EB and BB correlations are now available for testing the anisotropies. Especially, we have shown that when the gauge interaction dominates over  the interaction between the inflaton and the gauge field, the tensor sector can be more anisotropic than the scalar sector, resulting in enhanced correlations involving B-modes. For example, in Fig~\ref{fig:TBEB1} one can observe that the TB correlation in the model with the
 charged coupling can be order-of-magnitude larger compared to model with no electric charge coupling,  which is also considerably larger than the size of the BB correlation near the reionization bump at $l\lesssim 10$.

Having this said note that the sensitivities on different CMB correlations we studied here also crucially depend on the design of experiments and the proprieties of the detectors. It would be very interesting to perform experiment-specific investigations and dig into data to hunt for the anisotropic correlations.

%%%%%%%%%%%%%%%%%%%%%%%%%%%%%%%%%%%%%%%%%%%%%%%%%%%%%

\section*{Acknowledgment}
It is a pleasure to thank A. A. Abolhasani, S. Baghram,   N. Bartolo,  J. Chluba,  D. Jeong,  E. Komatsu,  M. Peloso  and  S.-C.~Su for helpful discussion. YW is supported by a Starting Grant of the European Research Council (ERC STG grant 279617), and the Stephen Hawking Advanced Fellowship.

%%%%%%%%%%%%%%%%%%%%%%%%%%%%%%%%%%%%%%%%%%%%%%%%%
\appendix

\section{ Interaction Lagrangians}
\label{int-lagrang}

In this Appendix we present the interaction Lagranians in Eqs. (\ref{Lzetah}) - (\ref{LhxzD}) in more details. Our starting point is the total Lagrangian
\ba
\label{interaction Lagrangian1}
L_{int} = -\frac{a^4}{4}f(\phi)^2 F_{\mu \nu} F^{\mu \nu} - \frac{a^4}{2} \e^2 \phi^2 A_{\mu} A^{\mu}
\ea
Expanding the above action around the background to second order in perturbations we get
\ba
\label{interaction Lagrangian2}
L_{int} &=& -2 f^2 A^{'2}_{x} \zeta h_{xx} + 4 f^2 A'_{x} \zeta \delta A'_{1} - 4f^2 A'_{x} \zeta \partial_{x}\delta
A_{0} - f^2 A'_{x} \delta A'_{1} h_{xx} + f^2 A'_{x} \partial_{x}\delta A_{0} h_{xx} + \frac{1}{2} f^2 \left( \partial_{x} \delta A_{0} \right)^2 \nonumber\\
&& -f^2 \delta A'_{1}\partial_{x}\delta A_{0} - f^2 A'_{x} \partial_{y} M' h_{xy} + f^2 A'_{x} h_{xy}\partial_{y} \delta A_{0} - f^2 A'_{x}h_{xz}D' + \frac{1}{2}f^2 \left( \partial_{y} \delta A_{0}\right)^2 -f^2\partial_{y}M' \partial_{y}\delta A_{0} \nonumber\\
&&+ \frac{1}{2}a^2 \e^2  \phi^2 \left(\delta A_{0} \right)^2 + a^2 \e^2 \phi^2 A_{x}\delta A_{1}h_{xx}
+ a^2 \e^2 \phi^2 A_{x} \partial_{y}M h_{xy} + a^2 \e^2 \phi^2 A_{x} D h_{xz} + a^2 \e^2 \phi A_{x}^2 \delta \phi h_{xx} \nonumber\\
&& - 2 a^2 \e^2 \phi A_{x} \delta \phi \delta A_{1}
\ea
Where we have used the following expression,
\ba
\left(\frac{\partial f^2}{\partial \phi}\right)\delta \phi = 4 f^2 \zeta
\ea
As we discussed in the main text,  we should integrate out the non-dynamical field $\delta A_{0}$. However, as we mentioned before, the resulting terms are sub-leading so we can safely neglect the contribution of $\delta A_{0}$ in Eq. (\ref{interaction Lagrangian2}).

Using the following useful formula
\ba
fA'_{x} &=& M_{P}\sqrt{3I\epsilon_{H}} (-\eta)^{-1} a \\
\e \phi A_x &=& M_{P}^2\e \sqrt{\frac{2I}{3}}\frac{a}{f} \\
\phi &=& M_{P}\sqrt{\frac{2}{\epsilon_{H}}} \\
f&=& \left(\frac{\eta^2}{\eta_{e}^2}\right)
\ea
In order to obtain the above equations, we have used the background attractor solution results which is given in Sec. \ref{background formulas}.

We calculate $L_{\zeta h_+}, L_{\zeta D_1}, L_{D_1 h_+}$ and $L_{D h_\times}$ in turn.
For   $L_{\zeta h_+}$ we have
\ba
L_{\zeta h_+} &=& -f^2 A^{'2}_{x} \left( \zeta^{*} h_{xx} + \zeta h^{*}_{xx} \right) - \frac{1}{2} \e^2 a^2 A_{x}^2\phi  \frac{\dot \phi}{H}  \left( \zeta^{*} h_{xx} + \zeta h^{*}_{xx} \right)\nonumber\\
&=& - \frac{3\sqrt{2}}{2} I \epsilon_{H} M_{P}^2\sin^2{\theta}a^2\left(-\eta\right)^{-2}\left( \zeta^{*} {h}_{+} + \zeta {h}^{*}_{+} \right)  + \frac{\e^2 \sqrt{2}}{6}  I \epsilon_{H}M_{P}^4 \sin^2{\theta} \left(\frac{a ^4}{f^2}\right)\left( \zeta^{*} {h}_{+} + \zeta {h}^{*}_{+} \right)
\ea

For $L_{\zeta D_1}$ we have
\ba
L_{\zeta D_1} &\simeq &  L_{\zeta \delta A_1}=
2f^2 A'_{x} \left( \zeta^{*} \delta A'_{1} + \zeta \delta A^{'*}_{1}\right) + \e^2 a^2 A_{x}\phi  \frac{\dot \phi}{H}  \left( \zeta^{*} \delta A_{1} + \zeta \delta A^{*}_{1} \right)  \nonumber\\
&=& -2 M_{P}\sqrt{3I \epsilon_{H}}\sin^2{\theta} \left(\frac{a f}{\eta}\right) \left( \zeta ^{*} D'_{1} + c.c.\right) - 2 \e^2 M_{P}^3 \sqrt{\frac{I \epsilon_{H}}{3}}\sin^2{\theta} \left(\frac{a^3}{f}\right) \left( \zeta ^{*} D_{1} + c.c.\right)
\ea
where we have neglected the longitudinal mode $D_2$ so  $\delta A_1 \simeq  D_1 \sin \theta^2$.

Similarly, to calculate $L_{D_1 h_+}$ we have to calculate $L_{h_{xx} \delta A_{1}}$  and $L_{h_{xy} \delta A_{2}}$
which respectively are
\ba
L_{h_{xx} \delta A_{1}} &=& -\frac{1}{2} f^2 A'_{x} \left( \delta A^{'*}_{1} h_{xx} +  \delta A^{'}_{1} h^{*}_{xx}\right) + \frac{1}{2}\e^2 a^2 \phi^2 A_{x} \left( \delta A^{*}_{1} h_{xx} +  \delta A_{1} h^{*}_{xx}\right) \nonumber\\
&=& \frac{M_{P}}{2} \sqrt{\frac{3I\epsilon_{H}}{2}}\sin^4{\theta}\left( \frac{fa}{\eta}\right) \left( D^{'*}_{1} {h}_{+} +  c.c. \right) + \sqrt{\frac{I}{6\epsilon_{H}}} \e^2 M_{P}^3 \sin^4{\theta}\left( \frac{a^3}{f}\right)\left( D^{*}_{1} {h}_{+} +  c.c. \right)
\ea
and
\ba
L_{h_{xy} \delta A_{2}} &=& -\frac{1}{2} f^2 A'_{x} k_{y} \left( i M^{'} h^{*}_{xy} -i   M^{'*} h_{xy}\right) + \frac{1}{2}\e^2 a^2 \phi^2 A_{x} k_{y}\left( i M h^{*}_{xy} -i   M^{*} h_{xy}\right) \nonumber\\
&=& \frac{M_{P}}{2} \sqrt{\frac{3I\epsilon_{H}}{2}} \sin^2{\theta}\cos^2{\theta}\left( \frac{fa}{\eta}\right)\left(  D^{'}_{1} {h}^{*}_{+} + c.c.\right) + \sqrt{\frac{I}{6\epsilon_{H}}} \e^2 M_{P}^3 \sin^2{\theta}\cos^2{\theta}\left( \frac{a^3}{f}\right)\left(  D_{1} {h}^{*}_{+} + c.c. \right)
\ea
where the relation $M \simeq (i/k) \cos \theta D_1$ have been used in the limit where
we neglect the longitudinal mode.  Combining  $L_{h_{xx} \delta A_{1}}$  and $L_{h_{xy} \delta A_{2}}$ we obtain   $L_{D_1 h_+}$
as in Eq. (\ref{Lh+D1}).

Finally, to calculate $L_{D h_\times}$ we have to calculate $L_{h_{xz} \delta A_3}$ which is
\ba
L_{D h_\times} &=&
L_{h_{xz} \delta A_{3}}= -\frac{1}{2} f^2 A'_{x} \left( D^{'} h^{*}_{xz} +   D^{'*} h_{xz}\right) +
\frac{1}{2}\e^2 a^2 \phi^2 A_{x} \left(D h^{*}_{xz} +   D^{*} h_{xz}\right) \nonumber\\
&=& \frac{M_{P}}{2} \sqrt{\frac{3I \epsilon_{H}}{2}}\sin{\theta} \left(\frac{fa}{\eta}\right)\left( i D^{'} {h}^{*}_{\times} + c.c.\right) + \sqrt{\frac{I}{6\epsilon_{H}}}\e^2 M_{P}^3 \sin{\theta} \left(\frac{a^3}{f}\right)\left( i D {h}^{*}_{\times} + c.c. \right)
\ea

%%%%%%%%%%%%%%%%%%%%%%%%%%%%%%%%%%%%%%%%%%%%%%%%%%%%%%%%%%%

\section{ Interaction Hamiltonian}
\label{int-hamilton}
In this appendix, we are going to calculate the interaction Hamiltonian in our model which is required to proceed with in-in formalism. Since in this model, we do have kinetically coupled fields, one can be worried about the relation $H_{int} = - L_{int}$. So it is worth to calculate it by bruce force. We skip the details and only mention the final result for the interaction Hamiltonian.
\begin{align}
\label{int Hamil}
H_{int} &=  H_{\zeta h_+} + H_{\zeta D_1} + H_{D_{1} h_{+}} + H_{D h_\times}
\end{align}
Where we have {\color{black}{
\begin{align}
\label{Hzetah}
H_{\zeta h_+} & = - \frac{3\sqrt{2}}{2} I \epsilon_{H} M_{P}^2\sin^2{\theta}a^2\left(-\eta\right)^{-2}\left( \zeta^{*} {h}_{+} + \zeta {h}^{*}_{+} \right)  -  \frac{\e^2 \sqrt{2}}{6}  I \epsilon_{H}M_{P}^4 \sin^2{\theta} \left(\frac{a ^4}{f^2}\right)\left( \zeta^{*} {h}_{+} + \zeta {h}^{*}_{+} \right)
\\
\label{HzetaD1}
H_{\zeta D_1} & = +2 M_{P}\sqrt{3I \epsilon_{H}}\sin^2{\theta} \left(\frac{a f}{\eta}\right) \left( \zeta ^{*} D'_{1} + c.c.\right) + 2 \e^2 M_{P}^3 \sqrt{\frac{I \epsilon_{H}}{3}}\sin^2{\theta} \left(\frac{a^3}{f}\right) \left( \zeta ^{*} D_{1} + c.c.\right) \nonumber\\
& = - L_{\zeta D_1}  \\
\label{HD1h}
H_{h_+  D_1} &= -\frac{M_{P}}{2} \sqrt{\frac{3I\epsilon_{H}}{2}}\sin^2{\theta}\left( \frac{fa}{\eta}\right) \left( D^{'*}_{1} {h}_{+} +  c.c.\right) - \sqrt{\frac{I}{6\epsilon_{H}}} \e^2 M_{P}^3 \sin^2{\theta}\left( \frac{a^3}{f}\right)\left( D^{*}_{1} {h}_{+} + c.c.\right)\nonumber\\
& = - L_{h_+ D_1}  \\
\label{HDh}
H_{h_\times  D} &= -\frac{M_{P}}{2} \sqrt{\frac{3I \epsilon_{H}}{2}}\sin{\theta} \left(\frac{fa}{\eta}\right)\left( i D^{'} {h}^{*}_{\times} + c.c.\right) - \sqrt{\frac{I}{6\epsilon_{H}}}\e^2 M_{P}^3 \sin{\theta} \left(\frac{a^3}{f}\right)\left( i D{h}^{*}_{\times} + c.c. \right)\nonumber\\
& = - L_{h_\times D}
\end{align}}}
As a result, we see that $H_{int} = - L_{int}$ is not generally true for all kinetically coupled interactions. Especially for $H_{\zeta h_+}$, we can not use $H_{int} = - L_{int}$. We should notice that in the above analysis, we have neglected the mass terms. Because it is shown in \cite{Emami:2013bk} to a very good approximation, all of the fields are nearly massless in this model.

%%%%%%%%%%%%%%%%%%%%%%%%%%%%%%%%%%%%%%%%%%%%%%%%%%%%%

\section{ The in-in analysis}

\label{in-in}

Here we present the integral form of the in-in integrals in more details.

%%%%%%%%%%%%%%%%%%%%%%%%%%%%%%%%%%%%%%%%%%%%%%%%%%%%%
\subsection{In-In integrals for anisotropic power spectrum}

For the anisotropy corrections in power spectrum we have
\ba
\label{delta-P-zeta-b}
\delta \langle \zeta_\bfk \zeta_\bfk^* \rangle  = - \int_{\eta_{0}}^{\eta_{e}} d\eta_{1} \int_{\eta_{0}}^{\eta_{1}}d\eta_{2} \left \langle
\bigg{[} L_{I}(\eta_{2}) , \bigg{[} L_{I}(\eta_{1}) ,
 \zeta_\bfk (\eta_e) \zeta_\bfk^*(\eta_e)  \bigg{]}\bigg{]} \right \rangle  \, .
\ea
where the leading interaction Lagrangian is   $L_{\zeta D_1} =   L^{(1)}_{\zeta D_1} +L^{(2)}_{\zeta D_1}$
with
\ba
L^{(1)}_{\zeta D_1} \equiv
 -2 M_{P}\sqrt{3I \epsilon_{H}}\sin^2{\theta} \left(\frac{a f}{\eta}\right) \left( \zeta ^{*} D'_{1} + c.c.\right)
  \quad , \quad
L^{(2)}_{\zeta D_1} \equiv
- 2 \e^2 M_{P}^3 \sqrt{\frac{I \epsilon_{H}}{3}}\sin^2{\theta} \left(\frac{a^3}{f}\right) \left( \zeta ^{*} D_{1} + c.c.\right)
\ea
As explained in the main text, depending on whether one chooses either $L^{(1)}_{\zeta D_1}$ or $L^{(2)}_{\zeta D_1}$ in place of  $L_I(\eta_1)$ and $L_I(\eta_2)$ in the   integral Eq. (\ref{delta-P-zeta-b}), there are four different terms  in  $\delta \langle \zeta_\bfk \zeta_\bfk^* \rangle$ denoted by $\delta \langle \zeta_\bfk \zeta_\bfk^* \rangle_{ij}$ where $i=1, 2$ and with the  assumption  that
$L_I(\eta_1) = L^{(i)}_{\zeta D_1}$  and  $L_I(\eta_2) = L^{(j)}_{\zeta D_1}$.
For example,    $\delta \langle \zeta_\bfk \zeta_\bfk^* \rangle_{12}$ means
$L_I(\eta_1) = L^{(1)}_{\zeta D_1}$  and  $L_I(\eta_2) = L^{(2)}_{\zeta D_1}$. In total  we have
\ba
\label{correction zeta}
\delta  \bigg{\langle} \zeta_{\bfk}(\eta_{e}) \zeta_{\bfk}^*(\eta_{e})\bigg{\rangle} =
\delta \bigg{\langle} \zeta_{\bfk}(\eta_{e}) \zeta_{\bfk}^*(\eta_{e})\bigg{\rangle} _{11}
+\delta \bigg{\langle} \zeta_{\bfk}(\eta_{e}) \zeta_{\bfk}^*(\eta_{e})\bigg{\rangle} _{12}
+ \delta \bigg{\langle} \zeta_{\bfk}(\eta_{e}) \zeta_{\bfk}^*(\eta_{e})\bigg{\rangle} _{21}
+ \delta \bigg{\langle} \zeta_{\bfk}(\eta_{e}) \zeta_{\bfk}^*(\eta_{e})\bigg{\rangle} _{22}
\ea
We calculate each of them in turn.
\ba
\delta  \bigg{\langle} \zeta_{\mathbf{k}}(\eta_{e}) \zeta_{\mathbf{k}}^*(\eta_{e})\bigg{\rangle}_{11}  &&=  384 I \eH M_P^2  \sin^4{\theta} \
\int_{\eta_{0}}^{\eta_{e}} d\eta_{1} \left(\frac{a f}{\eta} \right)_{\eta_1}
\im \left[ \zeta_k (\eta_{1})\zeta_k^{*}(\eta_{e})\right]  \\
&& ~~~~~~~~~~~~~~~~~~~~~~~\times \int_{\eta_{0}}^{\eta_{1}}d\eta_{2}
 \left(\frac{a f}{\eta} \right)_{\eta_2}
\im \left[ \zeta_k(\eta_{2})\zeta_k^{*}(\eta_{e})  D_{1k}^{' *} (\eta_{1}) D_{1k}^{'}(\eta_{2}) \right]
\ea
As discussed in the main text, expanding the integrand for small $k \eta$ arguments and assuming $k \eta_0 = -1$ and $k \eta_e =0$ the above integral can be calculated analytically and we get
\ba
\delta  \bigg{\langle} \zeta_{\mathbf{k}}(\eta_{e}) \zeta_{\mathbf{k}}(\eta_{e})\bigg{\rangle}_{11}&=& \frac{6 I N^2}{k^3 \epsilon_{H}} \left(\frac{H}{M_{P}} \right)^2 \sin^2{\theta}
\ea
Similarly
\ba
\delta  \bigg{\langle} \zeta_{\mathbf{k}}(\eta_{e}) \zeta_{\mathbf{k}}^*(\eta_{e})\bigg{\rangle}_{12}  &&=  128 I \eH M_P^2  \e^2  \sin^4{\theta} \
\int_{\eta_{0}}^{\eta_{e}} d\eta_{1} \left(\frac{a f}{\eta} \right)_{\eta_1}
\im \left[ \zeta_k (\eta_{1})\zeta_k^{*}(\eta_{e})\right]  \\
&& ~~~~~~~~~~~~~~~~~~~~~~~\times \int_{\eta_{0}}^{\eta_{1}}d\eta_{2}
 \left(\frac{a^3}{f} \right)_{\eta_2}
\im \left[ \zeta_k(\eta_{2})\zeta_k^{*}(\eta_{e})  D_{1k}^{ '*} (\eta_{1}) D_{1k}(\eta_{2}) \right] \\
&=& -\frac{31 }{490}  \frac{\e^2 I}{k^3 {\epsilon_{H}}} \sin^2{\theta}  \, ,
\ea
\ba
\delta  \bigg{\langle} \zeta_{\mathbf{k}}(\eta_{e}) \zeta_{\mathbf{k}}^*(\eta_{e})\bigg{\rangle}_{21}  &&=  128 I \eH M_P^2  \e^2  \sin^4{\theta} \
\int_{\eta_{0}}^{\eta_{e}} d\eta_{1}     \left(\frac{a^3}{f} \right)_{\eta_1}  \im \left[ \zeta_k (\eta_{1})\zeta_k^{*}(\eta_{e})\right]  \\
&& ~~~~~~~~~~~~~~~~~~~~~~~\times \int_{\eta_{0}}^{\eta_{1}}d\eta_{2}
   \left(\frac{a f}{\eta} \right)_{\eta_2}
\im \left[ \zeta_k(\eta_{2})\zeta_k^{*}(\eta_{e})  D_{1k}^{ *} (\eta_{1}) D_{1k}'(\eta_{2}) \right] \\
&=& -\frac{ I \e^2 N}{7 k^3 \epsilon_H}   \sin^2{\theta}  \, .
\ea
Finally
\ba
\delta  \bigg{\langle} \zeta_{\mathbf{k}}(\eta_{e}) \zeta_{\mathbf{k}}^*(\eta_{e})\bigg{\rangle}_{22}  &&=  \frac{128}{3} I \eH M_P^6  \e^4  \sin^4{\theta} \
\int_{\eta_{0}}^{\eta_{e}} d\eta_{1}     \left(\frac{a^3}{f} \right)_{\eta_1}  \im \left[ \zeta_k (\eta_{1})\zeta_k^{*}(\eta_{e})\right]  \\
&& ~~~~~~~~~~~~~~~~~~~~~~~\times \int_{\eta_{0}}^{\eta_{1}}d\eta_{2}
  \left(\frac{a^3}{f} \right)_{\eta_2}
\im \left[ \zeta_k(\eta_{2})\zeta_k^{*}(\eta_{e})  D_{1k}^{ *} (\eta_{1}) D_{1k}(\eta_{2}) \right] \\
 &=& \frac{9}{2156}  \frac{\e^4 I}{k^3}\left(\frac{M_{P}}{m} \right)^2 \sin^2{\theta} \, .
\ea

%%%%%%%%%%%%%%%%%%%%%%%%%%%%%%%%%%%%%%%%%%%%%%%%%%%%%
\subsection{In-in for tensor power spectra}

Now we calculate the anisotropy in tensor power spectra
 $\langle {h}_{\bfk \times } {h}_{\bfk \times }^*\rangle$ and $\langle {h}_{\bfk + } {h}_{\bfk + }^*\rangle$.

Let us  start with  $\langle {h}_{\times } {h}_{\times }^*\rangle$.  The interaction Lagrangian is
$L_{D {h}_{\times } } =  L_{D {h}_{\times } }^{(1)} +  L_{D {h}_{\times } }^{(2)} $ where
$ L_{D {h}_{\times } }^{(1)}$ and  $ L_{D {h}_{\times } }^{(2)}$ respectively are the first term and the second term in Eq. (\ref{LhxzD}).   Following the same convention as  in anisotropy analysis for curvature perturbation in power spectrum we have
\ba
\delta \bigg{\langle} {h}_{\times } {h}_{\times }^*\bigg{\rangle} &=& - \int_{\eta_{0}}^{\eta_{e}} d\eta_{1}\int_{\eta_{0}}^{\eta_{1}} d\eta_{2}\bigg{[}L_{D {h}_{\times }  } ,\bigg{[}L_{D {h}_{\times}}, {h}_{\times\mathbf{k}} {h}_{\times\mathbf{k}}\bigg{]}\bigg{]} \nonumber\\
&=& \delta \bigg{\langle} \widehat{h}_{\times }\widehat{h}_{\times }\bigg{\rangle}_{11} +  \delta \bigg{\langle} \widehat{h}_{\times }\widehat{h}_{\times }\bigg{\rangle}_{12} + \delta \bigg{\langle} \widehat{h}_{\times }\widehat{h}_{\times }\bigg{\rangle}_{21} + \delta \bigg{\langle} \widehat{h}_{\times }\widehat{h}_{\times }\bigg{\rangle}_{22}
\ea
The results for each contribution are
\ba
\label{hhtimes correction-b}
\label{tensor times 1}
\delta \bigg{\langle} {h}_{\times\mathbf{k}}  {h}_{\times\mathbf{k}}^*\bigg{\rangle}_{11}&=& -\left(12 I \epsilon_{H} M_{P}^2 \right) \sin^2{\theta} \int_{\eta_{0}}^{\eta_{e}} d\eta_{1}  \left(\frac{a f}{\eta}\right)_{\eta_{1}}\int_{\eta_{0}}^{\eta_{1}} d\eta_{2} \left(\frac{a f}{\eta}\right)_{\eta_{2}} {\rm Im}\left( {h}_{\times}(\eta_{1}){h}^{*}_{\times}(\eta_{e})\right) \times \nonumber\\
&& \times {\rm Im}\left({h}_{\times}(\eta_{2}) {h}_{\times}^{*}(\eta_{e}) D'(\eta_{2})D^{'*}(\eta_{1})\right) \nonumber\\
&=& \left(\frac{12}{k_1^3}\right)\left(\frac{H}{M_{P}}\right)^2 I \epsilon_{H} N^2\sin^2{\theta} \\
\label{tensor times 2}
\delta \bigg{\langle} {h}_{\times\mathbf{k}}  {h}_{\times\mathbf{k}}^*\bigg{\rangle}_{12}&=& -\left(8 I M_{P}^4 \right) \e^2 \sin^2{\theta} \int_{\eta_{0}}^{\eta_{e}} d\eta_{1}  \left(\frac{a^3 }{f}\right)_{\eta_{1}}\int_{\eta_{0}}^{\eta_{1}} d\eta_{2} \left(\frac{a f}{\eta}\right)_{\eta_{2}}{\rm Im}\left( {h}_{\times}(\eta_{1}){h}^{*}_{\times}(\eta_{e})\right) \times \nonumber\\
&& \times {\rm Im}\left( {h}_{\times}(\eta_{2}){h}_{\times}^{*}(\eta_{e})D'(\eta_{2})D^{*}(\eta_{1})\right) \nonumber\\
&=& -\left(\frac{4}{7 k^3}\right)N I\e^2\sin^2{\theta}\\
\label{tensor times 3}
\delta \bigg{\langle} {h}_{\times\mathbf{k}}  {h}_{\times\mathbf{k}}^*\bigg{\rangle}_{21}&=& -\left(8 I M_{P}^4 \right) \e^2 \sin^2{\theta} \int_{\eta_{0}}^{\eta_{e}} d\eta_{1}\left(\frac{a f}{\eta}\right)_{\eta_{1}}  \int_{\eta_{0}}^{\eta_{1}} d\eta_{2}\left(\frac{a^3 }{f}\right)_{\eta_{2}} {\rm Im}\left( {h}_{\times}(\eta_{1}){h}^{*}_{\times}(\eta_{e})\right) \times \nonumber\\
&& \times {\rm Im}\left( {h}_{\times}(\eta_{2}){h}_{\times}^{*}(\eta_{e})D(\eta_{2})D^{'*}(\eta_{1})\right) \nonumber\\
&=& -\left(\frac{62}{245 k^3}\right) I \e^2 \sin^2{\theta}
\ea
\ba
\label{tensor times 4}
\delta \bigg{\langle} {h}_{\times\mathbf{k}}  {h}_{\times\mathbf{k}}^*\bigg{\rangle}_{22}&=& -\left(\frac{16}{3} I M_{P}^6 \right)\e^4 \sin^2{\theta} \int_{\eta_{0}}^{\eta_{e}} d\eta_{1}\left(\frac{a^3 }{f}\right)_{\eta_{1}}\int_{\eta_{0}}^{\eta_{1}} d\eta_{2} \left(\frac{a^3}{f}\right)_{\eta_{2}}{\rm Im}\left( {h}_{\times}(\eta_{1}){h}^{*}_{\times}(\eta_{e})\right) \times \nonumber\\
&& \times {\rm Im}\left({h}_{\times}(\eta_{2}) {h}_{\times}^{*}(\eta_{e}) D(\eta_{2}) D^{*}(\eta_{1}) \right)\nonumber\\
&=& \left(\frac{6}{539 k^3 }\right) \left(\frac{M_{P}}{H}\right)^2 \left(\frac{I \e^4}{\epsilon_{H}}\right) \sin^2{\theta}
\ea
So finally we get,
\ba
\label{correction hh times final-b}
\delta \bigg{\langle} {h}_{\times\mathbf{k}_{1}}  {h}_{\times\mathbf{k}_{2}}\bigg{\rangle} \simeq \left( \left(12I \epsilon_{H}N^2\right)\left(\frac{H}{M_{P}}\right)^2 -\left(\frac{4}{7}N I\e^2\right) + \left(\frac{6}{539}\right) \left(\frac{M_{P}}{H}\right)^2 \left(\frac{I \e^4}{\epsilon_{H}}\right)\right)\left( \frac{\sin^2{\theta}}{k^3}\right)
\ea

In addition, we have to calculate   $\langle {h}_{+ }{h}_{+}^*\rangle$. In this case the relevant interaction Lagrangians are  $L_{D_{1} {h}_{+ } }$ and $ L_{\zeta {h}_{+ }  }$ so we have
\ba
\label{tensor plus}
\delta \bigg{\langle} {h}_{+} {h}_{+}\bigg{\rangle} &=& - \int_{\eta_{0}}^{\eta_{e}} d\eta_{1}\int_{\eta_{0}}^{\eta_{1}} d\eta_{2}\bigg{[}L_{D_{1} {h}_{+ }  } ,\bigg{[}L_{D_{1} {h}_{+}}, {h}_{+} {h}_{+}\bigg{]}\bigg{]} - \int_{\eta_{0}}^{\eta_{e}} d\eta_{1}\int_{\eta_{0}}^{\eta_{1}} d\eta_{2}\bigg{[}L_{\zeta {h}_{+ }  } ,\bigg{[}L_{\zeta {h}_{+}}, {h}_{+} {h}_{+}\bigg{]}\bigg{]}
\ea
As mentioned in the main text, comparing $L_{D_{1} {h}_{+ } }$ and $ L_{\zeta {h}_{+ }  }$ we see that $ L_{\zeta {h}_{+ }  }$ is suppressed compared to $L_{D_{1} {h}_{+ } }$ by a factor $\sqrt{I} \ll 1$
so to leading order in anisotropy we can neglect the contribution from  $ L_{\zeta {h}_{+ }  }$ in
$\langle {h}_{+ }{h}_{+}^*\rangle$. As a result the analysis is exactly the same as in  the case of  $\langle {h}_{\times }{h}_{\times}^*\rangle$ and therefore
\ba
\label{tensor plus final-b}
\delta \bigg{\langle} {h}_{+} {h}_{+}\bigg{\rangle} =
\delta \bigg{\langle} {h}_{\times} {h}_{\times}\bigg{\rangle} \, .
\ea
%%%%%%%%%%%%%%%%%%%%%%%%%%%%%%%%%%%%%%%%%%%%%%%%%%%%%%%%
\subsection{In-in for  scalar-tensor cross-correlation}
Here we present the in-in analysis for the cross-correlation $\langle \zeta h_{s} \rangle $
for $s =\pm$.  The corresponding in-in integrals are
\ba
\bigg{\langle} \zeta_{\mathbf{k}_{1}}(\eta_{e}) {h}_{+\mathbf{k}_{2}}(\eta_{e})\bigg{\rangle} &=& i \int_{\eta_{0}}^{\eta_{e}} d\eta_{1}\bigg{[} H_{\zeta {h}_{+}} , \zeta_{\mathbf{k}_{1}}{h}_{+\mathbf{k}_{2}}\bigg{]} - \int_{\eta_{0}}^{\eta_{e}} d\eta_{1}\int_{\eta_{0}}^{\eta_{1}} d\eta_{2}\bigg{[}L_{\zeta D_{1}} ,\bigg{[}L_{D_{1} {h}_{+}}, \zeta_{\mathbf{k}_{1}}{h}_{+\mathbf{k}_{2}}\bigg{]}\bigg{]} \nonumber\\
&& -\int_{\eta_{0}}^{\eta_{e}} d\eta_{1}\int_{\eta_{0}}^{\eta_{1}} d\eta_{2}\bigg{[} L_{D_{1} \widehat{h}_{+}},\bigg{[} L_{\zeta D_{1}}, \zeta_{\mathbf{k}_{1}}{h}_{+\mathbf{k}_{2}}\bigg{]}\bigg{]} \nonumber\\
&=& \bigg{\langle} \zeta_{\mathbf{k}_{1}}(\eta_{e}) {h}_{+\mathbf{k}_{2}}(\eta_{e})\bigg{\rangle}_{1} + \bigg{\langle} \zeta_{\mathbf{k}_{1}}(\eta_{e}) {h}_{+\mathbf{k}_{2}}(\eta_{e})\bigg{\rangle}_{2} + \bigg{\langle} \zeta_{\mathbf{k}_{1}}(\eta_{e}) {h}_{+\mathbf{k}_{2}}(\eta_{e})\bigg{\rangle}_{3}
\ea
In the following, we calculate the above cross-correlation step by step,
\ba
\label{cross1}
\bigg{\langle} \zeta_{\mathbf{k}}(\eta_{e}) {h}_{+\mathbf{k}}(\eta_{e})^*\bigg{\rangle}_{1} &=&2i M_{P}^2 \left(I \epsilon_{H} \sin^2{\theta} \right)\int_{\eta_{0}}^{\eta_{e}} d\eta_{1} \bigg{(}-3\sqrt{2}\left(\frac{a^2}{\eta^2} \right) -  \frac{\e^2M_{P}^2 }{3} \left(\frac{a^4}{f^2} \right) \bigg{)}_{\eta_{1}} {\rm Im}\left( \zeta(\eta_{1}) \zeta^{*}(\eta_{e}){h}_{+}(\eta_{1}){h}^{*}_{+}(\eta_{e})\right) \nonumber\\
&=& \frac{2I}{3k_1^3} \left( \frac{H}{M_{P}} \right)^2  \left( 3\sqrt{2} N + \frac{\e^2 M_{P}^2}{28 H^2} \right)\sin^2{\theta} \\
&&\nonumber\\
\label{cross2}
\bigg{\langle} \zeta_{\mathbf{k}}(\eta_{e}) {h}_{+\mathbf{k}}(\eta_{e})^*\bigg{\rangle}_{2} &=&
- \int_{\eta_{0}}^{\eta_{e}} d\eta_{1}\int_{\eta_{0}}^{\eta_{1}} d\eta_{2}\bigg{[} L_{\zeta D_{1}}
,\bigg{[}L_{D_{1} {h}_{+}}, \zeta_{\mathbf{k}_{1}}\widehat{h}_{+\mathbf{k}_{2}}\bigg{]}\bigg{]}\nonumber\\
&\equiv& \bigg{\langle} \zeta_{\mathbf{k}_{1}} {h}_{+\mathbf{k}_{2}}\bigg{\rangle}_{21} + \bigg{\langle} \zeta_{\mathbf{k}_{1}} {h}_{+\mathbf{k}_{2}}\bigg{\rangle}_{22} + \bigg{\langle} \zeta_{\mathbf{k}_{1}} {h}_{+\mathbf{k}_{2}}\bigg{\rangle}_{23} + \bigg{\langle} \zeta_{\mathbf{k}_{1}}
\widehat{h}_{+\mathbf{k}_{2}}\bigg{\rangle}_{24}\\
&&\nonumber\\
\label{cross3}
\bigg{\langle} \zeta_{\mathbf{k}}(\eta_{e}) {h}_{+\mathbf{k}}(\eta_{e})^*\bigg{\rangle}_{3} &=&
- \int_{\eta_{0}}^{\eta_{e}} d\eta_{1}\int_{\eta_{0}}^{\eta_{1}} d\eta_{2}\bigg{[} L_{D_{1} \widehat{h}_{+}}
,\bigg{[}L_{\zeta D_{1}}, \zeta_{\mathbf{k}_{1}}{h}_{+\mathbf{k}_{2}}\bigg{]}\bigg{]}\nonumber\\
&\equiv& \bigg{\langle} \zeta_{\mathbf{k}_{1}} {h}_{+\mathbf{k}_{2}}\bigg{\rangle}_{31} + \bigg{\langle} \zeta_{\mathbf{k}_{1}} {h}_{+\mathbf{k}_{2}}\bigg{\rangle}_{32} + \bigg{\langle} \zeta_{\mathbf{k}_{1}} {h}_{+\mathbf{k}_{2}}\bigg{\rangle}_{33} + \bigg{\langle} \zeta_{\mathbf{k}_{1}}
{h}_{+\mathbf{k}_{2}}\bigg{\rangle}_{34}
\ea
where we have defined $\bigg{\langle} \zeta_{\mathbf{k}_{1}} {h}_{+\mathbf{k}_{2}}\bigg{\rangle}_{2i}$ as,
\ba
\label{cross21}
\bigg{\langle} \zeta_{\mathbf{k}}(\eta_{e}) {h}_{+\mathbf{k}}(\eta_{e})^*\bigg{\rangle}_{21}&=& \left(24 I \epsilon_{H} M_{P}^2 \sqrt{2} \right) \sin^4{\theta} \int_{\eta_{0}}^{\eta_{e}} d\eta_{1}  \left(\frac{a f}{\eta}\right)_{\eta_{1}}\int_{\eta_{0}}^{\eta_{1}} d\eta_{2} \left(\frac{a f}{\eta}\right)_{\eta_{2}} {\rm Im}\left( \widehat{h}_{+}(\eta_{1})\widehat{h}^{*}_{+}(\eta_{e})\right) \times \nonumber\\
&& \times {\rm Im}\left( \zeta(\eta_{2})\zeta^{*}(\eta_{e})D'_{1}(\eta_{2})D^{'*}_{1}(\eta_{1})\right) \nonumber\\
&=& - \left(\frac{3\sqrt{2}}{k_1^3}\right)\left(\frac{H}{M_{P}}\right)^2 I N^2\sin^2{\theta} \\
\label{cross22}
\bigg{\langle} \zeta_{\mathbf{k}}(\eta_{e}) {h}_{+\mathbf{k}}(\eta_{e})^*\bigg{\rangle}_{22}&=& \left(16 I \sqrt{2}M_{P}^4 \right) \e^2 \sin^4{\theta} \int_{\eta_{0}}^{\eta_{e}} d\eta_{1}  \left(\frac{a^3 }{f}\right)_{\eta_{1}}\int_{\eta_{0}}^{\eta_{1}} d\eta_{2} \left(\frac{a f}{\eta}\right)_{\eta_{2}}{\rm Im}\left( \widehat{h}_{+}(\eta_{1})\widehat{h}^{*}_{+}(\eta_{e})\right) \times \nonumber\\
&& \times {\rm Im}\left( \zeta(\eta_{2})\zeta^{*}(\eta_{e})D'_{1}(\eta_{2})D^{*}_{1}(\eta_{1})\right) \nonumber\\
&=& \left(\frac{4\sqrt{2}}{3 \epsilon_{H}k_1^3}\right) \left( \frac{3}{28}N - \frac{361}{3920} \right)I\e^2\sin^2{\theta} \\
\label{cross23}
\bigg{\langle} \zeta_{\mathbf{k}}(\eta_{e}) {h}_{+\mathbf{k}}(\eta_{e})^*\bigg{\rangle}_{23}&=& \left(8 I \epsilon_{H} \sqrt{2}M_{P}^4 \right)\e^2 \sin^4{\theta} \int_{\eta_{0}}^{\eta_{e}} d\eta_{1}\left(\frac{a f }{\eta}\right)_{\eta_{1}}\int_{\eta_{0}}^{\eta_{1}} d\eta_{2} \left(\frac{a^3}{f}\right)_{\eta_{2}}{\rm Im}\left( \widehat{h}_{+}(\eta_{1})\widehat{h}^{*}_{+}(\eta_{e})\right) \times \nonumber\\
&& \times {\rm Im}\left( \zeta(\eta_{2})\zeta^{*}(\eta_{e})D_{1}(\eta_{2})D^{'*}_{1}(\eta_{1})\right) \nonumber\\
&=& \left(\frac{31 \sqrt{2}}{980 k_1^3}\right) I \e^2 \sin^2{\theta} \\
\label{cross24}
\bigg{\langle} \zeta_{\mathbf{k}}(\eta_{e}) {h}_{+\mathbf{k}}(\eta_{e})^*\bigg{\rangle}_{24}&=& \left(\frac{16}{3} I \sqrt{2} M_{P}^6 \right)\e^4 \sin^4{\theta} \int_{\eta_{0}}^{\eta_{e}} d\eta_{1}\left(\frac{a^3 }{f}\right)_{\eta_{1}}\int_{\eta_{0}}^{\eta_{1}} d\eta_{2} \left(\frac{a^3}{f}\right)_{\eta_{2}}{\rm Im}\left( \widehat{h}_{+}(\eta_{1})\widehat{h}^{*}_{+}(\eta_{e})\right) \times \nonumber\\
&& \times {\rm Im}\left( \zeta(\eta_{2})\zeta^{*}(\eta_{e})D_{1}(\eta_{2})D^{*}_{1}(\eta_{1})\right)\nonumber\\
&=& -\left(\frac{3\sqrt{2}}{2156 k_1^3 \epsilon_{H}}\right) \left(\frac{M_{P}}{H}\right)^2 I \e^4 \sin^2{\theta}
\ea
Where we have defined $\ln{\left(-k_1\eta_{e}\right)} = -N$.\\

In addition, we have defined $\bigg{\langle} \zeta_{\mathbf{k}_{1}} {h}_{+\mathbf{k}_{2}}\bigg{\rangle}_{3i}$ as,

\ba
\label{cross31}
\bigg{\langle} \zeta_{\mathbf{k}}(\eta_{e}) {h}_{+\mathbf{k}}(\eta_{e})^*\bigg{\rangle}_{31}&=& \left(24 I \epsilon_{H} \sqrt{2} M_{P}^2\right) \sin^4{\theta} \int_{\eta_{0}}^{\eta_{e}} d\eta_{1}  \left(\frac{a f}{\eta}\right)_{\eta_{1}}\int_{\eta_{0}}^{\eta_{1}} d\eta_{2} \left(\frac{a f}{\eta}\right)_{\eta_{2}} {\rm Im}\left( \zeta(\eta_{1})\zeta^{*}(\eta_{e})\right) \times \nonumber\\
&& \times {\rm Im}\left({h}_{+}(\eta_{2}){h}^{*}_{+}(\eta_{e})D'_{1}(\eta_{2})D^{'*}_{1}(\eta_{1})\right) \nonumber\\
&=& - \left(\frac{3\sqrt{2}}{k^3}\right)\left(\frac{H}{M_{P}}\right)^2 I N^2\sin^2{\theta} \\
\label{cross32}
\bigg{\langle} \zeta_{\mathbf{k}}(\eta_{e}) {h}_{+\mathbf{k}}(\eta_{e})^*\bigg{\rangle}_{32}&=& \left(16 I \sqrt{2}M_{P}^4 \right) \e^2 \sin^4{\theta} \int_{\eta_{0}}^{\eta_{e}} d\eta_{1}  \left(\frac{a f}{\eta}\right)_{\eta_{1}}\int_{\eta_{0}}^{\eta_{1}} d\eta_{2} \left(\frac{a^3 }{f}\right)_{\eta_{2}}{\rm Im}\left( \zeta(\eta_{1})\zeta^{*}(\eta_{e})\right)
 \times \nonumber\\
&& \times {\rm Im}\left( {h}_{+}(\eta_{2})\widehat{h}^{*}_{+}(\eta_{e}) D_{1}(\eta_{2})D^{'*}_{1}(\eta_{1})\right) \nonumber\\
&=& \left(\frac{31 \sqrt{2}}{490 k^3}\right) I \e^2 \sin^2{\theta} \\
\label{cross33}
\bigg{\langle} \zeta_{\mathbf{k}}(\eta_{e}) {h}_{+\mathbf{k}}(\eta_{e})^*\bigg{\rangle}_{33}&=& \left(8 I \epsilon_{H} \sqrt{2}M_{P}^4 \right)\e^2 \sin^4{\theta} \int_{\eta_{0}}^{\eta_{e}} d\eta_{1}\left(\frac{a^3 }{f}\right)_{\eta_{1}}\int_{\eta_{0}}^{\eta_{1}} d\eta_{2} \left(\frac{a f}{\eta}\right)_{\eta_{2}} {\rm Im}\left( \zeta(\eta_{1})\zeta^{*}(\eta_{e})\right) \times \nonumber\\
&& \times {\rm Im}\left( {h}_{+}(\eta_{2}){h}^{*}_{+}(\eta_{e}) D_{1}^{'}(\eta_{2})D^{*}_{1}(\eta_{1})\right) \nonumber\\
&=& \left(\frac{2\sqrt{2}}{3 \epsilon_{H}k^3}\right) \left( \frac{3}{28}N - \frac{361}{3920} \right)I\e^2\sin^2{\theta}
\ea
\ba
\label{cross34}
\bigg{\langle} \zeta_{\mathbf{k}}(\eta_{e}) {h}_{+\mathbf{k}}(\eta_{e})^*\bigg{\rangle}_{34}&=& \left(\frac{16}{3} I \sqrt{2} M_{P}^6 \right)\e^2 \sin^4{\theta} \int_{\eta_{0}}^{\eta_{e}} d\eta_{1}\left(\frac{a^3 }{f}\right)_{\eta_{1}}\int_{\eta_{0}}^{\eta_{1}} d\eta_{2} \left(\frac{a^3}{f}\right)_{\eta_{2}}
{\rm Im}\left( \zeta(\eta_{1})\zeta^{*}(\eta_{e})\right) \times \nonumber\\
&& \times {\rm Im}\left( {h}_{+}(\eta_{2}){h}^{*}_{+}(\eta_{e}) D_{1}(\eta_{2})D^{*}_{1}(\eta_{1})\right) \nonumber\\
&=& -\left(\frac{3\sqrt{2}}{2156 k^3 \epsilon_{H}}\right) \left(\frac{M_{P}}{H}\right)^2 I \e^4 \sin^2{\theta}
\ea

So adding these nine terms and assuming $N \gg 1$ we obtain
\ba
\label{final cross correlation-b}
\bigg{\langle} \zeta_{\mathbf{k}_{1}}(\eta_{e}) {h}_{+\mathbf{k}_{2}}(\eta_{e})\bigg{\rangle} &\simeq&
I \left(- 6\sqrt{2} \frac{ N^2 H^2 }{M_{P}^2}  + \frac{  \sqrt{2} \e^2 N}{7 \epsilon_{H} }  - \frac{3\sqrt{2} \e^4 }{1078 \epsilon_{H} } \frac{M_{P}^2}{H^2}  \right) \frac{\sin^2{\theta}}{k^3}
\ea

As for the other cross-correlation we have
\ba
\bigg{\langle} \zeta_{\mathbf{k}_{1}}(\eta_{e}) \widehat{h}_{\times\mathbf{k}_{2}}(\eta_{e})\bigg{\rangle}=0 \, .
\ea
This is because at the second order level $\zeta $ does not see $\widehat{h}_{\times}$.

\end{document}